%% file: VGGPhDThesisMainV4.tex
\documentclass[12pt]{report}
\usepackage{epsf}
\usepackage{graphicx}
\usepackage{amsfonts}
\usepackage{toc}

\begin{document}

\thispagestyle{empty}
\pagenumbering{roman}
\begin{center}
{\Large MIXED-SYMMETRY SHELL-MODEL CALCULATIONS \\}
{\Large IN NUCLEAR PHYSICS \\}
\vfill
A Dissertation \\
\quad \\
Submitted to the Graduate Faculty of the \\
Louisiana State University and \\
Agricultural and Mechanical College \\
in partial fulfillment of the \\
requirements for the degree of \\
Doctor of Philosophy \\
\quad \\
in \\
\quad \\
The Department of Physics and Astronomy \\
\vfill
by \\
Vesselin Gueorguiev Gueorguiev \\
M.S. Sofia University, 1992 \\
December, 2002 \\
\end{center}

\vfill
\noindent
\begin{center}
\vfill 
\vfill 
\vfill
\vfill
\end{center}
\pagebreak

\addcontentsline{toc}{chapter}{Dedication}
\chapter*{Dedication}

\quad 
At the end of my Ph.D. study program, I look back in time and think of
the events and people that have taught, encouraged and supported me in my
study of physics. I would never forget the events of Summer of 1982 that set
the direction of my profession. That year I finished middle school and had to
make a decision on a high school. Since I had shown interest in mathematics,
physics, and technology, my mother recommended that I apply to the National
Natural Science High School in Sofia, Bulgaria. I spent the whole summer
reviewing my school books in mathematics and physics. That was the first time
in my life that I had to concentrate on a broad range of information, extract
the essential elements, and commit them to memory. I discovered the joy and
satisfaction of learning, problem solving, and overcoming obstacles through
hard work. Due to the quality of the National Natural Science High School and
my superior performance, I was accepted in the Sofia University as a physics
student where I continued to acquire knowledge in physics and mathematics. I
am thankful to my school teachers, many of whom were professors in physics at
the Sofia University, for keeping me interested in physics and the natural
sciences. My interest in theoretical physics jelled during my final years at
the Sofia University where I attended many lectures on various subjects in
mathematical and theoretical physics. It was then that I became interested in
symmetries and group theory, and especially the newly emerging concept
of quantum (deformed) Lie algebras.

The next important event in my life was the choice of a professor for my
master thesis. I still remember going from one professor to another looking
for someone who was working on quantum algebras. Finally, I met my M.S.
advisor, Professor R. P. Roussev at the Institute of Nuclear Research and
Nuclear Energy of the Bulgarian Academy of Sciences, Sofia, Bulgaria. After a 
few short meetings with Professor Roussev and his coworkers, Professors P. P.
Raychev and A. I. Gueorguieva, I was given a paper, one of the fundamental
papers on the topic, to read and explain. Back then, I did not see this as
another test that I had to pass, rather I thought of it as an opportunity to
show what I had learned and what I was capable of doing. Not until many years
later did I appreciate that this was a defining time for me, one that enabled
me to continue working with and learning from Professor Roussev and his
colleagues, with all of whom I became a good friend. Working with them was
one of the best research experiences in my life. I am very thankful to them 
for their time, help, interesting conversations, and long hours spent in
lengthy calculations and hard work. I am also grateful to many other
colleagues at the Institute of Nuclear Research and Nuclear Energy of the
Bulgarian Academy of Sciences that I had a chance to meet and work with. My 
next opportunity came as a surprise to me. I was not planning on going abroad
as I was married to a wonderful wife, I had great colleagues, and I found my
work to be very rewarding. But the challenge I faced was not mine alone,
it was a problem for Bulgaria as it was for most other eastern European
countries -- limited opportunities due to political upheaval and difficult
economic times.

In the spring of 1994 I met Professor Jerry P. Draayer at the Annual Bulgarian
International Workshop on Nuclear Theory, Rila Mountains, Bulgaria. As a
result, I guess, of a conversation between Professors J. P. Draayer and A. I.
Gueorguieva -- a conversation that I know very little about -- I was offered
the chance to come to Louisiana State University as a Ph.D. student in
Professor J. P. Draayer's group. This was an honor I could not turn down. I
am very grateful to my advisor Professor J. P. Draayer who gave me the
opportunity to learn and work in a different international culture and
environment, and to experience and enjoy interactions with teachers, 
students, and participants at many workshops and conferences in the United
States and abroad. I am also very grateful to Professor J. P. Draayer and his
wife Lois for their hospitality and the valuable and pleasant time spent in
their home.

I also would like to thank the International Hospitality Foundation, and
especially my host family, Professor R. Imlay and his wife Dena Imlay, for
their warm hospitality and valuable introduction to the American Culture. I am
very grateful to Thomas Beuschel and Kenneth Bernstein, who helped me out in
my first days in Baton Rouge; to Jutta Escher, Gabriela Popa and Ivan
Chompalov for their continuing support through the years; as well as to the
other recent and former graduate students in the Nuclear Theory group at
Louisiana State University who contributed to a stimulating work environment.
I am also grateful to Professor A. I. Gueorguieva, Dr. Ulrich Eichmann, and
Kristina Sviratcheva for their friendship and support.

At last, but not least, I cannot find words and space to write my extreme
gratefulness to my beloved wife Petia and our precious children Anna and Alex
for their support and understanding, and to my mother, father, and sister,
and to my relatives for the support they have provided through the years.

\addcontentsline{toc}{chapter}{Acknowledgments}
\chapter*{Acknowledgments}

\quad 
First of all, I wish to express my sincere gratitude to my advisor,
Professor Jerry P. Draayer, who suggested my Ph. D. project and provided the
necessary environment for its realization through his guidance, patience, and
understanding. I would like also to acknowledge the help of Dr. W. E. Ormand
and C. Johnson with whom two major topics were studied, the structure of the
$^{24}$Mg nucleus, and the SU(3) symmetry breaking in the lower $pf$-shell;
discussions with Professor A. Rau with whom the toy model was worked out;
and interesting conversations and discussions with Dr. C. Bahri.

I would like to thank Professors E. Zganjar, R. Haymaker, A. Rau, and C.
Johnson from the Department of Physics and Astronomy, and A. Raman, from the
Department of Mechanical Engineering, for their comments and suggestions, and
for serving on my dissertation committee. I am grateful to the professors and
administrative personnel in the Department of Physics and Astronomy for
educational, technical, and administrative help, as well as for their friendly
and kindly attitude. My thanks go also to my recent and former colleagues and
friends in the Nuclear Theory group and in the Department of Physics and
Astronomy. I am grateful to the LSU Writing Center for providing help in the
preparation of my dissertation manuscript and especially to Dr. Joe Abraham
and Lauren Moise.

I wish to acknowledge the financial support from the Department of Physics and
Astronomy, a dissertation fellowship from the Louisiana State University
Graduate School, a travel grant from the Charles E. Coates Memorial Fund, and
the U. S. National Science Foundation support under Grant No. PHY-9970769 and
Cooperative Agreement No. EPS-9720652 that includes matching from the
Louisiana Board of Regents Support Fund.

\vfill

\pagebreak
\tableofcontents

\pagebreak
\addcontentsline{toc}{chapter}{List of Tables}
\listoftables

\pagebreak
\addcontentsline{toc}{chapter}{List of Figures}
\listoffigures

\pagebreak
\addcontentsline{toc}{chapter}{Abstract}
\chapter*{Abstract}

\quad
Advances in computer technologies allow calculations in ever larger model
spaces. To keep our understanding growing along with this growth in 
computational power, we consider a novel approach to the nuclear shell model.
The one-dimensional harmonic oscillator in a box is used to introduce the
concept of an oblique-basis shell-model theory. By implementing the Lanczos
method for diagonalization of large matrices, and the Cholesky algorithm for
solving generalized eigenvalue problems, the method is applied to nuclei. The
mixed-symmetry basis combines traditional spherical shell-model states with
SU(3) collective configurations. We test the validity of this mixed-symmetry
scheme on $^{24}$Mg and $^{44}$Ti. Results for $^{24}$Mg, obtained using the
Wilthental USD intersection in a space that spans less than 10\% of the
full-space, reproduce the binding energy within 2\% as well as an accurate
reproduction of the low-energy spectrum and the structure of the states --
90\% overlap with the exact eigenstates. In contrast, for an $m$-scheme 
calculation, one needs about 60\% of the full space to obtain compatible
results. Calculations for $^{44}$Ti support the mixed-mode scheme although the
pure SU(3) calculations with few irreps are not as good as the standard
$m$-scheme calculations. The strong breaking of the SU(3) symmetry results in
relatively small enhancements within the combined basis. However, an
oblique-basis calculation in 50\% of the full $pf$-shell space is as good as a
usual $m$-scheme calculation in 80\% of the space. Results for the lower
$pf$-shell nuclei $^{44-48}$Ti and $^{48}$Cr, using the Kuo-Brown-3
interaction, show that SU(3) symmetry breaking in this region is driven by the
single-particle spin-orbit splitting. In our study we observe some
interesting coherent structures, such as coherent mixing of basis states,
quasi-perturbative behavior in the toy model, and enhanced B(E2) strengths
close to the SU(3) limit even though SU(3) appears to be rather badly broken.
The results suggest that a mixed-mode shell-model theory may be useful in
situations where competing degrees of freedom dominate the dynamics, and
full-space calculations are not feasible.

\pagebreak

\pagenumbering{arabic}

\input{VGGPhDThesisCh1and2.tex}

\input{VGGPhDThesisCh3.tex}

\input{VGGPhDThesisCh4.tex}

\input{VGGPhDThesisCh5.tex}

\input{VGGPhDThesisCh6.tex}

\input{VGGPhDThesisCh7.tex}

\input{VGGPhDThesisReferences.tex}
\input{VGGPhDThesisAppendix.tex}

\chapter*{Vita}
\addcontentsline{toc}{chapter}{Vita}

\quad
Vesselin Gueorguiev was born on May 27, 1967, in Sofia, Bulgaria. He
started his physics career at age 15 while attending the National Natural
Science High School in Sofia, Bulgaria, where he received a diploma with a
physics major as a Semiconductor Production Operator. He continued his studies
in physics at the University of Sofia, St. Kliment Ohridski in Sofia,
Bulgaria, where he majored in nuclear and elementary particle physics and
received a master of  science degree in 1992. He was then employed as a
research-assistant at the Department of Theoretical Physics in the Institute
of Nuclear Research and Nuclear Energy of the Bulgarian Academy of Sciences,
Sofia, Bulgaria. His Doctor of Philosophy degree in nuclear physics will be
awarded by Louisiana State University in December, 2002.

During his graduate school years, he attended many workshops and conferences
and presented his research results at the 1998 and 2000 International 
Symposia in Nuclear Physics at Oaxtepec, M\'{e}xico, the April 2000 and 2001
annual meetings of the American Physical Society, the 2000 and 2002
International Workshops on Nuclear Theory in Bulgaria, the 2001 annual
meeting of the Division of Computational Physics of the American Physical
Society, the 2002 Nuclear Structure Conference: Mapping the Triangle in
Wyoming, and the 2002 International Colloquium on Group Theoretical Methods in
Physics, Paris, France. He was a visiting scientist at the Institute for
Nuclear Theory at the University of Washington in Seattle in October of 2000.
He is the recipient of several awards from Louisiana State University,
including a Graduate School Dissertation Fellowship, Coates Travel Award, and
Graduate School Tuition Waiver. He is the author of 14 publications,
including four abstracts, and three submitted papers.

\end{document}

%% file: VGGPhDThesisCh1and2.tex
\chapter{Introduction}

\quad
Selecting the right basis to perform calculations is a very essential step
in analyzing any eigenvalue problem; it is especially true for many body
quantum mechanical problems. When performing calculations, symmetries are
also very important. Each of the fundamental quantities, such as energy ($E$
), linear momentum ($p$), and angular momentum ($L$), is conserved due to an
exact symmetry of the physical space. It is well known that energy
conservation is due to time translational symmetry, linear momentum
conservation is due to space translational symmetry, and angular
momentum conservation is related to rotational symmetry. Any
mathematical description used in physics takes advantage of these symmetries
and incorporates them explicitly. For example, in the Lagrangian (Hamiltonian)
formalism, the Lagrangian (Hamiltonian) of the system is explicitly
invariant with respect to the fundamental symmetries, such as time
translation, space translation and space rotation. Most examples of exactly
solvable problems come from systems with some type of symmetry \cite
{Elliott-Symmetries}.

The notion of a stable equilibrium state, which is often related to an
energy minimum, is another very important concept in physics. If $\vec{x}
_{0} $ is an equilibrium point of the Hamiltonian function ($H$) for a
classical particle, then $H$ can conveniently be expressed in a Taylor
series around $\vec{x}_{0}$: 
\[
H=\frac{1}{2m}\vec{p}^{2}+\frac{1}{2}k(\vec{x}-\vec{x}_{0})^{2}+\mathcal{O}
\left( \Delta x^{4}\right). 
\]
This way, the harmonic oscillator described by the Hamilton $H=\frac{1}{2m} 
\vec{p}^{2}+\frac{1}{2}k\vec{x}^{2}$ turns out to be one of the most
important model systems in physics with many applications \cite
{MoshinskyBookOnHO}. $SU\left( n\right) $ is the symmetry group of the
$n$-dimensional harmonic oscillator, while $Sp(2n,R)$ is the corresponding
dynamical group. Thus, the $SU(3)$ symmetry of the three-dimensional harmonic
oscillator is a very important approximate symmetry of a system near equilibrium.

Symmetries are very useful in the construction of shell-model structures in
nuclear physics as well as in atomic physics. A shell-model structure is
based on some exactly solvable limit of an effective interaction potential.
An exactly solvable system allows for a well defined set of basis states. In
particular, if bound states exist, they can be considered as single-particle
levels of the system. Usually, a shell model assumes a mean field with which
the particles of the system interact. For example, in atomic physics, the
shell structure is mainly due to the Coulomb field of the nucleus, while in
nuclear physics the mean field is often taken to be the Hartree-Fock mean
field. In this approach, the particle-particle interaction is assumed to be
incorporated as much as possible in the average mean field. In particular,
the nuclear spherical shell model is very successful in the description of
nuclei \cite{Heyde's-shell model}. Despite the enormous success of the
spherical shell model, it is generally difficult to deal with nuclei in the
middle of the shell (mid-shell nuclei) using this model. For such nuclei the
collective degrees of freedom are very essential and the shell-model
configuration space is very big. Therefore, for these nuclei, a shell model
based on the collective degrees of freedom is more appropriate. Elliott's $SU(3)$
model is useful for understanding the collectivity in light nuclei, up to $A<28$
($sd$-shell) \cite{Elliott's SU(3) model}. For heavier nuclei with $A>80$, the
pseudo-$SU(3)$ version of Elliott's model is very successful in the description of
the collective modes \cite {pseudo SU(3) symmetry}. For these nuclei ($A>80,$), the
deformed Nilsson model is more accurate in the description of the single-particle
levels than the simple spherical shell model
\cite{Nilsson model}.

At least in principle, collective phenomena, such as rotational spectra
with strong $B(E2)$ transitions, should be reproduced by the microscopic
models. However, to do so using the spherical shell model, one needs
sufficiently many particle configurations. Unfortunately, the dimensionality
of the space grows combinatorially with the number of particles placed in
the allocated levels. This binomial growth is a major computational problem.
On the other hand, the $SU(3)$ model allows for a good understanding of the
collective nuclear properties in light and heavy mid-shell nuclei. However,
for nuclei near closed shells, the spherical shell model is more favorable due
to the dominance of the single-particle phenomena in these nuclei \cite{VGG
SU(3)andLSinPF-ShellNuclei}. Therefore, it seems plausible to consider a
hybrid-type calculation that uses these two models. In general, the two
bases, the spherical shell-model basis and the $SU(3)$ shell-model basis,
will not be orthogonal to each other. Such a calculation can be considered
as an ``\textbf{oblique}'' basis shell-model calculation \cite{VGG
24MgObliqueCalculations}.

The oblique-basis calculation for nuclei is the subject of the research
presented here. Oblique-basis calculations are expected to be of a practical
value in systems with competing degrees of freedom. For example, our study
shows the relevance of the oblique calculation in the case of $^{24}$Mg. For
this nucleus, the single particle excitations described by the spherical
shell model and the collective excitations described by the $SU(3)$ shell
model are important. When we combine the two bases, we obtain a significant
gain in the convergence of the low-energy spectra towards the full space
result. In particular, the addition of the leading-$SU(3)$ irreducible
representations (irreps) yields the right placement of the $K=2$ band and
the correct order for most of the low-lying levels. Indeed, an even more
detailed analysis shows that the structure of the low-lying states is
significantly improved through the addition of a few $SU(3)$ irreps.

The oblique-basis calculation will be an unnecessary numerical complication
for systems where one of the excitation modes is dominant. For example, in
the lower $pf$-shell nuclei $^{44}$Ti and $^{48}$Cr, the spherical shell
model gives a significant part of the low-energy wave functions within a few
spherical shell-model configurations, while in the $SU(3)$ shell-model
basis one will need more than a few $SU(3)$ irreps. This fact is mainly due
to the strong breaking of the $SU(3)$ in the lower $pf$-shell induced by the
spin-orbit interaction \cite{VGG SU(3)andLSinPF-ShellNuclei}. In spite of
the results in the lower $pf$-shell, it is expected that in the mid-shell
region some sort of $SU(3)$ collective structure will gain importance 
\footnote{It was pointed by Chairul Bahri that the deformed Nilsson diagram for the
$pf$-shell suggest a pseudo $SU(3)$ symmetry. Another alternative could be a
quasi-$SU(3)$ symmetry.}. If this is to happen, then the oblique-basis calculation
will be an important alternative for calculating the structure of nuclei, such as
$^{56}$Fe and $^{56}$Ni.

Results of the shell-model calculations for lower $pf$-shell nuclei show
that $SU(3)$ symmetry breaking in this region is driven by the
single-particle spin-orbit splitting. However, even though states of the
yrast band exhibit $SU(3)$ symmetry breaking, the results also show that the
yrast band $B(E2)$ values are insensitive to this fragmentation of the $
SU(3) $ symmetry; specifically, the quadrupole collectivity as measured by $
B(E2)$ transition strengths between low-lying members of the yrast band
remain high even though $SU(3)$ appears to be broken. Results for $^{44,46,48}$Ti
and $^{48}$Cr using the Kuo-Brown-3 two-body interaction \cite {KB3 interaction}
are given to illustrate these observations.

\chapter{The Nuclear Shell Model}

\quad
In some sense, the shell structure of nuclei is more complicated than the
shell structure of atoms. The shell structure of atoms is due to the Coulomb
force between the nucleus and the electrons. It may be a nice coincidence,
but it is a fact that the Coulomb potential problem in quantum mechanics is
an exactly solvable problem \cite{MoshinskyBookOnHO}. In the case of nuclei,
the situation is more complicated. The reason is that there is no single
source of a central potential. Instead, all nucleons are considered to act
together, generating a mean field. Within this mean field, the problem is
more tractable \cite{Heyde's-shell model}. Here, we do not consider the
problem of how to obtain the mean-field potential. Instead, we just use some
general symmetry properties that a phenomenological potential and a
realistic effective interaction should obey. These symmetry properties
provide insight about the relevant single-particle basis within which one
can consider the problem.

\section{Magic Numbers in Nuclei}

\quad
Maria G. Mayer's discussion of the magic numbers in nuclei has clearly
demonstrated the nuclear shell structure associated with the
independent-particle model for nuclei \cite{Mayer-1948 Magic Numbers}. In
this model, each closed-shell configuration provides a convenient first
approximation. In this approximation, one can assume that the system under
consideration consists of a closed-shell core plus valence particles in a
valence shell. This approach very successfully explains the ground state
properties of nuclei \cite{Mayer-1950-I IPNSM}.

In order to understand and obtain qualitatively good results for the
structure of the excited states, one has to consider a configuration mixing
in the valence space. This usually leads to a very big model space.
Therefore, a further truncation scheme is required. In this chapter, we will
discuss two main approaches used in the nuclear shell model, namely the
spherical shell-model truncation scheme and the $SU(3)$ shell-model
truncation schemes.

\section{The Nuclear Interaction}

\quad
From a fundamental point of view, the problem of the relevant
nucleon-nucleon interaction is very important. However, it is outside the
scope of the research presented here. Even when one is provided with a good
phenomenological nucleon-nucleon interaction, there is a lot of hard work to
be done before one can finally set things up and calculate some
experimentally meaningful results. Usually, a Hartree-Fock procedure is
employed to reduce the many-particle Schr\"{o}dinger equation to a
single-particle Schr\"{o}dinger equation with a self-consistent mean field.
Once the single-particle states and energies are defined, then the $n$-particle
configurations are formed using Slater determinants. Finally, a configuration
mixing is used to take into account some of the residual interaction. This process
may be simplified by using a phenomenological single-particle potential and a
realistic interaction with a set of parameters adjusted to fit the experimental
data.

In this section, we consider a phenomenological interaction that contains
some effective one-body and two-body potentials that are obtained from the
original two-body nucleon-nucleon interaction: 
\[
H=\sum_{i=1}^{A}T_{i}+\frac{1}{2}\sum_{i\neq j}^{A}V\left( \left|
r_{i}-r_{j}\right| \right) \rightarrow \sum_{s\in \left\{ {valence\quad
particles}\right\} }\left( t_{s}+U_{s}\right) +V_{res}. 
\]
$T_{i}$ is the kinetic energy of the $i$-th nucleon, $V\left( \left|
r_{i}-r_{j}\right| \right) $ is the two-body nucleon-nucleon interaction, $
t_{s}$ is an effective one-body kinetic energy of the valence particles, $
U_{s}$ is the effective mean-field potential, and $V_{res}$ is the effective
residual two-body interaction between the valence particles \cite
{Heyde's-shell model}. The effective one-body interaction $H^{1b}=t+U$
provides a set of single-particle states: 
\[
H^{1b}\phi _{i}\left( x\right) =\left( t+U\right) \phi _{i}\left( x\right)
=\varepsilon _{i}\phi _{i}\left( x\right). 
\]

The many-body wave function for a fermion system has to obey the Pauli
principle. Thus, a fully antisymmetric combination, a Slater determinant,
has to be constructed: 
\[
\Psi \left( \vec{x}_{1},....,\vec{x}_{n}\right) =\det \left| 
\begin{array}{cccc}
\phi _{1}\left( \vec{x}_{1}\right) & \phi _{1}\left( \vec{x}_{2}\right) & 
\cdots & \phi _{1}\left( \vec{x}_{n}\right) \\ 
\phi _{2}\left( \vec{x}_{1}\right) & \phi _{2}\left( \vec{x}_{2}\right) & 
\cdots & \phi _{2}\left( \vec{x}_{n}\right) \\ 
\vdots & \vdots & \vdots & \vdots \\ 
\phi _{n}\left( \vec{x}_{1}\right) & \phi _{n}\left( \vec{x}_{2}\right) & 
\cdots & \phi _{n}\left( \vec{x}_{n}\right)
\end{array}
\right|. 
\]
Here, the single-particle wave functions $\phi _{m}\left( \vec{x}_{s}\right) 
$correspond to the $s$-th particle in the $m$-th single-particle state with
quantum numbers depending on the exact symmetries of the single-particle
problem. Usually, these quantum numbers include angular momentum ($j$) and
parity ($\pi $).

\section{Hamiltonian in Second Quantized Form}

\quad
Given the single-particle levels, one can simplify the notation by going
from the coordinate representation of the single-particle levels to an
occupation representation. This process is often called a second
quantization since the wave functions are constructed from appropriate
creation/annihilation tensor operators acting on a vacuum state: 
\[
\phi ^{\alpha jm}(x)\rightarrow \left| \alpha jm\right\rangle =a_{\alpha
jm}^{+}\left| 0\right\rangle. 
\]
Here, $\alpha $ stands for other quantum numbers, such as harmonic-oscillator
shell numbers, spin and isospin labels. The vacuum state $\left|
0\right\rangle $ is a reference state on which everything else is built. The
vacuum state $\left| 0\right\rangle $ may have a different meaning depending
on the quantum labels of the annihilation operators. The annihilation
operators usually define the vacuum as follows: 
\[
a_{\alpha jm}\left| 0\right\rangle =0. 
\]
For example, if $a_{\alpha jm}^{+}$ and $a_{\alpha jm}$ represent some real
particles, such as fermions, then clearly the vacuum state is a state of no
particles at all. If $a_{\alpha jm}^{+}$ and $a_{\alpha jm}$ represent the
valence nucleons, then the vacuum state $\left| 0\right\rangle $ would
represent the closed-shell core. In the forthcoming chapters, we consider 
$\left| 0\right\rangle $ to represent closed-shell nuclei. For example, $ ^{16}O$
is the closed-shell nucleus when we study nuclei in the valence $sd$-shell;
$^{40}Ca$ is the closed-shell nucleus when we study nuclei in the valence
$pf$-shell.

In this second quantized form, the effective Hamiltonian is:

\begin{equation}
H=\sum_{i}\varepsilon _{i}a_{i}^{+}a_{i}+\frac{1}{4}
\sum_{i,j,k,l}V_{ij,kl}a_{i}^{+}a_{j}^{+}a_{k}a_{l}. \label{H=aa+aaaa}
\end{equation}
Here, $\varepsilon _{i}$ are single-particle energies derived from
excitation spectra of one valence particle system, i.e. $^{17}O$ in the case
of the $sd$-shell. The $V_{kl,ij}$ are two-body matrix elements derived from
an initial approximation, which are improved by a data fitting across the
range of nuclei in consideration. For example, in the case of the $sd$-shell
we would use the $63$ two-body matrix elements obtained by Wildenthal \cite
{Wildenthal}.

\section{Spherical Shell Model for Nuclei}

\quad
We have already mentioned the independent-particle model \cite{Mayer-1950-I
IPNSM}. This model uses the harmonic-oscillator potential as an effective
single-particle potential for nucleons \cite{Mayer-1950-II-IPNSM} plus a
spin-orbit interaction that provides for the correct shell closure \cite
{Haxel-1949 IPNSM}. In addition, there is a strong pairing part in the
two-body interaction. The pairing interaction and the quadrupole-quadrupole
interaction \cite{Haxel-1949 IPNSM} are essential parts of the two-body
interaction.

\subsection{Single-Particle Basis}

\quad
In computations based on the independent-particle basis, we use a
phenomenological Hamiltonian (\ref{H=aa+aaaa}) with single-particle levels
labeled by the harmonic-oscillator quantum numbers $nljm$ as follows:

\begin{itemize}
\item $n$ is the harmonic-oscillator shell,

\item $l$ is the angular momentum quantum number,

\item $j=l\pm \frac{1}{2}$ is the total angular momentum of the nucleon
with spin $1/2,$

\item $m$ is the third projection of the total spin $\vec{j}.$
\end{itemize}

Within the above labeling scheme, the single-particle wave functions in the
coordinate representation have the form: 
\begin{equation}
\phi _{nlsjm}\left( x\right) =\left\langle x|nlsjm\right\rangle
=\sum_{m=m_{l}+m_{s}}\left\langle lm_{l},sm_{s}|jm\right\rangle R_{nl}\left(
r\right) Y_{lm_{l}}\left( \theta,\varphi \right) \chi _{m_{s}}.
\label{spherical wave functions}
\end{equation}
Here, $\left\langle lm_{l},sm_{s}|jm\right\rangle $ stand for the
Clebsch-Gordan coefficients of $SU\left( 2\right)$, $R_{nl}\left( r\right) $
are the radial wave functions, $Y_{lm_{l}}\left( \theta,\varphi \right) $
are the spherical harmonics, and $\chi _{m_{s}}$ are the internal spin$\frac{
1}{2}$ wave functions for nucleons.

\subsection{Many-Particle Basis}

\quad
In the occupation number representation, Slater determinant states are
constructed from $n_{1}...n_{k}$ nucleons by means of the fermion particle
creation operators $a_{i}^{+}:$ 
\begin{equation}
\left| n_{1}...n_{k}\right\rangle =\prod\limits_{s=1}^{k}\left(
a_{s}^{+}\right) ^{n_{s}}\left| 0\right\rangle,
\label{SSM-ManyParticleBasis}
\end{equation}
where the operators $a_{i}^{+}$ and $a_{i}$ obey a Fermi algebra: 
\begin{eqnarray*}
a_{i}^{+}a_{j}^{+}+a_{j}^{+}a_{i}^{+} &=&0, \\
a_{i}a_{j}+a_{j}a_{i} &=&0, \\
a_{i}^{+}a_{j}+a_{j}a_{i}^{+} &=&\delta _{ij}.
\end{eqnarray*}
Here, the labels of the operators $a_{i}^{+}$ and $a_{i}$ correspond to some
specific quantum labels $nlsjm$ of the spherical single-particle wave
functions (\ref{spherical wave functions}).

\subsection{Configuration Truncation and the M-scheme Basis}

\quad
Based on the independent-particle model, one can make an initial
approximation to the wave functions of nuclei. This approximation uses the
lowest energy configuration $\left[ n_{1,}...,n_{k}\right]$, where $n_{i}$
is the number of identical particles placed in the $i$-th orbital subject to
the condition $0\leq n\leq 2j+1$. The energy of such a configuration is given
by the expression $E_{\left[ n_{1,}...,n_{k}\right] }=\sum_{i}\varepsilon
_{i}n_{i}.$ It is immediately clear that in general there would be some
degeneracy. Thus, the proper description of the excitation spectrum would
need the two-body part of the interaction to lift this degeneracy. However,
even then, using only the few lowest energy configurations is not sufficient
to describe properly collective excitons in the mid-shell nuclei.

For heavy mid-shell nuclei, one needs to include a significant number of
configurations. One way to proceed and include many configurations is to
consider many-particle states with good $J$ and $M_{J}$ via $SU(2)$ coupling
within each configuration. Codes based on this approach usually rely heavily
on $3j$, $6j,$ and higher $SU(2)$ symbols \cite{French's Oak-Ridge Code,
NATHAN}. Since these $j$-symbols are calculated repeatedly, an efficient $
SU(2)$ package and a smart way to store often used coefficients are very
essential. Recently, an $SU(3)$ code using the same strategy has been
successfully developed \cite{Bahri-RME}. This code relies on a very
efficient data storage technique \cite{Park-WST}.

An alternative computational method is the $M$-scheme approach \cite{the
M-scheme approach}. In this approach, instead of using states with good $J$
and $M_{J},$ one uses only states with good $M_{J}$ and lets the Hamiltonian
select the states of good $J$. Diagonalizing the Hamiltonian in such a basis
results in a few of the lowest energy eigenstates. The $M$-scheme set of
states is convenient since $M_{J}$ is an additive quantum number. In order
to provide for good total angular momentum ($J$)$,$ one has to include all
states of fixed $M_{J}$ within a given configuration. This method relies
heavily on large matrix diagonalization algorithms. One such algorithm is
the Lanczos algorithm which is very fast and efficient \cite{Van Loan
Cullum-Lanczos}. The Lanczos algorithm is a cornerstone of the modern $M-$
scheme shell-model codes \cite{Whitehead-shell model}.

To illustrate the spherical shell-model truncation scheme, we consider
$^{24}$Mg. For this nucleus, the lowest configuration providing the initial
approximation to the ground state is
$0s^{4}0p^{12}0d_{5/2}^{8}1s_{1/2}^{0}0d_{3/2}^{0}$. Here,
$0s^{4}0p^{12}$ is the core nucleus $^{16}O$; the valence space is $0d_{5/2}$
$1s_{1/2}$ $ 0d_{3/2}$ with the lowest configuration of $8$ particles, $4$ protons
+ $4$ neutrons, in the $0d_{5/2}.$ If we explicitly write down a $jj$ coupled
state with good $J$ and $M_{J}$ within the $0d_{5/2}^{8}$ configuration,
then we would see that all the states with a fixed total $M_{J}$ within the $
0d_{5/2}^{8}$ configuration contribute to this state with good $J$ and $
M_{J}.$ Since the Hamiltonian ($H$) respects the rotational symmetry, its
eigenvectors must have good $J$ and $M_{J}$ values. Therefore, diagonalizing 
$H$ in the space of all the states with fixed $M_{J}$ within the $
0d_{5/2}^{8}$ configuration will automatically produce eigenstates with
different $J$ values and same $M_{J}$ values.

Usually, one has to include many configurations by using some selection
principle. Often, the selection scheme uses the energy of the
configurations. In this scheme, one includes only configurations that are
within some range $\Delta E$ relative to the lowest energy configuration.
Another selection scheme, which we use for the present study, considers the
number of particles excited out of the lowest energy configuration into the
full harmonic-oscillator shell. This selection scheme takes into account
possible collective pair excitations when applied with two and four particle
excitations outside of the lowest energy configuration.\footnote{Recently, it has
been shown that one can successfully extrapolate some observables, such as energy
eigenvalues, quadrupole moments, B(E2) transition strengths and Gamow-Teller
transition strengths, using successively bigger truncation spaces. For more
details see nucl-th/0203012 by Mizusaki and Imada and nucl-th/0112014 by
Zelevinsky and Volya.}

\section{The SU(3) Shell Model for Nuclei}

\quad
If one considers a system near equilibrium, then it is possible to
approximate its potential with a harmonic-oscillator potential. Since the
symmetry group of the three-dimensional harmonic oscillator is $SU\left(
3\right) $, it is plausible to use $SU(3)$ basis states. In this section we
discuss the $SU(3)$ shell model. We begin with a review of Elliott's $SU(3)$
model \cite{Elliott's SU(3) model}. In particular, we present two
single-particle labeling schemes, the spherical and cylindrical labeling
scheme. Then, the structure of a general $SU(3)$ irrep in the cylindrical
labeling scheme is given. Next, we describe the $SU(3)$ truncation scheme
which is based on $SU(3)$ invariant two-body interactions. We conclude the
section with a brief discussion of the $SU(3)$ breaking interactions.

\subsection{Labeling of the States in Elliott's SU(3) Model}

\quad
In this section we review group theoretical concepts that are important to
the development of the theory and introduce $SU(3)$ conventions adopted in
our discussion. We consider the physical reduction, $SU(3)\supset SO(3),$
and the canonical group reduction, $SU(3)\supset U(1)\otimes SU(2),$ with
their respective labels.

First we consider the physical group reduction $SU(3)\supset SO(3)$. This
reduction yields a convenient labeling scheme for the generators of $SU(3)$
in terms of $SO(3)$ tensor operators. The commutation relations for these $
SU(3)\supset SO(3)$ tensor operators are given in terms of ordinary $SO(3)$
Clebsch-Gordan coefficients (CGC) $(jm,j^{\prime }m^{\prime }|j^{\prime
\prime }m^{\prime \prime })$ \cite{Elliott's SU(3) model}: 
\begin{eqnarray}
\lbrack L_{m},L_{m^{\prime }}] &=&-\sqrt{2}(1m,1m^{\prime }|1m+m^{\prime
})L_{m+m^{\prime }}, \nonumber \\
\lbrack Q_{m},L_{m^{\prime }}] &=&-\sqrt{6}(2m,1m^{\prime }|2m+m^{\prime
})Q_{m+m^{\prime }}, \label{LQ - Elliott I} \\
\lbrack Q_{m},Q_{m^{\prime }}] &=&3\sqrt{10}(2m,2m^{\prime }|1m+m^{\prime
})L_{m+m^{\prime }}. \nonumber
\end{eqnarray}
Here, $L_{m}$ are generators of the angular momentum and $Q_{m}$ is an
algebraic quadrupole operator.

Within this reduction scheme, states of an $SU(3)$ irrep $(\lambda,\mu )$
have the following labels:

\begin{itemize}
\item $(\lambda,\mu )$ -- $SU(3)$ irrep labels,

\item $l$ -- total orbital angular momentum, which corresponds to the
second order Casimir operator of $SO(3)$,

\item $m_{l}$ -- projection of the angular momentum along the laboratory $z$-axis,

\item $k$ -- projection of the angular momentum in a body-fixed frame,
which is related to multiple occurrences of $SO(3)$ irreps with angular
momentum $l$ in the $(\lambda,\mu )$ irrep.
\end{itemize}

\noindent Unfortunately, this scheme has only one additive label, namely $
m_{l}$, and in addition, there are technical difficulties associated with
handling the $k$ label.

The labeling scheme for our study is the canonical group reduction, $
SU(3)\supset U(1)\otimes SU(2)$ \cite{VGG-1998 su3 good M}. In this scheme $
Q_{0}$ is the $U(1)$ generator and the $SU(2)$ generators are proportional
to $L_{0}$, $Q_{+2}$, and $Q_{-2}$ \cite{Elliott's SU(3) model}. Under the
action of the generators of these $U(1)$ and $SU(2)$ groups, the remaining
four generators of $SU(3)$ transform like two conjugate spin $[\frac{1}{2}]$ 
$SU(2)$ tensors with $\varepsilon =\pm 3$ values for $Q_{0}$. In this
scheme, states of a given $SU(3)$ irrep $(\lambda,\mu )$ have the following
labels:

\begin{itemize}
\item $(\lambda,\mu )$ -- $SU(3)$ irrep labels,

\item $\varepsilon $ -- eigenvalue of the quadrupole moment ($Q_{0}$),

\item $m_{l}$ -- projection of the orbital angular momentum along the $z$-axis
($L_{0}$),

\item $n_{\rho }$ -- related to the second order Casimir operator of $SU(2)$,
which for symmetric $(\lambda,0)$ irreps is simply the number of oscillator
quanta in the $(x,y)$ plane.
\end{itemize}

\begin{figure}[tbp]
\begin{center}
\leavevmode
\epsfxsize = 4.2in
\centerline {\includegraphics[width= 4.2in]{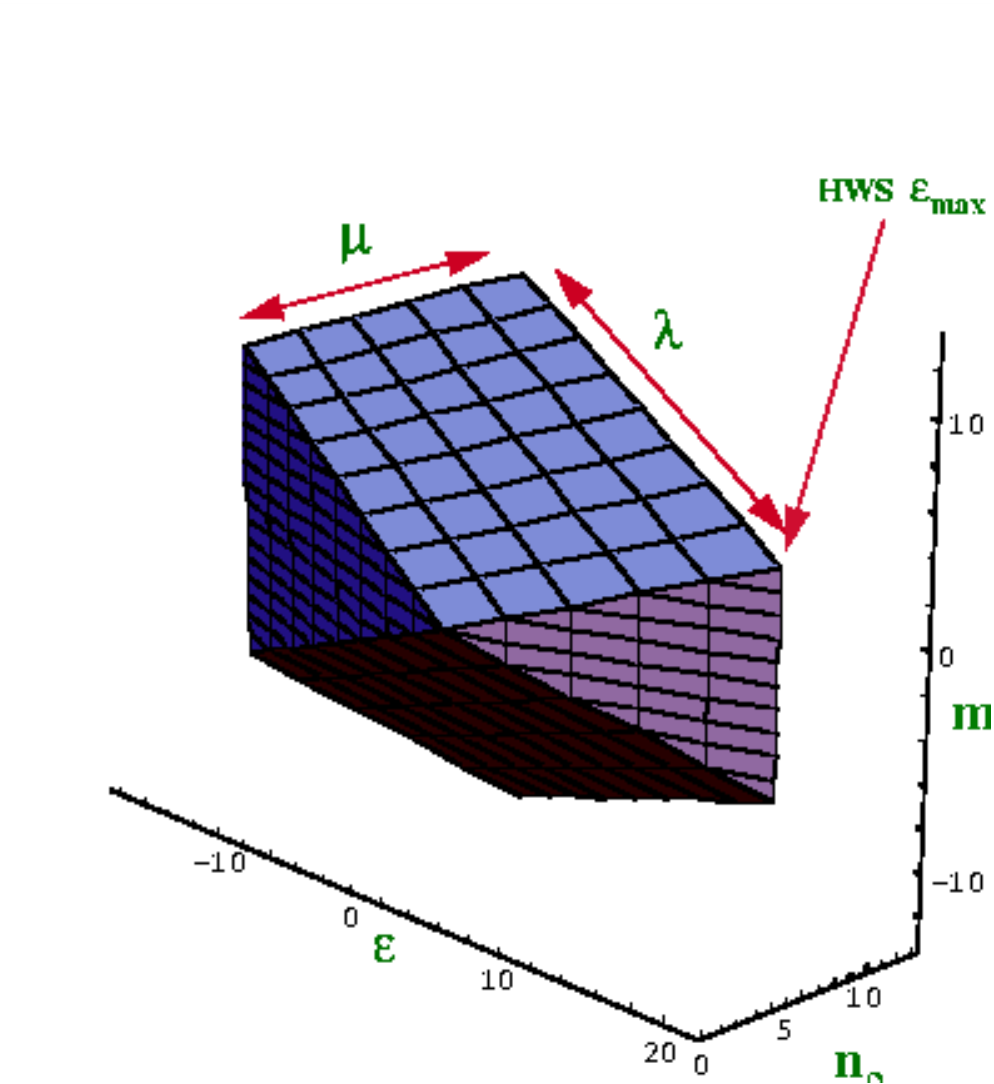}}
\end{center}
\caption{Three-dimensional view of the $(\lambda,\mu )$ $SU(3)$ irrep.}
\label{3D-view of SU(3) irrep}
\end{figure}

This canonical reduction, $SU(3)\supset U(1)\otimes SU(2)$, has two additive
labels, $\varepsilon $ ($Q_{0}$) and $m_{l}$ ($L_{0}$) and the allowed
values of these labels for fixed $SU(3)$ irrep $(\lambda,\mu )$ are given
by \cite{Hecht}:

\begin{eqnarray}
\varepsilon &=&2\lambda +\mu -3(p+q) \label{pqm-parametriztion} \\
n_{\rho } &=&\mu +(p-q) \nonumber \\
m_{l} &=&n_{\rho }-2m \nonumber
\end{eqnarray}

where $0\leq p\leq \lambda $, $0\leq q\leq \mu $, and $0\leq m\leq n_{\rho}$.

\subsection{SU(3) Truncation Scheme}

\quad
It should be pointed out that the quadrupole operator $Q$ used in (\ref{LQ -
Elliott I}) is actually an algebraic quadrupole operator 
\[
Q_{2\mu }^{a}=\sqrt{\frac{4\pi }{5}}\sum_{i}\left( \frac{r_{i}^{2}}{b^{2}}
Y_{2\mu }\left( \hat{r}_{i}\right) +b^{2}p_{i}^{2}Y_{2\mu }\left( \hat{p}
_{i}\right) \right), 
\]
with $b^{2}=\frac{\hbar }{m\omega }.$ However, the matrix elements of $Q^{a}$
reduce to the matrix elements of the physical collective quadrupole operator $
Q^{c}$ within a single harmonic-oscillator shell. 
\[
Q_{2\mu }^{c}=\sqrt{\frac{16\pi }{5}}\sum_{i}\frac{r_{i}^{2}}{b^{2}}Y_{2\mu
}\left( \hat{r}_{i}\right). 
\]
In general, the set of operators $Q^{c}$ and $L$ are part of a $Sp\left(
6,R\right)$ Lie algebra. Within the $Sp\left( 6,R\right) $ model, the $Q^{c}$
operators connect same parity harmonic-oscillator neighbor shells \cite{Juta
Escher's thesis}.

The algebraic realization of the $SU(3)$ model has the advantage that one
can easily connect the important collective operators with the algebraically
significant $SU(3)$ operators. An example of a significant $SU(3)$ operator
is the second order Casimir operator of $SU\left( 3\right)$:
\begin{equation}
C_{2}^{SU(3)}=\frac{1}{4}(3L^{2}+Q\cdot Q). \label{C2su3=LL+QQ}
\end{equation}

By using the generators of $SU(3)$ as labeled by the physical reduction $
SU(3)\supset SO(3),$ we can easily write a general algebraic $SU(3)$
Hamiltonian: 
\begin{equation}
H=H_{osc}+\chi Q\cdot Q+\frac{1}{2\mathcal{J}}L\cdot
L+aC_{3}^{SU(3)}+bL\cdot Q\cdot L+c(L\cdot Q)\cdot (Q\cdot L)+dL\cdot S
\label{Hsu3}
\end{equation}
Here, $H_{osc}$ is the harmonic-oscillator Hamiltonian with single-particle
energies $\varepsilon _{n}=\hbar \omega \left( n+\frac{3}{2}\right)$. The
strength of the quadrupole-quadrupole interaction is $\chi $. The `bare'
classical moment of inertia is $\mathcal{J}$, when the effective moment of
inertia will depend on $a,b$, $c$ and $d$. The parameter $a$ is related to
the third order Casimir operator $C_{3}^{SU(3)}$ of $SU(3).$ The strengths
of the other $SO(3)$ invariant interactions, denoted by $b$ and $c,$ contain
third and fourth order products of the $SU(3)$ generators relevant to the
multiplicity of the $SO(3)$ irreps within the physical reduction $
SU(3)\supset SO(3)$ \cite{Thomas Beuschel's thesis}$.$ If $b$ and $c$ are
such that $bL\cdot Q\cdot L+c(L\cdot Q)\cdot (Q\cdot L) \sim \gamma
K^{2}+L^{2}$ where $\gamma $ is the strength of the $K$-band splitting, then
the collective states in the $SU(3)\supset SO(3)$ chain labeled by $\left|
N[f](\lambda \mu )\kappa LSJM_{J}\right\rangle $ would provide a basis in
which $H$ is diagonal.

The main advantage of using an algebraic Hamiltonian, such as (\ref{Hsu3}),
is its $SU(3)$ symmetry. Therefore, an $SU(3)$ invariant Hamiltonian ($H$) does
not connect states from different $SU(3)$ irreps. Since the $Q\cdot Q$
interaction is proportional to the $C_{2}$ of $SU(3),$ it can be used as an
essential $SU(3)$ truncation scheme. This scheme prescribes $C_{2}$-ordered
importance of the $SU\left( 3\right) $ irreps. In this scheme, one selects $SU(3)$
irreps $(\lambda,\mu )$ with $C_{2}=\lambda ^{2}+\mu ^{2}+\lambda \mu
+3(\lambda +\mu )$ values close to the biggest possible $C_{2}$ value. The
irrep with the biggest possible $C_{2}$ value is called the leading $SU(3)$
irrep. The leading irrep often corresponds to a total spin $S=0$
configuration. This way the leading irrep becomes also the dominant irrep
for the low-lying energy states because the strength of the $L\cdot S$
interaction is usually expected to be small. This is due to the strong
spin-pairing which tends to bring $S=0$ lower in energy. However, the
one-body part of the $\sum_{i}l_{i}\cdot s_{i}$ interaction can cause
significant deviation in the dominance of the leading irrep.

Expressing the $SU(3)$ Hamiltonian (\ref{Hsu3}) in a second quantized form
(\ref{H=aa+aaaa}) gives: 
\begin{equation}
H=\hbar \omega \left( n+\frac{3}{2}\right) \sum_{i}a_{i}^{+}a_{i}+\frac{1}{4}
\sum_{i,j}(\chi \left\langle ij\left| Q\cdot Q\right| kl\right\rangle
+...)a_{i}^{+}a_{j}^{+}a_{k}a_{l}. \label{Hsu3N+aaaa}
\end{equation}
Here, the labels $i$ are shorthand notation for the single-particle labels
in the $SU(3)$ shell-model scheme, that is, $i\rightarrow \tau \tau
_{0}(\eta,0)\kappa lsjm_{j}$ in the $SU(3)\supset SO(3)$ chain or
respectively $i\rightarrow \tau \tau _{0}(\eta,0)n_{\rho }\varepsilon
m_{l}sm_{s}$ in the $SU(3)\supset SU(2)$ chain. As usual, $\tau =1/2$ is the
isospin quantum number with $\tau _{0}=\pm 1/2$ for protons/neutrons
respectively, and $(\eta,0)$ is the SU(3) irrep corresponding to a given
harmonic-oscillator shell $n$ ($\eta =n$). The remaining labels were
discussed in the previous section on the SU(3) shell model.

\subsection{Interactions that Break the SU(3) Symmetry}

\quad
Degenerate single-particle energies are an essential ingredient for good $
SU(3)$ symmetry; this is clear from our discussion on the general algebraic $
SU(3)$ Hamiltonian (\ref{Hsu3}) and its second quantized form (\ref
{Hsu3N+aaaa}). However, we already discussed that the breaking of the
single-particle degeneracy by the spin-orbit interaction is essential for
the description of the correct nuclear shell closures in terms of the
independent-particle model. Therefore, in case of a significant
single-particle splitting, which is due to the orbit-orbit interaction $
\sum_{i}l_{i}^{2}$ and the spin-orbit interaction $\sum_{i}l_{i}\cdot s_{i}$,
there would be a significant disturbance in the $SU(3)$ truncation scheme. In this
case, the spherical shell model described earlier would work and its truncation
scheme could be used.

Another SU(3) breaking factor is the pairing interaction. This interaction
is a very essential short-range two-body nuclear interaction that can have
significant impact on any $SU(3)$-based calculations as well as on the
spherical shell-model type calculations. Although we have studied some
effects of the pairing interaction in the $sd$-shell as well as in the $pf$-shell,
we would rather not engage in this matter. We only mention that effects of the
pairing in the context of the pseudo-$SU(3)$ model have been studied before by C.
Bahri \cite{Chairul Bahri's Thesis}, and currently we are considering
incorporating the pairing effects within an oblique-basis type calculation via the
broken pair model \cite{Heyde's-shell model}.

%% file: VGGPhDThesisCh3.tex
\chapter{Toy Model of a Two-Mode System}

\quad
The study of $^{24}$Mg, which will be discussed later in more detail, has
successfully demonstrated the oblique-basis concept
\cite{VGG 24MgObliqueCalculations}. The quality of the results for
$^{24}$Mg are due to the near equal importance of the two basis sets
used. On the one hand, the spherical shell-model basis is well-suited for
description of the single-particle excitations; on the other hand, the
$SU(3)$ shell model puts an emphasis on the collective excitations in
nuclei. These two modes are crucial for the $^{24}$Mg example.

In general, determining the relevant excitations is a cornerstone in the
study of any system; in some sense this is the art of physics. Usually one
basis works well for one system, but fails for another system. The reason is
that in any general method, such as the variational method, perturbation
theory, or fixed-basis matrix diagonalization, one needs to start with a
good guess about the Hamiltonian and the states that describe the relevant
excitation modes \cite{Skyrme-1957 CinQM}.

When applying perturbation theory, one is often concerned with a small
perturbation of an exactly solvable limit of the full Hamiltonian \cite
{Fernandez-2000,Arteca-1990}. However, there are many examples when the
relevant Hamiltonian has more than one exactly solvable limit \cite
{Rau-1987,Rau-2002}. This is a common situation when a dynamical symmetry
group is used in the construction of the Hamiltonian \cite{Iachello-1987}, 
\cite{Arima and Iachello}, \cite{Cheng-Li Wu et al}. \textit{What shall we
do if the system described by such a Hamiltonian is nowhere near any of the
exact limits?} In these situations, the problem may be better approached by
using states associated with both limits. This set of states will form
an oblique--mixed-mode--basis for the calculation.

Taking into account the importance of the relevant energy scale of a problem
and the wave function localization with respect to the range of the
potential, the oblique-basis method can be taken beyond the idea of using two
orthonormal basis sets. Specifically, one can consider a
variationally-improved basis set starting with some initially guessed basis
states. In the occupation number representation for the nuclear shell model,
this variationally-improved basis method seems inapplicable.\footnote{In the
occupation-number representation one assumes a fixed single-particle structure
and then expands the states in the Slater determinants provided by this
basis. From this point of view, there is no room for variationally-improved
basis states since each Slater determinant is a single-integer machine word.}
However, the method seems interesting because of its possible relevance to
multi-shell \textit{ab-initio} nuclear and atomic physics calculations. The
method may also be related to some renormalization-type techniques.
Therefore, a brief discussion of the variationally-improved basis and its
possible applications is given in the Appendix.

In this chapter some relevant mathematical notation and concepts used in the
oblique-basis method are introduced. Specifically, we demonstrate the
concept of the oblique basis on a simple two-mode system, the
one-dimensional harmonic oscillator in a box. First, we discuss the concept
and then the two exactly solvable limits of our toy model are briefly
summarized. A qualitative discussion of the expected spectrum of the
one-dimensional harmonic oscillator in a box is given. This is followed by an
example spectrum and quantitative estimates. Some specific problems related
to the structure of the Hilbert space will be addressed. Finally, the main
results will be discussed, especially a quasi-perturbative behavior and a
coherent structure within the strong mixing region.

\section{Harmonic Oscillator in a One-Dimensional Box}

\quad
Let us start with an abstract two-mode system. For simplicity, we assume
that the Hamiltonian for the system under investigation has two exactly
solvable limits, for example: 
\begin{equation}
H=(1-\lambda )H_{0}+\lambda H_{1}+\lambda (1-\lambda )H_{2}.
\label{2-mode system H}
\end{equation}
Here, $H_{0}$ and $H_{1}$ are two exactly solvable Hamiltonians. This way,
we have $H$ $\rightarrow H_{0}$ in the limit $\lambda \rightarrow 0$ and $H$ 
$\rightarrow H_{1}$ when $\lambda \rightarrow 1$. In the vicinity of these
two limits we can approach the problem using standard perturbation theory.
However, for $\lambda \approx \frac{1}{2}$ we have a very mixed system with
unclear behavior which could be complicated further by an interaction $H_{2}$
between the natural modes of $H_{0}$ and $H_{1}.$

In the expression (\ref{2-mode system H}), $\lambda $ is introduced to
simplify the discussion. In general, we have more than one parameter in the
Hamiltonian. Often the exactly solvable limits are described as
hypersurfaces in the full parameter space. It could even be that there are
three or more exactly solvable limits. For example, the Interacting Boson
Model (IBM) has three exactly solvable limits \cite{MoshinskyBookOnHO}.
Another example with three exactly solvable limits is the commonly used
nuclear schematic interaction. It has nondegenerate single-particle energies
($\varepsilon _{i}$), pairing ($P^{+}P$) two-body interaction, and
quadrupole-quadrupole ($Q\cdot Q$) two-body interaction: 
\[
H=\varepsilon _{i}N_{i}+GP^{+}P-\chi Q\cdot Q. 
\]

Here, we consider the simplest two-mode system that is sufficiently close to
the problem we have to solve for nuclei. The system under consideration
consists of a one-dimensional harmonic oscillator in a one-dimensional box
of size $2L$ \cite{Armen and Rau}: 
\begin{equation}
H=\frac{1}{2m}p^{2}+V_{L}(q)+\frac{m\omega ^{2}}{2}q^{2}.
\label{H-ho-1Dbox}
\end{equation}
where $V_{L}(q)$ is the confining potential which is zero for $\left|
q\right| <L$ and $\infty $ for $\left| q\right| \geq L$. This system has two
exactly solvable limits. A more realistic model might consist of a
three-dimensional harmonic oscillator and a square-well potential since
these two potentials are known to be good starting points in the nuclear
shell model \cite{Heyde-1994}.

The one-dimensional harmonic oscillator in a one-dimensional box model has
been used as an example by Barton, Bray, and Mackane in their discussion on
the effects of the distant boundaries on the energy levels of a
one-dimensional quantum system \cite{Barton-Bray-Mckane-1990}. Also, some
studies have already been done for the cylindrical symmetric system of a
three-dimensional harmonic oscillator between two impenetrable walls \cite
{Marin and Cruz-1988}. However, the bi-modal structure of the problems has
not been discussed in these studies. The essential two-mode regime of such
problems has been studied in the context of a two-dimensional confinement of
a particle in an external magnetic field by Rosas \textit{et al.} \cite{Rosas
et al-2000}. Some authors have generalized the one-dimensional harmonic
oscillator by introducing time dependent parameters in the Hamiltonian \cite
{Lejarreta-1999} and have recognized the two limiting cases of a free
particle and harmonic oscillator. The infinite square well and the harmonic
oscillator have been considered as the two limiting cases of a power-law
potential within the context of wave packet collapses and revivals \cite
{Robinett-2000 AJP,Robinett-2000 JMP}. Here, we focus our study on the
bi-modal structure of the one-dimensional harmonic oscillator in a
one-dimensional box.

The first limit of the toy model (\ref{H-ho-1Dbox}) is $\omega =0.$ This is
a free particle in a one-dimensional box with size $2L:$ 
\begin{equation}
H_{0}=\frac{1}{2m}p^{2}+V_{L}(q).  \label{1D box Hamiltonian}
\end{equation}
The eigenvectors and energies are labeled by $n=0,1,...$ and are given by
the expressions: 
\begin{eqnarray}
\Phi _{n}(q) &=&\left\{ 
\begin{tabular}{lll}
$\sqrt{\frac{1}{L}}\cos \left( (n+1)\frac{\pi }{2}\frac{q}{L}\right) $ & if
& n is even \\ 
$\sqrt{\frac{1}{L}}\sin \left( (n+1)\frac{\pi }{2}\frac{q}{L}\right) $ & if
& n is odd
\end{tabular}
\right. ,  \label{1D box FW and En} \\
E_{n} &=&\frac{1}{2m}\left( (n+1)\frac{\pi }{2}\right) ^{2}\left( \frac{
\hbar }{L}\right) ^{2}.  \nonumber
\end{eqnarray}
This limit corresponds to extreme nuclear matter when the short range
nuclear force produces an effective interaction well represented by a
square-well potential \cite{Heyde-1994}. We can think of this limit as a
one-dimensional equivalent of a three-dimensional model where nucleons are
confined within a finite volume of space representing the nucleus.

The other exactly solvable limit of the toy model (\ref{H-ho-1Dbox}) is the
harmonic oscillator in one dimension: 
\begin{equation}
H_{1}=\frac{1}{2m}p^{2}+\frac{m\omega ^{2}}{2}q^{2}.
\label{Harmonic oscillator Hamiltonian}
\end{equation}
In dimensionless coordinates 
\[
q\rightarrow \tilde{q}\sqrt{\frac{\hbar }{m\omega }},\quad p\rightarrow 
\tilde{p}\sqrt{m\hbar \omega }, 
\]
we have: 
\[
H_{1}=\hbar \omega \frac{1}{2}\left( \tilde{p}^{2}+\tilde{q}^{2}\right) . 
\]
Thus the eigenvectors and energies are labeled by $n=0,1,...$ and are given by
the expressions: 
\begin{eqnarray}
\Psi _{n}(q) &=&\sqrt{\frac{1}{bn!2^{n}\sqrt{\pi }}}H_{n}\left( \frac{q}{b}
\right) \exp \left( -\frac{1}{2}\frac{q^{2}}{b^{2}}\right),\quad b=\sqrt{
\frac{\hbar }{m\omega }}  \label{Harmonic oscillator WF and En} \\
E_{n} &=&\hbar \omega \left( n+\frac{1}{2}\right) .  \nonumber
\end{eqnarray}
Where $H_{n}$ are the Hermite polynomials. This limit corresponds to the
three-dimensional harmonic oscillator model for nuclei.

In a one-dimensional toy model, the anharmonic oscillator with a quartic
anharmonicity would be the appropriate counterpart of the $Sp(6,R)$ shell
model since the quadrupole-quadrupole interaction $Q\cdot Q$ goes as $\sim
r^{4}$ and $Q$ connects same parity harmonic oscillator shells. If we
restrict the model space to only one harmonic oscillator shell, then we can
use the algebraic quadrupole moment $\tilde{Q}$ of Elliott \cite{Elliott's
SU(3) model} because within a single shell $\tilde{Q}$ is the same as $Q$ 
\cite{MoshinskyBookOnHO}. Thus for our study it is appropriate to consider
the one-dimensional harmonic oscillator to correspond to the $SU\left(
3\right) $ shell model for nuclei.

\section{Spectral Structure at Different Energy Scales}

\quad
Often in physics the spectrum of a system is different for different
energy scales. This usually reflects the existence of different excitation
modes of the system. For the toy model Hamiltonian (\ref{H-ho-1Dbox}) we can
clearly define three spectral types:

\begin{itemize}
\item  Spectrum of a particle in a one-dimensional box (\ref{1D box FW and
En}) with quadratic dependence on $n$ ($E_{n}$ $\sim $ $n^{2}$),

\item  Spectrum of the one-dimensional harmonic oscillator (\ref{Harmonic
oscillator WF and En}) with linear dependence on $n$ ($E_{n}\sim $ $n$),

\item  Intermediate spectrum that is neither of the above two types.
\end{itemize}

\begin{figure}[tbp]
\begin{center}
\leavevmode
\epsfxsize = \textwidth

\centerline {\includegraphics[width= \textwidth]{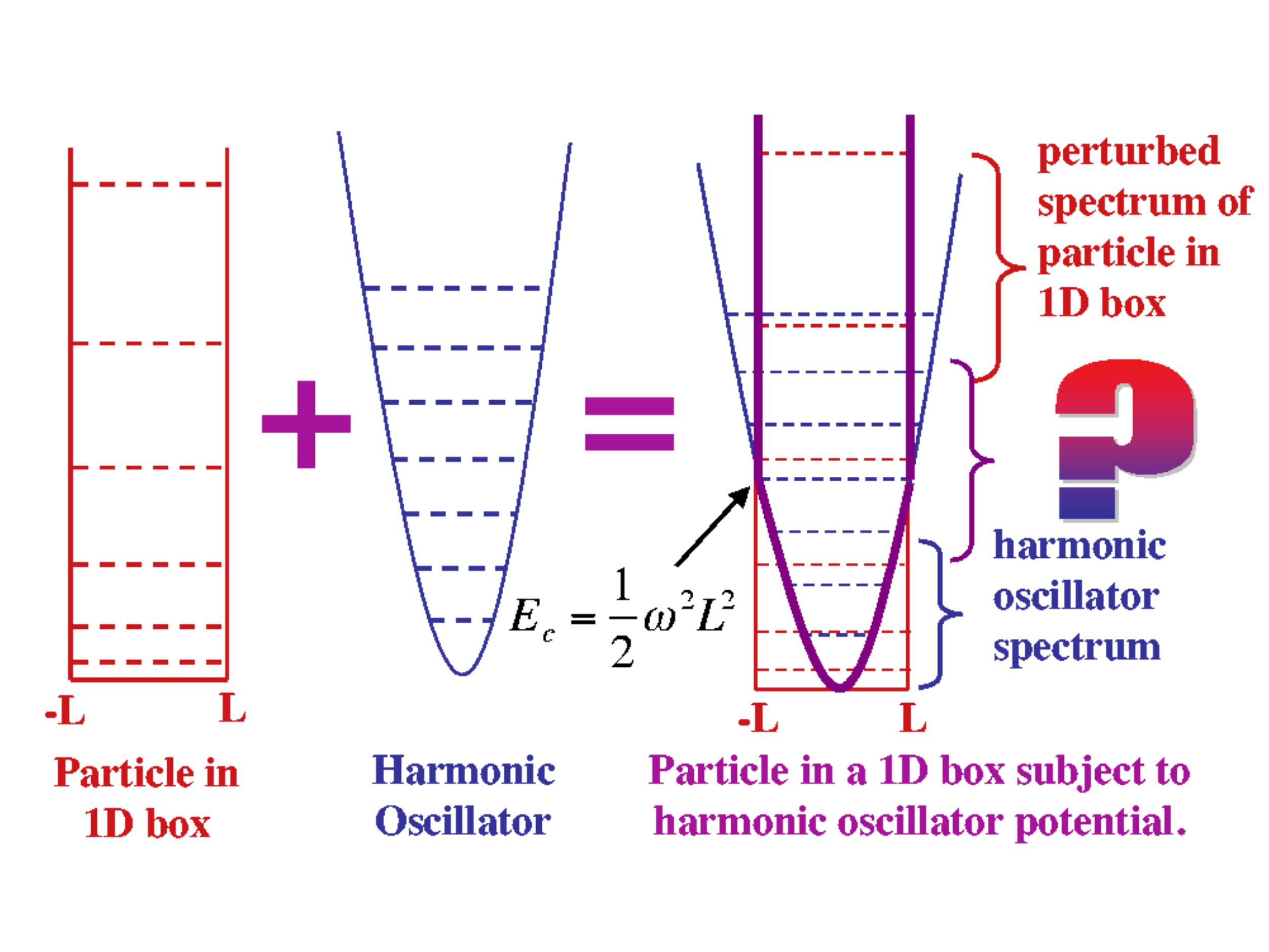}}

\end{center}
\caption{Two--mode toy system. The structure of the interaction potential of
a particle in a one-dimensional box subject to a harmonic oscillator
restoring force towards the center of the box.}
\label{1D+HO-potental}
\end{figure}

\begin{figure}[tbp]
\begin{center}
\leavevmode
\epsfxsize = \textwidth

\centerline {\includegraphics[width= \textwidth]{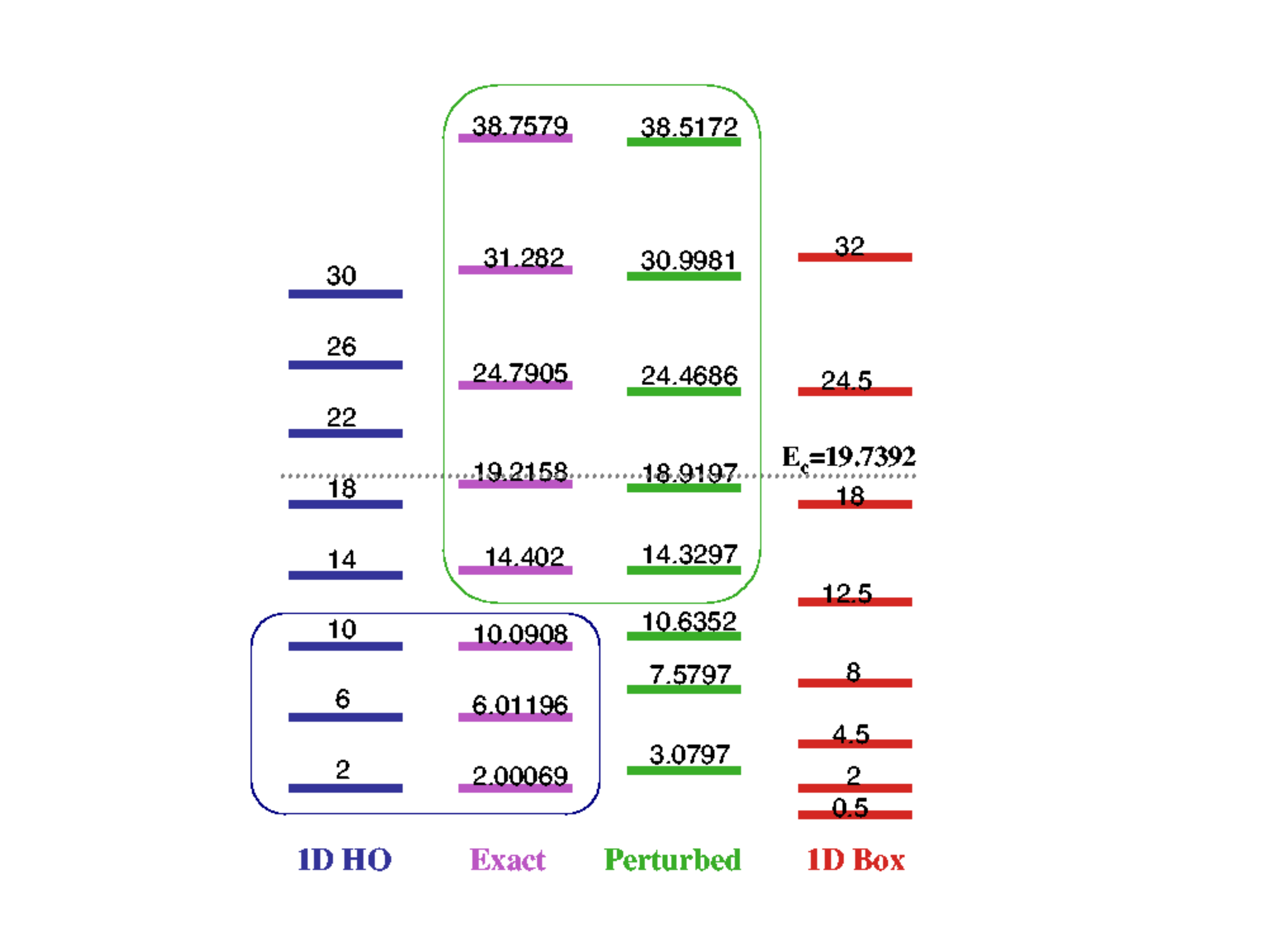}}

\end{center}
\caption{Spectral structure of the two--mode system for $m=\hbar =2L/\pi =1$
and $\omega =4.$}
\label{w4SpectralStructure}
\end{figure}

From Fig. \ref{1D+HO-potental} we expect that the particle in a box spectrum
should be operative at high energies. These energies are energies where the
box boundaries dominate over the harmonic oscillator potential. In this
regime one can use standard perturbation theory to calculate the energy
for a particle in a box perturbed by a harmonic oscillator potential. It
can be shown that perturbation theory will give better results for
higher energy levels. For $n\rightarrow \infty $ the first
correction ($\delta E_{n}^{1}$) approaches the constant value of $m\omega
^{2}L^{2}/6.$ An estimate on when the perturbation calculations are feasible
using $E_{n+1}^{0}-E_{n}^{0}>>\left\langle n\left| V\right| n\right\rangle $
gives: 
\begin{equation}
n>>2m^{2}\omega ^{2}L^{4}/(3\hbar ^{2}\pi ^{2}).
\label{1D box spectrum begins}
\end{equation}
This analysis is confirmed by the numerical calculations shown in Fig. \ref
{w4SpectralStructure} where the perturbed particle in a box spectrum is
really operative at $n>3$ for the case of $m=\hbar =2L/\pi =1$ and $\omega
=4.$

The intermediate spectrum should be observed when the harmonic oscillator
turning points coincide with the walls of the box. Therefore, the critical
energy scale that separates the two extreme spectral structures is given by: 
\begin{equation}
E_{c}=\frac{m\omega ^{2}}{2}L^{2}  \label{Ec for 1D box and HO }
\end{equation}
Notice that the constant energy shift $m\omega ^{2}L^{2}/6$ in the energy of
the high energy levels $\delta E_{n>>1}^{1}$ is one-third of the critical
energy ($E_{c}/3$).

At low energies, where the one-dimensional harmonic oscillator determines
the classical turning points to be far from the boundaries, we expect to see
the harmonic oscillator spectrum as shown in Fig. \ref{w4SpectralStructure}.
The number of harmonic oscillator states that will be observed is easily
estimated using: 
\begin{equation}
E_{c}>E_{n}^{ho}\Rightarrow n_{\max }^{ho}=\frac{1}{2}\frac{m\omega L^{2}}{
\hbar }-\frac{1}{2}  \label{HO spectrum ends}
\end{equation}
It should be pointed out that there is a compatible number of levels,
usually bigger than $n_{\max }^{ho}$, below the $E_{c}$ corresponding to a
free particle in a box: 
\begin{equation}
E_{c}>E_{n}^{1D}\Rightarrow n_{\max }^{1D}=\frac{2}{\pi }\frac{m\omega L^{2} 
}{\hbar }-1.  \label{1Dbox spectrum ends}
\end{equation}
However, these states are mixed by the harmonic oscillator potential toward
the corresponding harmonic oscillator wave functions.

Using the ratio of the ground state energies,
$E_{g.s.}^{HO}/E_{g.s.}^{1D}=4m\omega L^{2}/(\hbar \pi ^{2})$, together with
(\ref{HO spectrum ends}) and (\ref{1Dbox spectrum ends}), the following
spectral situations apply:

\begin{itemize}
\item  For $\frac{m\omega L^{2}}{\hbar }>\left( \frac{\pi }{2}\right) ^{2}$
there are levels below $E_{c}$ corresponding to the harmonic oscillator and
the free particle in a box such that $E_{g.s.}^{HO}>E_{g.s.}^{1D}.$
However, only the harmonic oscillator levels are seen in the low energy
spectrum.

\item  For $\left( \frac{\pi }{2}\right) ^{2}>\frac{m\omega L^{2}}{\hbar }>
\frac{\pi }{2}$ there are only the ground states $E_{g.s.}^{1D}$ and
$E_{g.s.}^{HO}$ below $E_{c}$ and $E_{g.s.}^{1D}>E_{g.s.}^{HO}$

\item  For $\frac{\pi }{2}>\frac{m\omega L^{2}}{\hbar }>1$ there is only the
ground state of the harmonic oscillator $E_{g.s.}^{HO}$ below $E_{c}$.
\end{itemize}

Therefore, the smallest number of states\footnote{For simplicity we usually
fix the parameters as follows: $m=\hbar =1$, $L=\pi /2$.} to illustrate the
two mode spectra is the case of $m=\hbar =1$, $L=\pi /2$ and $\omega =4.$
With these parameters, formula (\ref{HO spectrum ends}) gives $n_{\max
}^{HO}=4.\,5348.$ Thus one should see no more than $4$ equidistant states as
shown in Fig. \ref{w4SpectralStructure}. In Fig. \ref {w4SpectralStructure}
there are three clear equidistant energy levels that correspond to a
harmonic oscillator spectrum.

With respect to the critical energy $E_{c},$ there is a more explicit
classification of the spectral structure:

\begin{itemize}
\item  Perturbed particle in a one-dimensional box spectrum for energies
$E>>E_{c}$ such that (\ref{1D box spectrum begins}) holds,

\item  One-dimensional harmonic oscillator spectrum (\ref{Harmonic
oscillator WF and En}) for energies $E_{c}>>E$ such that (\ref{HO spectrum
ends}) holds,

\item  Intermediate spectrum for energies $E\approx E_{c}$.
\end{itemize}

\section{Toy Model Calculations and Results}

\quad
Despite the simplicity of the toy model (\ref{H-ho-1Dbox}), the harmonic
oscillator in a box exhibits some of the essential characteristics of a more
complex system. Our main interest is in problems associated with the use of
fixed-basis calculations. In particular, one such problem is the slow
convergence of the calculations \cite{Armen and Rau}. If one can implement
an exact arithmetic, one may not worry too much about the slow convergence
when enough time, storage, and other resources are provided. However,
numerical calculations are plagued with numerical errors that may grow
significantly and render the results meaningless. From this point of view, a
calculation that converges slowly may be compromised by accumulated
numerical error.

\subsection{On the Hilbert Space of the Basis Wave Functions}

\quad
Before discussing the toy model using an oblique basis, it is instructive
to discuss briefly the harmonic oscillator problem (\ref{Harmonic oscillator
Hamiltonian}) using the wave functions for a free particle in a
one-dimensional box (\ref{1D box FW and En}); and vice versa, solving the
problem of a free particle in a one-dimensional box (\ref{1D box Hamiltonian})
using the wave functions for a particle in the harmonic oscillator potential
(\ref{Harmonic oscillator WF and En}).

\begin{figure}[tbp]
\begin{center}
\leavevmode
\epsfxsize = \textwidth

\centerline {\includegraphics[width= \textwidth]{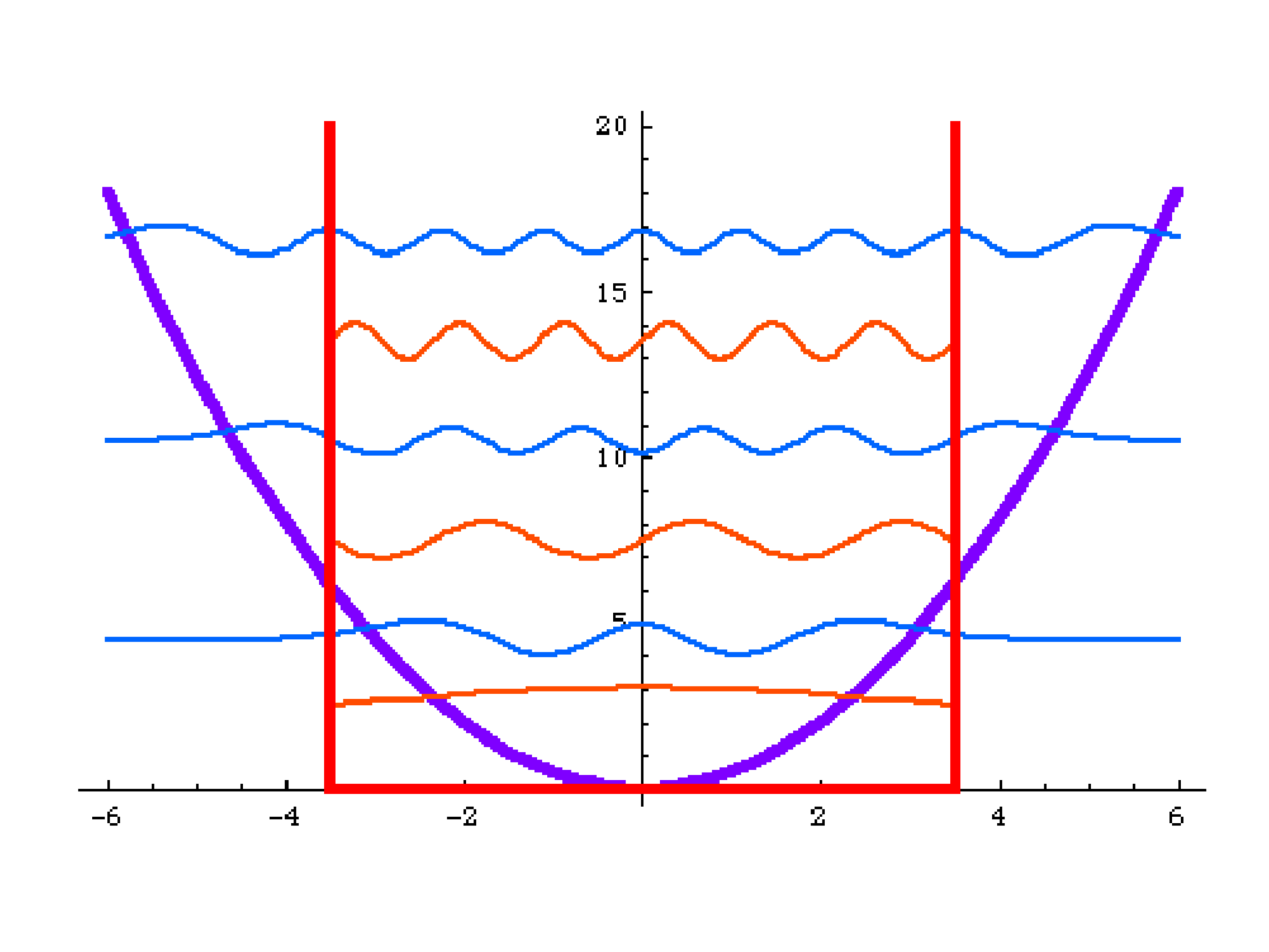}}

\end{center}
\caption{Spreading of the wave functions for the harmonic oscillator (blue)
and particle in a box (red).}
\label{wf-spread}
\end{figure}

Due to the structure of the wave functions, there are some specific problems
that need to be addressed. For example, using wave functions for a free
particle in a one-dimensional box to solve the harmonic oscillator problem
may not be appropriate especially for high energy states $E>>$ $E_{c}$. The
problem is that any linear combination of wave functions with the same
localized support, in our case the wave functions are localized within the
box, will still be a function with the same localized support (see Fig.
\ref{wf-spread}). That is, any linear combination of wave functions that are
zero outside of the box is a function that is zero outside of the box too. 
Because the harmonic oscillator potential gets wider for higher and higher
energies, any higher energy wave function must spread more than the previous
one. Similarly, the spreading of the harmonic oscillator wave functions is
responsible for the troubles that arise in solving the problem of a free
particle in a one-dimensional box using the harmonic oscillator wave
functions. The essence of these problems is in the structure of the
corresponding Hilbert spaces.

The influence of the boundary conditions on the properties of a quantum
mechanical system has been recognized from the dawn of quantum mechanics.
It is well known that some separable problems may re-couple due to the
boundary conditions \cite{Tanner-1991}. Some recent studies on the problem of
confined one-dimensional systems using equations for relevant cut-off
functions have been pioneered by Barton, Bray, and Mackane \cite
{Barton-Bray-Mckane-1990}. Their method has been further developed in a more
general setting by Berman \cite{Berman-1991}. Other authors aim at
variational procedures using simple cut-off functions
\cite{Marin and Cruz-1991 AJP, Marin and Cruz-1991 JPB} or derive asymptotic
estimates for multi-particle systems using the Kirkwood-Buckingham
variational method \cite {Pupyshev and Scherbinin-1999}. Somewhat different
approaches focus on shape-invariant potentials and use supersymmetric partner
potentials to derive energy shifts and wave function approximations \cite
{Dutt-Mukherjee-Varshni-1995}, as well as sample-size dependence of the
ground-state energy \cite{Monthus et al. -1996}. In the next few
paragraphs we discuss the structure of the relevant Hilbert spaces when
confinement is present.

\subsubsection{$\bullet$ Harmonic Oscillator in the One-Dimensional Box Basis}

\quad
Now we consider the harmonic oscillator problem (\ref{Harmonic oscillator
Hamiltonian}) using the wave functions for a free particle in a
one-dimensional box (\ref{1D box FW and En}). There are no difficulties for
energies $E<<$ $E_{c}$ (\ref{Ec for 1D box and HO }) where the harmonic
oscillator potential is still within the box. However, for energies $E>>$
$E_{c}$ the basis wave functions are localized only on the interval $[-L,L]$.
Thus they cannot provide the necessary spread over the potential width
(Fig. \ref{wf-spread}). This situation would be appropriate for the toy model
(\ref {H-ho-1Dbox}) but not for the pure harmonic oscillator problem (\ref
{Harmonic oscillator Hamiltonian}).

One simple solution of the spreading problem is to continue the basis wave
functions by periodicity. This way the necessary spread of the basis wave
functions can be achieved and the new basis will stay orthogonal but must be
re-normalized.\footnote{If one continues the wave functions to infinity, then
there is a normalization problem. However, if the continuation is on a finite
interval, then the functions can still be normalized.} However, these basis
wave functions do not decay to zero in the classically forbidden zone. This
means that some significant number of basis wave functions will be needed to
account for the necessary behavior within the classically forbidden zone.

Another alternative is to change the  support domain corresponding to non-zero
values of the function by stretching or squeezing it through a scaling of the
argument of the basis wave functions, $x\rightarrow x\alpha _{n}/L$. This way
the support becomes $[-L,L]$ $\rightarrow $ $[-\alpha _{n},\alpha _{n}]$.
Here, $\alpha _{n}$ is a scale factor for the $n$-th basis wave function
(\ref{1D box FW and En}) estimated either from the width of the harmonic
oscillator potential\footnote{The initial idea is to use basis states that
have a spread compatible with the width of the potential in the energy region
of interest, thus resolving the spectra only within that energy scale without
calculating the lower energy states. Unfortunately, it does not seem to work
since interference causes reduction of the wave spread and therefore drives
the solutions towards the lowest eigenstate.}, or determined by variational
minimization. Either way, the new set of basis functions will be
non-orthogonal. In general, there may be even a linear dependence. However,
for the basis functions discussed here, linear dependence may not appear due
to the different number of nodes for each wave function. The number of nodes
(zeros) is not changed under the re-scaling procedure. While the potential
width scaling is simpler, its applicability is more limited than the
variationally-determined one. In general, the variational approach can be
extended for much more general situations as discussed in the Appendix.

\subsubsection{$\bullet$ Particle in a Box in the Harmonic-Oscillator Basis}

\begin{figure}[tbp]
\begin{center}
\leavevmode
\epsfxsize = \textwidth

\centerline {\includegraphics[width= \textwidth]{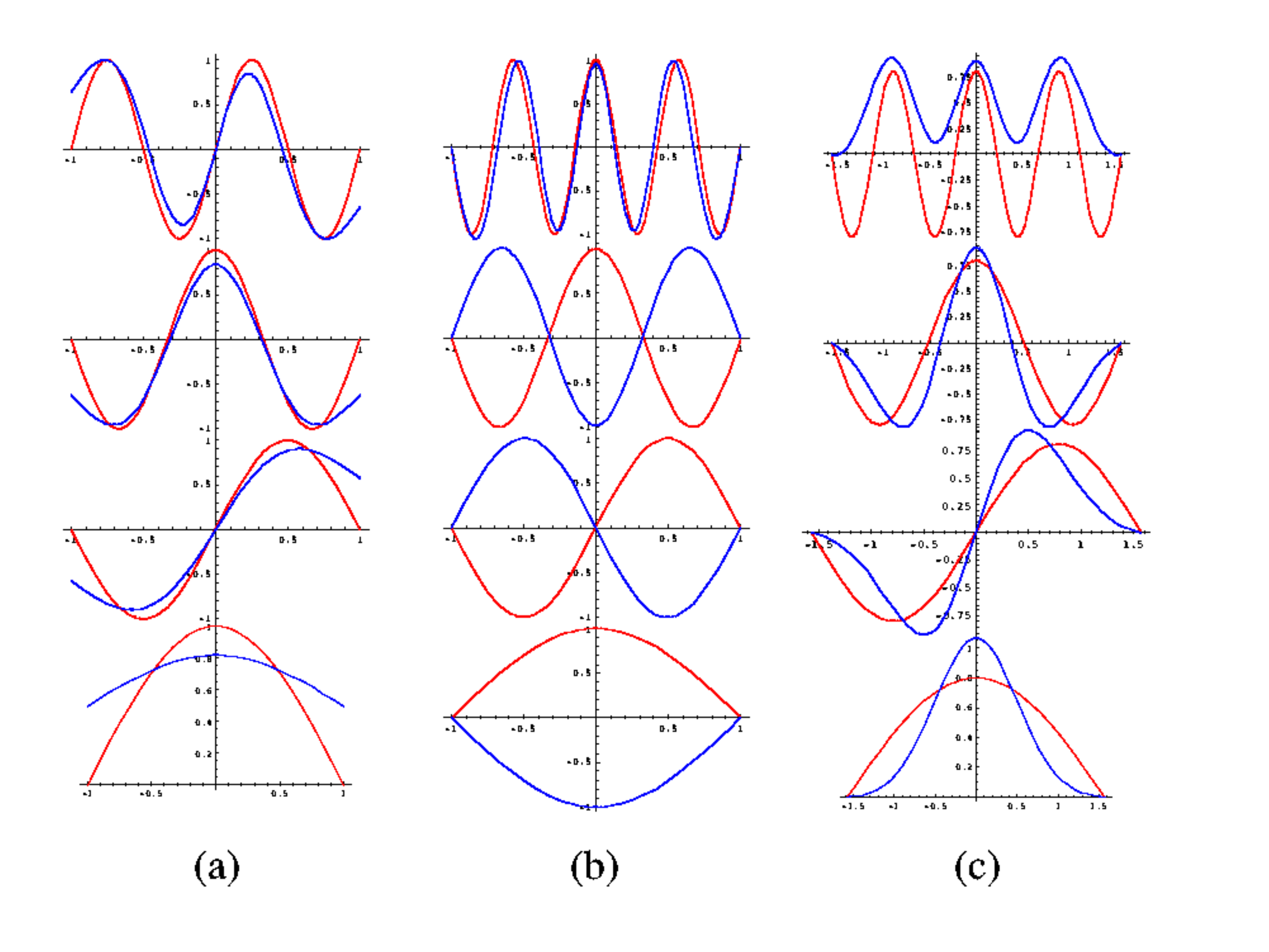}}

\end{center}
\caption{Harmonic-oscillator trial wave functions adjusted with respect to
the one-dimensional box problem: (a) adjusted according to the potential
width $E_{n}^{1Dbox}=\omega _{n}^{2}L^{2}/2\Rightarrow \omega _{n}=\frac{
\hbar }{L^{2}}\left( 1+2n\right) $, (b) nodally adjusted, (c) boundary
adjusted using $\Psi (q)\rightarrow \Psi (q)-\Psi \left( L\right)
(1+q/L)/2-\Psi \left( -L\right) (1-q/L)/2$}
\label{regularized-fw}
\end{figure}

\quad
Next, suppose we want to solve the problem of a free particle in a
one-dimensional box $[-L,L]$ (\ref{1D box Hamiltonian}) using the harmonic
oscillator wave functions (\ref{Harmonic oscillator WF and En}). The first
thing to do is to change the inner product of the wave functions: $\left(
f,g\right) =\int_{-\infty }^{\infty }f^{*}\left( x\right) g\left( x\right)
dx\rightarrow \int_{-L}^{L}f^{*}\left( x\right) g\left( x\right) dx. $ Then,
it is immediately clear that the set of orthonormal harmonic oscillator wave
functions $\Psi _{n}(q)$ (\ref{Harmonic oscillator WF and En}) will lose its
orthonormality and even its linear independence.\footnote{The set of
functions $\Psi _{n}\left( q\right)$ with support domain restricted to
$[-L,L]$ and denoted by $\Psi _{n}(q;[-L,L])$ may become linearly dependent
if $L$ is so small that there are more than one $\Psi _{n}(q;[-L,L])$ with
the same number of nodes within $[-L,L]$.} However, this is not the actual
trouble in such an approach.\footnote{The oblique basis type calculations
described later can successfully remove the linearly dependent basis states
in the process of handling the non-orthogonality of the basis.} Neither the
variational nor the potential-width wave function scaling will help to cure
the loss of hermiticity of the physically significant differential operators,
such as the momentum operator ($p=-i\hbar \frac{\partial }{\partial x}$) and
the Hamiltonian operator ($H=\frac{1}{2m}p^{2}$). This non-hermiticity is due
to the behavior of the basis states at the boundary, mainly the non-vanishing
of the wave functions at $-L$ and $L.$ For detailed analysis on the loss of
hermiticity, we refer the reader to the Appendix. In order to recover the
hermiticity of the differential operator $i\frac{\partial }{ \partial x}$, it
is sufficient\footnote{Wave functions with the same value at $\pm$ L is the
necessary condition; the wave functions should be zero only for an infinite
potential at $\pm$ L.} to make sure that our basis wave functions vanish
at the boundary points $-L$ and $L.$ For this purpose one can look at the
nodes of each basis wave function and scale it so that its outer nodes are at
the boundary points.\footnote{From the nodal structure of the harmonic
oscillator wave functions, given by the Hermite polynomials, it is clear that
the first two wave functions ($\Psi _{0}$ and $\Psi _{1}$) cannot be used
since they have less than two nodes.} Since the physical requirement that
the wave functions have to be zero at the boundary is the cornerstone in
quantizing the free particle in a one-dimensional box (\ref{1D box FW and
En}), it is not surprising that the nodally adjusted harmonic oscillator wave
functions are very close to the exact wave functions for the free particle in
a one-dimensional box as shown in Fig. \ref{regularized-fw}.

In general, calculating the nodes of a function may become very complicated.
To avoid problems with finding the roots, one can use the following
technique\footnote{ This technique has been suggested by Professor A. R. P.
Rau (private communications).}: the idea is to evaluate the value of the wave
function at the boundary points, then shift the wave function by a constant
to get zeros at the boundary, $\Psi (q)\rightarrow \Psi (q)-\Psi \left(
L\right) $. This idea works well for even wave functions, but has to be
generalized for odd wave functions by adding a linear term, $\Psi
(q)\rightarrow \Psi (q)-(\Psi\left( L\right) /L)q$. Thus for a general
function we can have: $\Psi (q)\rightarrow \Psi (q)-(1+q/L)\Psi \left(
L\right) /2-(1-q/L)\Psi \left( -L\right) /2$. In Fig. \ref{regularized-fw} we
have shown some of the resulting wave functions. Notice that this procedure
gives a new wave function $\Psi $ that is well behaved inside the interval
$[-L,L]$ and grows linearly with $q$ outside the interval $[-L,L]$. This is
in contrast to the behavior of the cut-off function $f(q)$ obtained by Barton
et al \cite {Barton-Bray-Mckane-1990}. The function $f(q)$ has $L/q$
singularity at the origin ($q=0$). The use of a cut-off function to enforce
boundary conditions has been developed by Barton et al \cite
{Barton-Bray-Mckane-1990} and Berman \cite{Berman-1991} and provides an
interesting integral equation for the cut-off function. On the other hand, a
simple cut-off function supplemented by a variational method seems to be very
effective \cite{Marin and Cruz-1991 AJP,Marin and Cruz-1991 JPB,Pupyshev and
Scherbinin-1999}.

It should be pointed out that by using the above process one can set up and
successfully run a modification of the usual Lanczos algorithm, to be
discussed later, to solve for the few lowest eigenvectors of the free
particle in a one-dimensional box through an arbitrarily chosen initial
wave function. The major modification is to project every new function, $\Psi
_{n+1}=$ $H\Psi _{n}$, into the appropriate Hilbert space and subtract the
components along any previous basis vectors. Only then should one attempt to
evaluate the matrix elements of $H$ related to the new basis vector that is
clearly within the correct Hilbert space. This way, one has to double the
number of scalar product operations compared to the usual algorithm where the
matrix elements of $H$ are calculated along with the complete
re-orthogonalization of the basis vectors.

\subsection{Discussion of the Toy Model Results}

\quad
Having considered the main problems one may face in studying the simple toy
model (\ref{H-ho-1Dbox}), 
\[
H=\frac{1}{2m}p^{2}+V_{L}(q)+\frac{m\omega ^{2}}{2}q^{2}, 
\]
we close the discussion with a sample spectrum for the case of $m=\hbar
=2L/\pi =1$ and $\omega =4$.

As one can see in Fig. \ref{w4SpectralStructure}, the first three
energy levels are really equally distant from one another and coincide with
the harmonic oscillator levels as expected from (\ref{HO spectrum ends}).
For these states, the wave functions are also the harmonic oscillator wave
functions. The intermediate spectrum is almost missing. After the $E_{c}$,
the spectrum is that of a free particle in a 1D box perturbed by the
harmonic oscillator potential. The oblique-basis type calculation reproduces
the first eight low energy states within a 14-dimensional calculation, seven
nodally adjusted harmonic oscillator states and seven states of a free
particle in a box, while the fixed-basis calculation, using only the wave
functions of a free particle in a one-dimensional box, requires 18 basis
states.

Due to the simplicity of the toy model, one does not find any big numerical
advantage of the oblique-basis calculation compared to the calculations
using the fixed basis of the 1D box wave functions. There are two main
reasons for this: (1) there is a sharp energy scale $E_{c}$ that separates
the two modes, (2) the spectrum above the energy $E_{c}$ has a nice regular
structure.

The nice regular structure above the energy $E_{c}$ results in a very
favorable situation for the usual fixed-basis calculations since the
dimension of the space needed to obtain the $n$-th eigenvalue grows as
$n+\alpha$. The parameter $\alpha$ is relatively small and does not change
much in a particular region of interest. For example, the $\omega =16$
calculations need only $\alpha =15$ extra basis vectors when calculating any
of the eigenvectors up to the hundredth vector. The relatively constant value
of $\alpha $ can be understood by considering the harmonic oscillator
potential as an interaction that creates excitations out of the $n$-th
unperturbed 1D box state. Therefore, $\alpha $ is the number of 1D box states
with energies in the interval $E_{n}^{0}$ and $E_{n}^{0}+\omega
^{2}/2\left\langle \Phi _{n}\right| x^{2}\left| \Phi _{n}\right\rangle $
where $E_{n}^{0}$ is the $n$-th unperturbed 1D box state energy. There is a
fast de-coupling of the higher energy states from any finite excitation
process that starts out of the $n$-th state. The fast de-coupling is due to
the increasing energy spacing of the 1D box spectrum. This results in a
finite number of states mixed by the presence of the harmonic oscillator
potential. Using the upper limit $E_{c}/3$ on $\delta E_{n}^{1}$, one can
easy estimate $\alpha$:
 
\[
\alpha \approx \frac{1}{3}n_{\max }^{1D}. 
\]

The sharp separation of the two modes allows for a safe use of the harmonic
oscillator states without any rescaling. This is especially true when
$\omega$ is very big since then the low energy states are naturally localized
within the box. Therefore, there is a clear shortcut: instead of
diagonalizing the Hamiltonian in some 1D box wave-function basis, one can
just use the harmonic oscillator wave functions.

\begin{figure}[tbp]
\begin{center}
\leavevmode
\epsfxsize = 5.2 in 

\centerline {\includegraphics[width= 5.2 in]{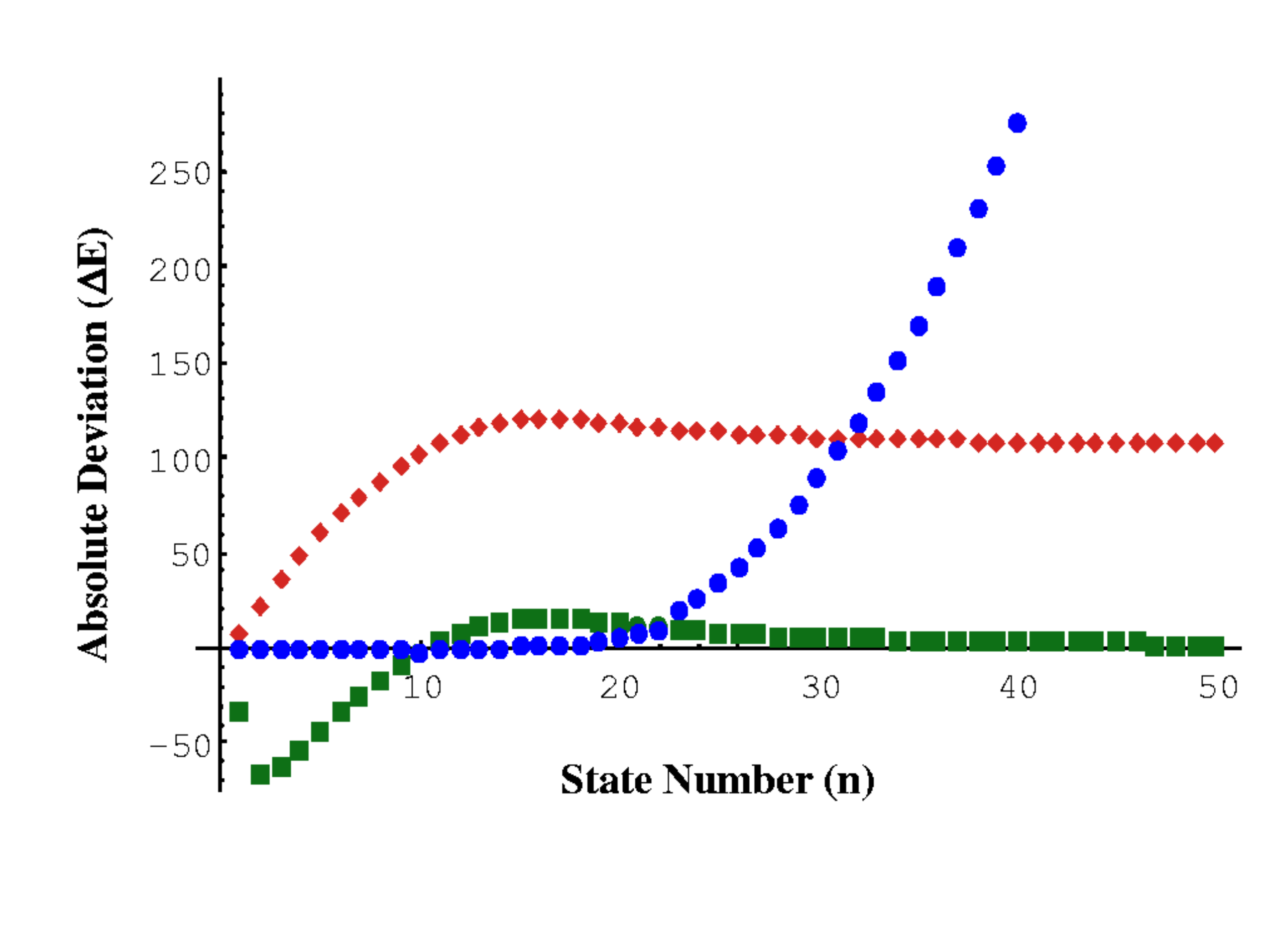}}

\end{center}
\caption{Absolute deviations from the exact energy eigenvalues for $\omega
=16$, $L=\pi /2$, $\hbar =m=1$ as a function of $n$. Blue circles represent
deviation of the exact energy eigenvalue from the corresponding harmonic
oscillator eigenvalue ($\Delta E=E^{exact}_n-E^{HO}_n$), the red diamonds are
the corresponding deviation from the energy spectrum of a particle in a 1D box
($\Delta E=E^{exact}_n-E^{1D}_n$), and the green squares are the first-order
perturbation theory results.}
\label{DeltaEn}
\end{figure}

\begin{figure}[tbp]
\begin{center}
\leavevmode
\epsfxsize =  5.2 in 

\centerline {\includegraphics[width= 5.2 in]{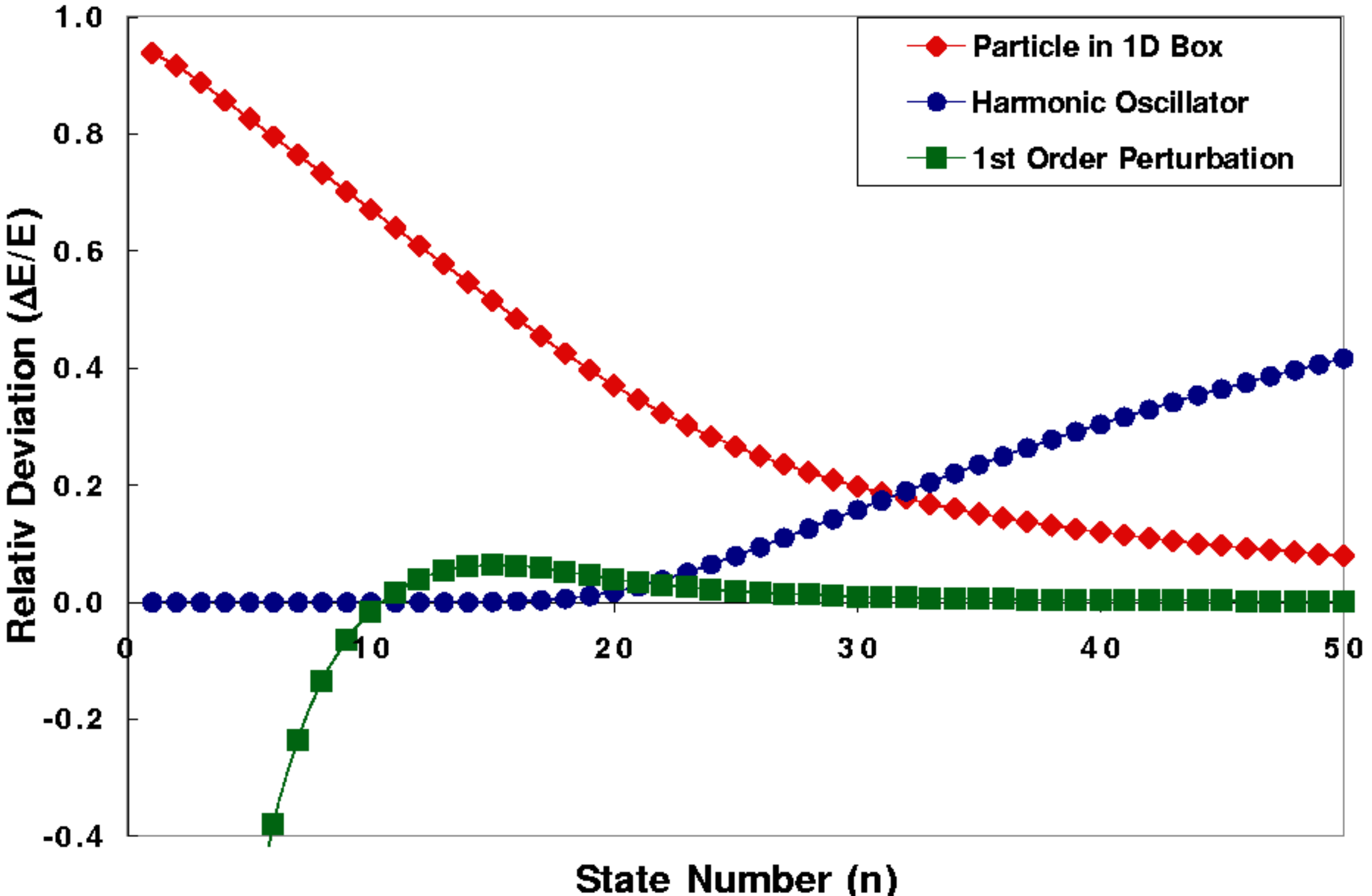}}

\end{center}
\caption{Relative deviations from the exact energy eigenvalues for $\omega
=16$, $L=\pi /2$, $\hbar =m=1$ as a function of $n$. The blue circles
represent deviation of the exact energy eigenvalue from the corresponding
harmonic oscillator eigenvalue (${\Delta E}/E=1-E^{HO}_n/E^{exact}_n$), the
red diamonds are the corresponding relative deviation from the energy spectrum
of a particle in a 1D box (${\Delta E}/E=1-E^{1D}_n/E^{exact}_n$), and the
green squares are the first-order perturbation theory results.}
\label{DeltaEoverE}
\end{figure}

Fig. \ref{DeltaEn} shows the absolute deviation ($\Delta
E=E_{n}^{exact}-E_{n}^{estimate}$) of the exact energy spectrum for the case
of $\omega =16$, $L=\pi /2$, $\hbar =m=1$. Here, $E_{n}^{estimate}$ refers to
the three energy estimates one cam make: the harmonic oscillator $E^{HO}_n$,
particle in a 1D box $E^{1D}_n$, and the first order perturbation theory
estimate considering the harmonic oscillator potential as a perturbation
($E_{n}^{1D}+\omega ^{2}/2\left\langle \Phi _{n}\right| x^{2}\left| \Phi
_{n}\right\rangle $). There are about 19 states that match a harmonic
oscillator spectrum which is consistent with the expected value from (\ref{HO
spectrum ends}). After the $n=20$ level, the perturbation theory gives
increasingly better results for the energy eigenvalues. Fig.
\ref{DeltaEoverE} shows the relative deviation
($1-E_{n}^{estimate}/E_{n}^{exact}$) of the exact energy spectrum for the
case of $\omega =16$, $L=\pi /2$, $\hbar =m=1.$

From these graphs, it seems that the transition region is somewhat absent
since the first-order perturbation theory takes on immediately after the
breakdown of the harmonic oscillator spectrum. Even though the first-order
perturbation theory gives good estimates for the energy levels in this
transition region, this is not a manifestation of a proper perturbation
theory. Rather, it is a manifestation of a coherent behavior \cite{Adiabatic
mixing}. What actually happens in this region is a coherent mixing of 1D box
states by the harmonic oscillator potential in the sense of a quasi-symmetry
discussed in the Appendix.

\begin{figure}[tbp]
\begin{center}
\leavevmode
\epsfxsize = \textwidth

\centerline {\includegraphics[width= \textwidth]{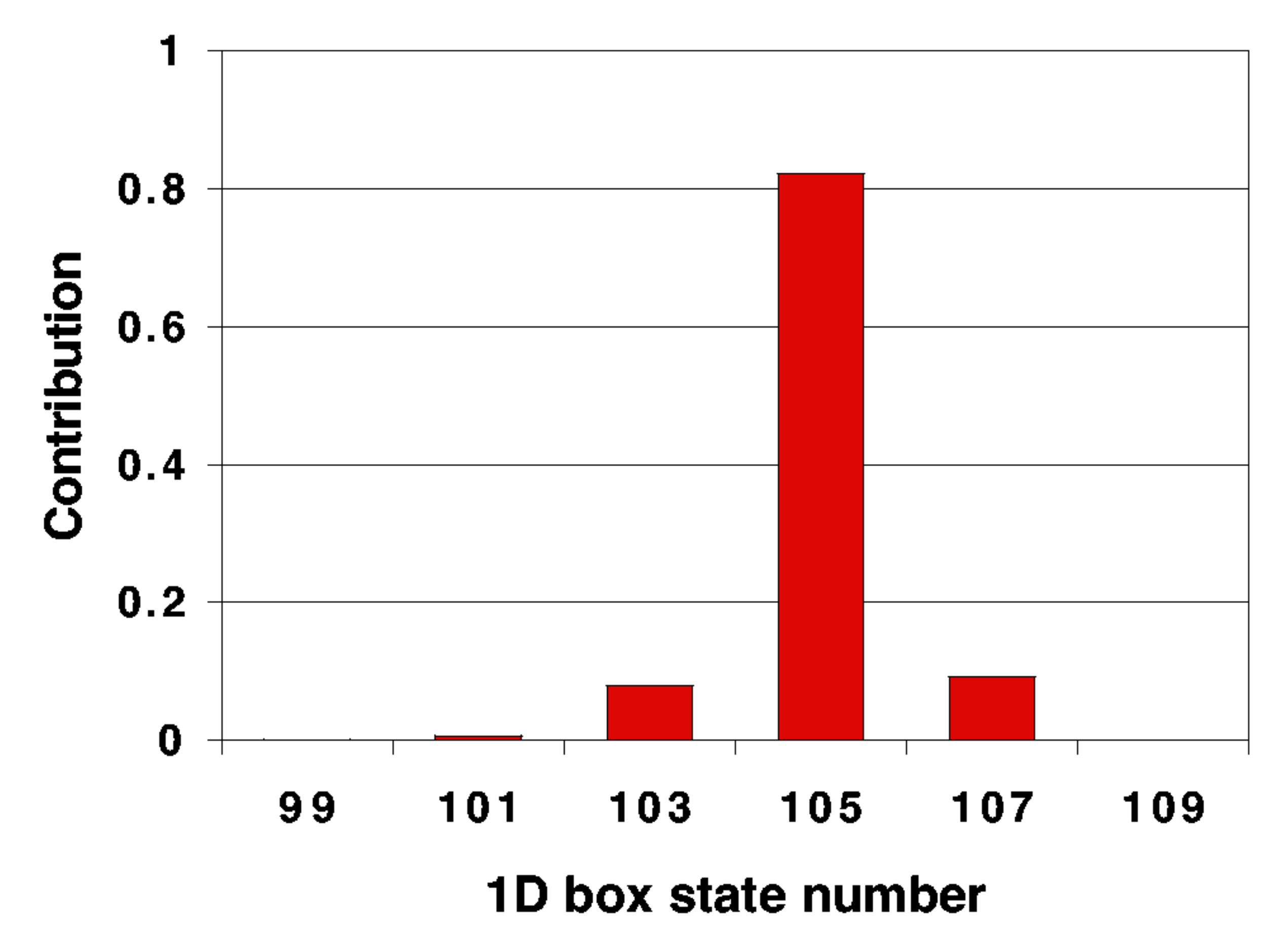}}

\end{center}
\caption{Non-zero components of the 105th exact eigenvector in the basis of
a free particle in a one-dimensional box. Parameters of the Hamiltonian are
$\omega =16$, $L=\pi /2$, $\hbar =m=1$.}
\label{State105}
\end{figure}

Notice that perturbation theory is valid, as expected, for high energy
states determined by the expression (\ref{1D box spectrum begins}). For the
high energy spectrum the harmonic oscillator potential acts as a small
perturbation. Thus the first-order corrections in the energy and the wave
function are small. Fig. \ref{State105} shows that the main component of the
105th exact wave function comes from the 105th 1D box wave function, as it
should for small perturbations.

\begin{figure}[tbp]
\begin{center}
\leavevmode
\epsfxsize = \textwidth

\centerline {\includegraphics[width= \textwidth]{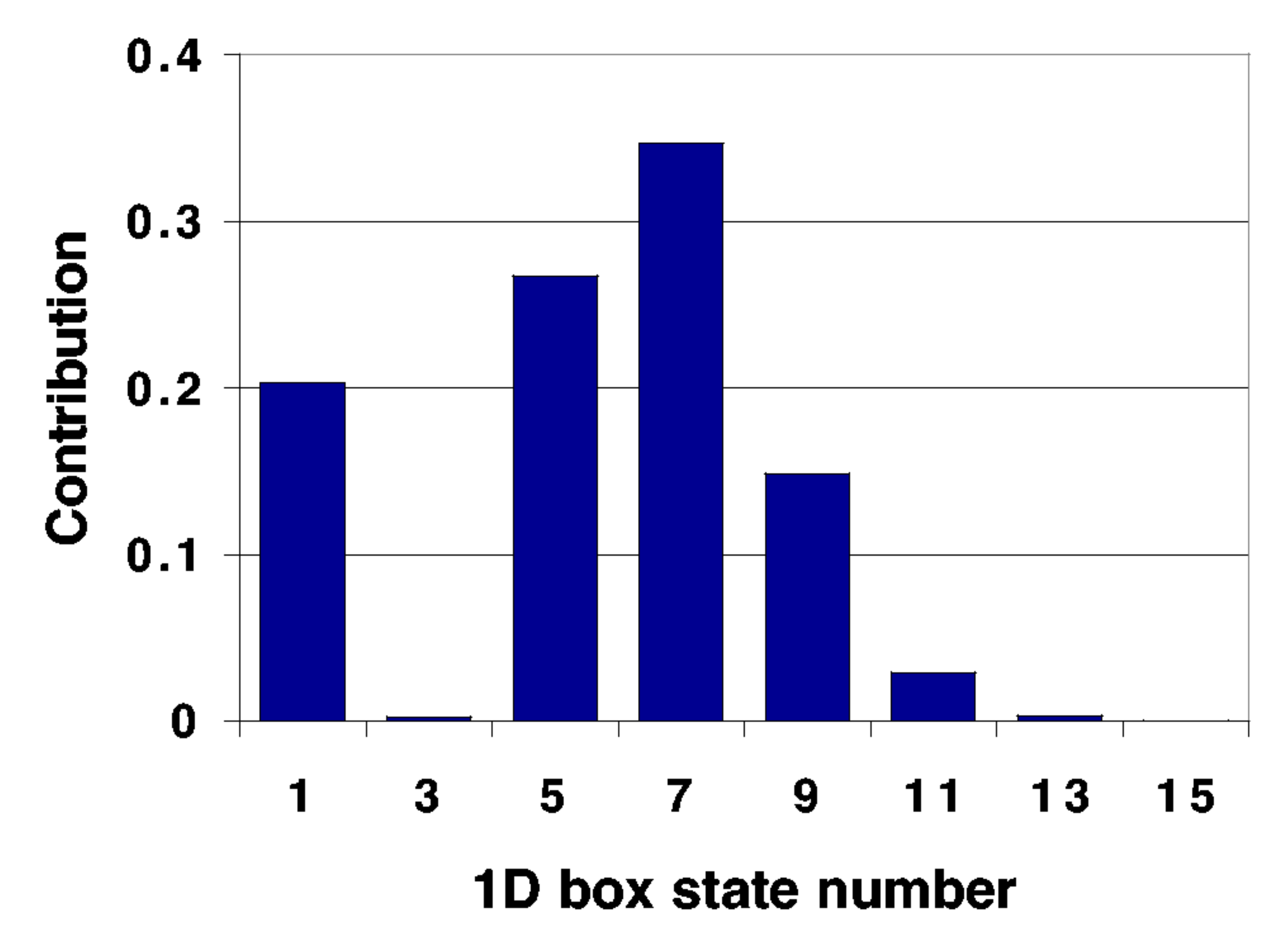}}

\end{center}
\caption{Non-zero components of the third harmonic oscillator eigenvector as
expanded in the basis of a free particle in a one-dimensional box.
Parameters of the Hamiltonian are $\omega =16$, $L=\pi /2$, $\hbar =m=1$.}
\label{State3}
\end{figure}

For low energy states, perturbation theory around the 1D box states is not
appropriate since the harmonic oscillator states are the true states in this
region. Specifically, for $m=\hbar =2L/\pi =1$ and $\omega =16,$ the first
ten states are exactly the harmonic oscillator states with a very high
accuracy. The next ten states have high overlaps with the corresponding
harmonic oscillator wave functions. For example, starting from 0.999999 at
the tenth state, the overlaps go down to 0.880755 at the twentieth state;
after that the overlaps get small very quickly. Fig. \ref{State3} shows the
structure of the third exact eigenvector when expanded in the 1D box basis.
Notice that the third 1D box wave function is almost missing from the
structure of the third harmonic oscillator wave function. Such small overlap
can happen at particular values of the parameter $\omega L^{2}$ relevant for
the problem at hand.

\begin{figure}[h]
\begin{center}
\leavevmode
\epsfxsize = 5in 

\centerline {\includegraphics[width= 5in]{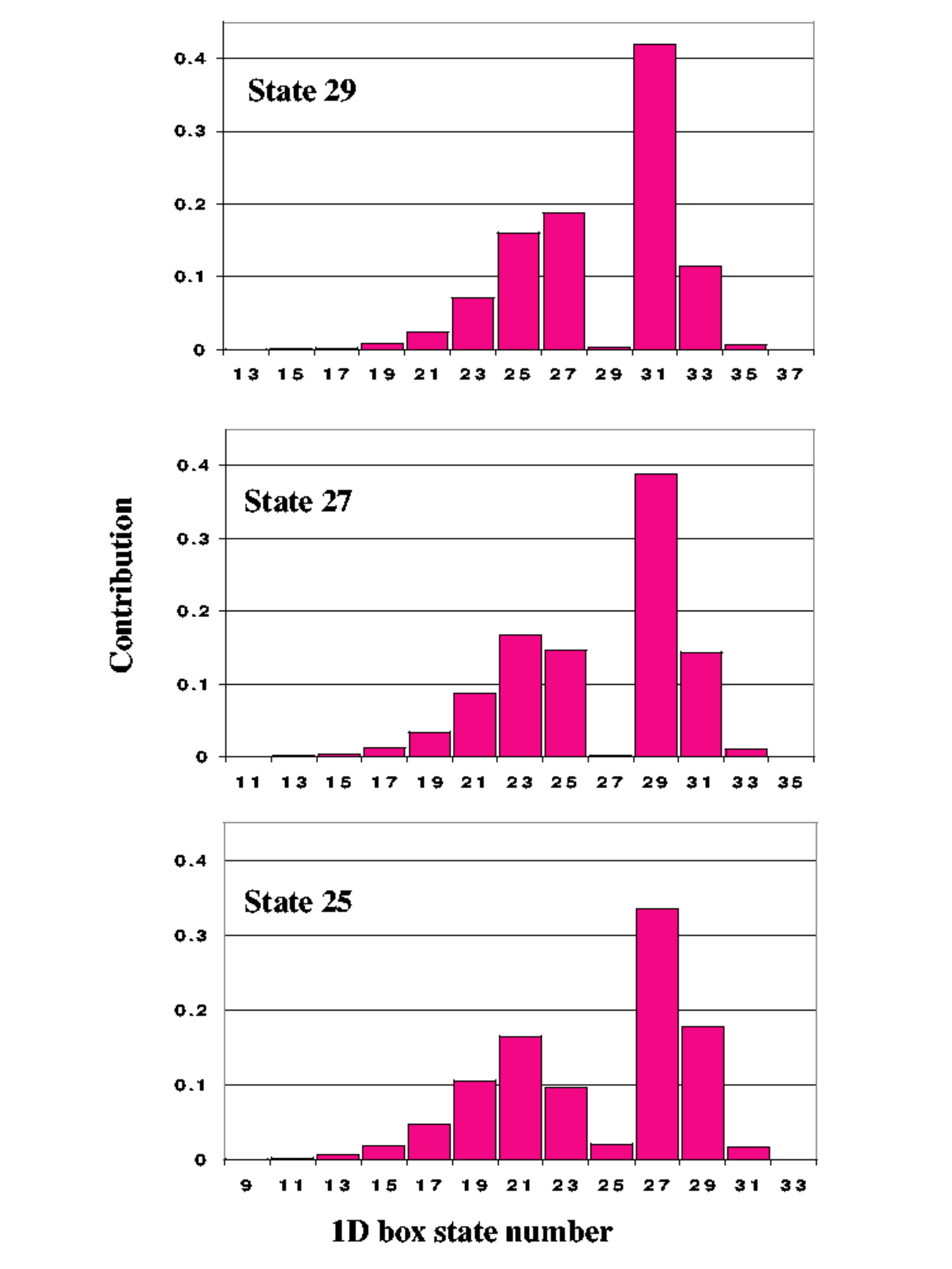}}

\end{center}
\caption{Coherent structure with respect to the non-zero components of the
25th, 27th and 29th exact eigenvector in the basis of a free particle in a
one-dimensional box. Parameters of the Hamiltonian are $\omega =16$,
$L=\pi/2$, $\hbar =m=1$.}
\label{States25to29}
\end{figure}

This pattern of having a small component of the exact wave function along
the corresponding 1D box wave function continues to persist into the
transition region. This is an unexpected behavior considering the fact that
the first order estimates of the energy levels are relatively good. Thus we
are confronted with a situation where perturbation theory is not appropriate
since level spacing is smaller than the magnitude of the ``perturbing
potential'' but the expectation values of the full Hamiltonian are
relatively close to the exact eigenvalues\footnote{A simple explanation of
this effect is that the unperturbed energies $E^0_n$ are such that $E^0_n >
\delta E^1_n$.}, even thought the corresponding 1D box wave functions are not
at all present in the exact wave function as shown in Fig. \ref{States25to29}.

In conclusion, there is a clear shortcut when using the oblique-basis idea.
This allows one to use the correct wave functions in the relevant low and
high energy regimes relative to $E_c$. There is a clear coherent mixing in the
transition region. Such a phenomenon has been also observed in the lower
$pf$-shell nuclei $^{44-48}$Ti and $^{48}$Cr which will be discussed in the
next chapters. Due to the simplicity of the model, there is a small numerical
gain in using oblique-basis calculations. However, there could be other
cases with a significant gain in using the oblique-basis type calculations.
Nuclear physics provides one such example as demonstrated in our study of
$^{24}$Mg \cite{VGG 24MgObliqueCalculations}. Another two-mode system of
interest is a particle confined in two dimensions by an external magnetic
field. This system is interesting because there is a lifting of the infinite
degeneracy of Landau like states due to the confinement \cite{Rosas et
al-2000}.

%% file: VGGPhDThesisCh4.tex
\chapter{Oblique Shell-Model Basics}

\quad
Some modern shell-model codes are based on the so-called $m$-scheme logic,
namely, the model space is spanned by many-particle configurations (Slater
determinants) with \textbf{good third component of the total angular momentum
($M_{J}$)} \cite{ANTOINE,OXBASH}. A good total angular momentum $J$, which is a
conserved symmetry due to the isotropy of the space, is obtained either by
angular momentum projection before, after, or as a consequence of the
diagonalization of the Hamiltonian. Codes of this type normally achieve a good
description of nuclear phenomena dominated by the single-particle effects. In
these codes the basis consists of single machine words representing the
many-particle configurations $\left| n_{1}...n_{k}\right\rangle
=\prod\limits_{s=1}^{k}\left(a_{s}^{+}\right) ^{n_{s}}\left| 0\right\rangle $.
Unfortunately, an equally good description of collective phenomena within the
framework of this approach is difficult to obtain due to the computational
problems associated with the size of the needed model space. On the other hand,
the $SU(3)$-based shell-model scheme is designed to give a simple
interpretation of the collective nuclear phenomena. An ideal scenario would
incorporate both the single-particle degree of freedom and the collective
degree of freedom, allowing the Hamiltonian of the system to ``choose'' the
admixture that is most appropriate. In this chapter, we discuss some of the
computational methods and techniques used in our calculations.

\section{Generalized Eigenvalue Problem}

\quad
The usual procedure for solving an eigenvalue problem $\hat{H}\vec{v}
=\lambda \vec{v}$ is to cast it into a matrix equation. In a non-orthogonal
basis \cite{Fox Lin. Alg. -nonortogonal}, this matrix equation includes an
overlap matrix ($\Theta _{ij}=\langle i|j\rangle $) and has the form
\begin{equation}
\sum_{j}\left(H_{ij}v_{j}-\lambda \Theta _{ij}v_{j}\right) =0.
\label{generalized eigenvalue problem - matrix eq.}
\end{equation}
For an orthonormal basis the overlap matrix becomes the identity matrix ($
\Theta _{ij}\rightarrow \delta _{ij}$), and the matrix form of the
eigenvalue problem is
\begin{equation}
\sum_{j}H_{ij}v_{j}=\lambda v_{i}.
\label{standart eigenvalue matrix equation}
\end{equation}
When the overlap matrix $\Theta $ is positive-definite, the Cholesky
algorithm \cite{Press -NR book Cholesky}, which decomposes $\Theta $ into the
product of an upper diagonal matrix ($U$) and its transposed ($U^{T}$), $
\Theta \rightarrow UU^{T}$, can be used to cast the generalized eigenvalue
problem (\ref{generalized eigenvalue problem - matrix eq.}) back into the
standard matrix equation (\ref{standart eigenvalue matrix equation}):
\begin{equation}
H^{\prime}\vec{v}^{\prime}=\lambda \vec{v}^{\prime},\quad H^{\prime}
=U^{-1}H\left(U^{-1}\right) ^{T},\quad \vec{v}^{\prime}=U^{T}\vec{v}.
\label{effective eigenvalue problem}
\end{equation}
The use of the Cholesky algorithm is essential for identifying the linearly
dependent vectors within the oblique basis. For large spaces, the effective
eigenvalue problem (\ref{effective eigenvalue problem}) can be solved
efficiently by using an appropriately modified Lanczos algorithm which we will
discuss in a later section.

For the calculations that will be discussed later, we use two basis sets.
The first set consists of spherical shell-model states (ssm-states)
expressed in spherical single-particle coordinates ($nlj$). The second set
has a good SU(3) structure (su3-states) which track nuclear deformation
\cite{Draayer SU3-(beta-gamma)}; this basis set is given in cylindrical
single-particle coordinates. By construction, both sets have the third
projection $M_{J}$ of the total angular momentum $J$ as a good quantum
number \cite{VGG-1998 su3 good M, the M-scheme approach}. Schematically, these
basis vectors and their overlap matrix can be represented in the following way:

\begin{eqnarray}
\mathrm{basis\quad vectors} &:&\mathrm{\quad}\left(\begin{array}{l}
e_{\alpha} :\mathrm{ssm\ - \ basis} \\
E_{i} :\mathrm{su3\ - \ basis}
\end{array}
\right), \label{Basis vectors} \\
\mathrm{overlap\quad matrix} &:&\mathrm{\quad}\Theta =\left(\begin{array}{ll}
\mathbf{1} & \Omega \\
\Omega ^{+} & \mathbf{1}
\end{array}
\right),\qquad \Omega _{\alpha i}=e_{\alpha}\cdot E_{i},
\label{Overlap matrix} \\
\mathrm{Hamiltonian\quad matrix} &:&\mathrm{\quad}H=\left(\begin{array}{ll}
H_{ssm \times ssm} & H_{ssm \times su3} \\
H_{su3 \times ssm} & H_{su3 \times su3}
\end{array}
\right) =\left(\begin{array}{ll}
H_{\alpha \beta} & H_{\alpha j} \\
H_{i\beta} & H_{ij}
\end{array}
\right). \label{hamiltonian Matrix}
\end{eqnarray}
In the above, $\alpha$ and $i$ span the following ranges: $\alpha = 1$,...,
dim(ssm-basis) and $i = 1$,..., dim(su3-basis).

Calculations in a nonorthogonal oblique basis require an evaluation of the
matrix elements of physical operators plus a knowledge of the scalar product
($e_{\alpha}\cdot E_{i}$) related to the overlap matrix. While it may be
desirable to have an analytical expression for the overlap matrix, as we
have for the single-particle overlap matrix \cite{Chacon overlap}, 
for practical purposes it suffices either to know the representation of
each basis state in a common set that spans the full space, which is counter
to the overall objective of reducing the number of basis states to a
manageable subset, or to expand one set in terms of the other. For the
present work, the $e_{\alpha}$, which can be represented by a single
machine word in a spherical single-particle scheme, were expanded in a
cylindrical basis, which is the representation for our collective SU(3)
basis vectors. This transformation is handled by an efficient routine that
exploits two computational aids: bit manipulation via logical operations and
a weighted search tree for fast data storage and retrieval \cite{Park-WST}.
Transformation of this type has to be done at least once per ssm basis state
($e_{\alpha}$). We transform the ssm-basis states since the result is
usually a vector with fewer components than a typical SU(3) basis state.
There is a simple way to calculate the overlap between states in different
single-particle bases \cite{Lang dot(a.b)=Det(A'B)}. However, for the
calculation of matrix elements of the Hamiltonian, it is better to transform
each $ e_{\alpha}$ vector in the basis used by the SU(3) states.

Matrix elements of the one-body and two-body Hamiltonian
\[
H=\sum_{i}\varepsilon _{i}a_{i}^{+}a_{i}+\frac{1}{4}
\sum_{i,j}V_{kl,ij}a_{i}^{+}a_{j}^{+}a_{k}a_{l}
\]
have to be evaluated in each subspace ($H_{\alpha \beta}$ and $H_{ji}$), as
well as between the two spaces ($H_{\alpha i}$ and $H_{j\beta}$), see (\ref
{hamiltonian Matrix}). The $H_{\alpha \beta}$ part is normally given and
evaluated in a spherical single-particle basis. By transforming the
Hamiltonian to a cylindrical single-particle basis one can obtain the $H_{ji}$
part of $H$. In order to compute the off-diagonal blocks ($H_{\alpha i}$,
$H_{j\beta}$, and overlap matrix elements between SU(3) and ssm-basis states),
both basis sets are expanded in a basis of Slater determinants using
cylindrical single-particle states. For example, any vector within the two
irreps (8,4) and (9,2) of $^{24}$Mg has at most 2120 cylindrical Slater
determinants; each ssm state, which itself is a single spherical Slater
determinant, typically expands into less than 1296 cylindrical Slater
determinants. We do not expand the SU(3) states into spherical-basis Slater
determinants because that would require a significant fraction of the entire
spherical shell-model space, defeating the rationale of our approach. Taking
into account the significant number of Hamiltonian matrix elements ($H_{ij}$
and $H_{i\beta}$) between multi-component states, it should be clear that this
is the most time consuming part of the calculation. The extra labels associated
with the intrinsic quadrupole moment $\varepsilon $ of each basis state is used
to produce well-structured band-like matrices and to speed up the calculation.
Specifically, basis states are pre-ordered according to their deformation as
reflected by $\varepsilon,$ and during the evaluation of $H$ a $\Delta
\varepsilon $ selection rule is applied.

It is important to point out that knowledge of the overlap matrix $\Theta $
and the matrix elements of $H$ in the two spaces ($H_{\alpha \beta}$, $
H_{ij}$) is not enough to obtain the correct off-diagonal block $H_{\alpha
i}.$ This is clear from the following explicit expression for $H_{\alpha i}$
which contains a summation along ($\bar{\beta}$) that lies outside of the
oblique model space ($\beta,i$):
\[
H_{\alpha i}=\sum_{\beta}H_{\alpha \beta}\Theta _{\beta i}+
\sum_{\bar{\beta}}H_{\alpha \bar{\beta}}\Theta _{\bar{\beta}i}.
\]
Thus a direct evaluation of $H_{\alpha i}$ is required.

\section{Geometrical Visualization of the Oblique Basis}

\quad
It is instructive to consider a geometrical visualization of the
oblique-basis concept. Since a set of vectors defines a hyperplane, it is
natural to ask the question: ``What is the angle between hyperplanes defined
by the bases under consideration?'' To answer this question, first consider
the angle $\theta $ between a normalized SU(3) basis vector $\vec{v}$ and
the subspace $V$ spanned by the spherical shell-model basis vectors. The
length of the projected vector $\vec{v}_{V}\in V$ is given by
$\cos(\vec{v},V)=\cos \theta =\left| \vec{v}_{V}\right| $. The space $V$ of the
spherical shell-model basis vectors induces a natural basis
$\vec{n}_{\varepsilon}$ in the SU(3) space
($\vec{n}_{\varepsilon}=n_{\varepsilon}^{i}\vec{E}_{i}$). The angle between each
new basis vector $\vec{n}_{\varepsilon}$ and the space $V$ will again be the
length of its projection into the space $V,$ but it has the nice property that
this set of orthogonal basis vectors stays orthogonal after the projection into
the space $V$:
\begin{eqnarray*}
\cos \theta _{\varepsilon} &=&\cos (\vec{n}_{\varepsilon},V)=\left| \vec{n}
_{\varepsilon V}\right|, \\
\vec{n}_{\varepsilon V} &=&\sum_{i,\alpha}n_{\varepsilon}^{i}(\vec{E}_{i}
\cdot \vec{e}_{\alpha})\vec{e}_{\alpha}= \sum_{i,\alpha}n_{\varepsilon}^{i}\Theta
_{i\alpha}\vec{e}_{\alpha}, \\
\left| \vec{n}_{\varepsilon V}\right| ^{2} &=
&\sum_{\alpha}(\sum_{i}n_{\varepsilon}^{i}\Theta
_{i\alpha})^{2}=\sum_{\alpha,i,j}n_{\varepsilon}^{i}\Theta
_{i\alpha}n_{\varepsilon}^{j}\Theta _{j\alpha}.
\end{eqnarray*}
In matrix notation this reads
\[
\left| \vec{n}_{\varepsilon V}\right| ^{2}=\vec{n}_{\varepsilon}\cdot
\hat{\Theta}\cdot \hat{\Theta}^{T}\cdot \vec{n}_{\varepsilon},
\]
where the natural basis vectors $\vec{n}_{\varepsilon}$ are eigenvectors of
the symmetric matrix $\hat{\Theta}\cdot \hat{\Theta}^{T}$
\begin{equation}
\hat{\Theta}\cdot \hat{\Theta}^{T}\cdot \vec{n}_{\varepsilon}=\varepsilon
^{2}\vec{n}_{\varepsilon}. \label{natural basis}
\end{equation}
It follows that $\left| \vec{n}_{\varepsilon V}\right| ^{2}=\vec{n}
_{\varepsilon}\cdot \hat{\Theta}\cdot \hat{\Theta}^{T}\cdot \vec{n}
_{\varepsilon}=\varepsilon ^{2}\vec{n}_{\varepsilon}\cdot \vec{n}
_{\varepsilon}=\varepsilon ^{2}$, and thus the matrix $\hat{\Theta}\cdot
\hat{\Theta}^{T}$ is positive definite ($\left| \vec{n}_{\varepsilon
V}\right| ^{2}=\varepsilon ^{2}\geq 0$) with eigenvalues determined by the $
\cos \theta $. This construction allows for a simple visualization of the
space spanned by the oblique basis: Choose the $x$-axis to correspond to
the space $V$ of all the spherical shell-model basis vectors and represent
the SU(3) space as a collection of unit vectors each at an angle $\cos
\theta =\varepsilon $ with respect to the $x$-axis. This construction will
be applied later to the geometry of oblique-basis space calculations to
demonstrate the relative orthogonality of the two vector sets, $e_{\alpha}$
and $E_{i}$.

\section{The Lanczos Algorithm}

\quad
The Lanczos algorithm is an essential scheme for obtaining a small
number of eigenvectors corresponding to the lowest or highest eigenvalues
\cite{Van Loan Cullum-Lanczos, Lanczos-1950}. It has been applied successfully
to spatial dimensions on the order of $10^{6}$ and even pushed up to
$10^{8}$ \cite{Ur et al}. This algorithm is a simple and very efficient method
to build a basis of the Hilbert space associated with an eigenvalue problem for
an operator $H$. In its simplest form, one starts with a trial state and
applies $H$ over and over to generate new states; the process can be applied
as many times as desired. This way one generates an orthonormal basis in
which the corresponding matrix of the operator $H$ is tri-diagonal. The
method is recursive and could be used in numerical, as well as in analytic
calculations \cite{Kaluza - Analytical Lanczos}. For our toy model described
earlier, we have used analytic realization (coordinate representation) of
the algorithm while for the calculations in nuclei a numerical matrix
realization has been more suitable due to the Fock representation of the
states.

In brief, the algorithm starts with the choice of a first normalized vector $
\vec{v}_{1}$ ($\left\langle \vec{v}_{1}|\vec{v}_{1}\right\rangle =1$). In
matrix calculations this vector is often chosen randomly. Then $H$ is applied
on $\vec{v}_{1}$ and a new vector orthogonal to $\vec{v}_{1}$ is constructed,
$\vec{v}_{2}=H\vec{v}_{1}-\left\langle \vec{v}_{1}|H\vec{v}_{1}\right\rangle
\vec{v}_{1}$. Next $\vec{v}_{2}$ is normalized and used to generate a new
vector and so on. It can be shown that the basis $\{\vec{v}_{n}\}$ generated
this way is orthonormal, and $H$ is tri-diagonal in this basis$.$ However,
numerical noise destroys the orthogonality and requires one to do full
reorthogonalization of each newly generated vector with all previous
vectors. One important feature of the Lanczos algorithm is that at each new
iteration it provides the vector that has the next most important
contribution within the model space. However, this is true only if our first
vector is a good trial guess to an exact state. A trial vector with some
bad components can cause problems.

Another feature of the Lanczos algorithm is that it preserves the symmetry of
the initial vector when this symmetry is a symmetry of the Hamiltonian as
well.\footnote{The Lanczos algorithm should in principle conserve symmetries;
however, machine round-off error often mixes in different states. The round-off
error is also the reason for performing a complete re-orthogonalization for each
newly generated state.} For example, if $H$ is invariant under parity
transformation, then the algorithm will produce only even parity states.
Similarly, if we start with a state with good $J$ and $M_{J}$, then applications
of $H$ will only produce states of the same symmetry. While this can be viewed
as an advantage, sometimes it can be a problem especially when the symmetry has
a finite irreducible sub-space. For example, if we start with a vector from a
finite irreducible sub-space, then after a finite number of iterations the
algorithm will exhaust the sub-space and any new vector will be linearly
dependent on the previously generated vectors. This breakdown of the algorithm
is generally overcome by introducing a new guess vector. In large matrix
diagonalizations, the new vector could be a random Gaussian vector, or it could
be ``the next vector'' from a prior given set. In our toy model, we carried out
Lanczos type calculations using the harmonic-oscillator basis as a prior given
set.

An interesting variation to the Lanczos algorithm aiming at a particular $k$-th
eigenvector has been suggested by Davidson \cite{Davidson}. In the Davidson
algorithm, one tries to increase the speed of convergence by modifying the way
a new vector is generated. For example, the Lanczos algorithm uses
$\vec{w}_{n}=H\vec{v}_{n}$ as a seed for a new vector $\vec{v} _{n+1}$ while
the Davidson algorithm uses the vector $\vec{w}_{n}=
\left(\lambda _{k}I-diag\left(H\right) \right) ^{-1}\left(He_{k}-\lambda
_{k}e_{k}\right) $ with enhanced components along the $k$-th eigenvector.
In the Davidson expression for $\vec{w}_{n}$, $e_{k}$ and $\lambda _{k}$
are the approximate $k$-th eigenvector and eigenvalue after the $n$-th
iteration, and $diag\left(H\right) $ is the diagonal part of $H$.

To solve the generalized eigenvalue problem (\ref{effective eigenvalue
problem}), we have to modify the Lanczos algorithm. In doing so, it is
more efficient to perform consecutive action of the matrices $U$ and $H$ on
the resulting vectors. The computational time in this case grows like the
dimensionality of the space to the second power ($n^{2}$). This is to be
compared to the case when one would first fully multiply these three
matrices: $(U^{-1})^{T},$ $H,$ and $U^{-1},$ which grows as the third power
of $n$ ($n^{3}$), and then act on vectors. 

In closing this section, we would like to point out some possible future
applications of the Lanczos algorithm and its modifications. Although
recently the algorithm is mostly used in huge matrix diagonalizations \cite
{Van Loan Cullum-Lanczos, Hausman - Cornelius and Vladimir}, it can also
have some applications to constrained problems as mentioned in our toy model
discussion of a particle in a box in the harmonic-oscillator basis. For such
problems, one would have to project $\vec{w}_{n}=H\vec{v}_{n}$ in the space
determined by the constraints, and only after that proceed with the
calculation. Another interesting application is related to the long standing
problem of doing \textit{ab-initio} calculations in nuclear physics with
effective interactions derived from a NN-interaction. Some current advances in
this field which takes advantage of the Lanczos algorithm has been reported by
Haxton \textit{et al.} \cite{Haxton and Song, Haxton and Luu}. A link between
the Lanczos method and space projection techniques, such as Brueckner, Feshbach,
and Bloch-Horowitz projection treatments \cite{Armen and Rau, Feshbach-1962,
Bloch and Horowitz} also seems intriguing \cite{Haxton and Song}.

\section{Mixed-Symmetry Basis for the Nuclear Shell Model}

\quad
Even though the purpose of this work has been reiterated several times in
different contexts, we feel strong motivation to state it again, but this
time in the context of a pure practical curiosity. Specifically, can we do
calculations using two or more important basis sets as usually employed in
different shell models? In particular, can we use spherical shell-model
states, which are related to the single-particle $j$-shell symmetry ($
\otimes ^{2j+1}U(1)=U(1)\otimes...\otimes U(1)\otimes U(1)$), together with
$SU(3)$ shell-model states, which are related to the $Q\cdot Q$ interaction?
With these questions in mind, which will be answered in the positive sense
in the next chapter, we continue our examination of the oblique basis concept
by focusing on the structure of the two basis sets used in our mixed-symmetry
shell-model calculations. In the rest of the chapter, we briefly go over the
spherical single-particle basis, which is then followed by a discussion of
the $SU(3)$ symmetry-adapted basis.

\subsection{Spherical Basis for Single-Particle Excitations}

\quad
We have already introduced the main concepts and notations in a previous
chapter, as well as the basic idea for configuration truncation in the
$m$-scheme basis. Here, we would like to touch upon some details related to
the specifics of our calculations. First of all the spherical basis states are
taken from an old $m$-scheme code \textit{NUCK} (\textit{GLASGOW)} \cite
{Whitehead-shell model}. This code has been used as a benchmark and a testing
ground of our oblique code results. The output from NUCK containing the
spherical shell-model states (ssm-states) is used as input for
\textit{GlsgwBasis2Redstick} which gives a binary form of the proton-neutron
basis states for use by the oblique code \textit{su3pn}. In order to speed
the calculations when $SU(3)$ states are also included, there is an option to
use $\varepsilon$--sorted ssm--states (sorting is provided by
\textit{EpsSorting}).

\subsection{SU(3) Basis for Collective Excitations}

\quad
We have already reviewed Elliott's $SU(3)$ shell model \cite{Elliott's
SU(3) model} and some of the single-particle labeling schemes in a previous
chapter. In this section, we briefly discuss the $SU(3)$ package used to
generate $SU(3)$ symmetry-adapted states with good third component of the total
angular momentum. Next the $SU(3)$ single-particle shell-model basis states
are discussed, and the action of the $SU(3)$ generators is explained. It is
then followed by a discussion of the structure of the \textit{Extreme Weight
State(s)} (EWS) of an $SU(3)$ irrep for protons (neutrons), and especially the
\textit{Highest Weight State(s)} (HWS) of the so-called leading $SU(3)$ irreps.
Once a HWS is known, then all states of the corresponding irrep can be
constructed using $SU(3)$ step operators \cite{Hecht}. Proton (neutron)
states with good third component of the angular momentum ($M_{J}$) are obtained
by considering the direct product $SU(3)\otimes SU_{S}(2)$ using spin highest
weight states, as well as spin \textit{Lowest Weight States(s)} (LWS) \cite
{VGG-1998 su3 good M}. Once the proton and neutron highest weight states are
constructed, then they can be coupled to different possible proton-neutron
highest weight states:
\[
(SU(3)\otimes SU_{S}(2))_{p}\otimes (SU(3)\otimes SU_{S}(2))_{n}\rightarrow
(SU(3)\otimes SU_{S}(2))_{pn}.
\]

This is the so-called strong coupling scheme. This scheme is used to couple
the proton and neutron irreps to some final proton-neutron irreps. Each
extreme weight state can be used for the generation of all states of good $
M_{J}$ within a given $SU(3)$ irrep. Another possible coupling scheme
extends $SU_{S}(2)$ to $SU(4)\supset SU_{S}(2) \otimes SU_{T}(2)$.
This is the supermultiplet scheme which is a good symmetry for light nuclei.
Since the goodness of the supermultiplet scheme does not extend to heavy
nuclei, we will not considered it further in this study.

\subsubsection{$\bullet$ The SU(3) Basis Generator}

\quad
The package for generation of $SU(3)$ symmetry-adapted states with good
third component of the total angular momentum consists of two major codes:
(1) $ SU(3)$ proton-neutron HWS generator (\textit{SU3\_HWS\_GEN}) and (2)
proton-neutron generator of good $M_{J}$ (\textit{PNGGMJ}). The \textit{\
SU3\_HWS\_GEN} routine provides the input for $PNGGMJ$ which generates the
basis states needed for our oblique calculations.

The overall algorithm has four basic components: 1) definition of the
single-particle levels and matrix elements of the $SU(3)$ generators for a
given proton (neutron) shell; 2) generation of the HWS of $SU(3)\otimes
SU_{S}(2)$ for a given spin $S$ and number of protons (neutrons) $N$; 3)
coupling of the proton HWS and neutron HWS to the desirable proton-neutron $
SU(3)$ HWS and $SU_{S}(2)$ LWS; 4) generation of all proton-neutron $
SU(3)\otimes SU_{S}(2)$ states with good third component of the total angular,
$ M_{J}$.

The \textit{PNGGMJ} code accepts any pn-HWS and generates all the states with a
given $M_{J}$ value. However, the \textit{\ SU3\_HWS\_GEN} code does not generate
all possible pn-HWS since one purpose of the current project is to include a few
essential $SU(3)$ basis states in an $m$-scheme type calculation. Thus the
\textit{SU3\_HWS\_GEN} code is set to generate only the leading proton and
neutron configurations and their possible couplings. Therefore, an additional
code (\textit{SU3Lister}) is needed to allow for a quick look at all the
proton-neutron $SU(3)$ irreps that can be generated by the current version of the
\textit{SU3\_HWS\_GEN} code. As a reference to a complete list of proton-neutron
$SU(3)$ irreps, one can use the $SU(3)$ reduced matrix elements package
\cite{Bahri-RME}. If other HWS are desired, then the \textit{SU3\_HWS\_GEN} code
can be modified to generate non-leading HWS. This could be done either by using 
Bahri's method \cite{Bahri-RME} or perhaps by using another more direct approach
tailored to the particular application at hand. The generation of non-leading HWS
is important when one wishes to include states that are not maximally deformed in
their intrinsic configuration. For example, non-leading HWS will be important if
the non-$Q\cdot Q$ parts of the interaction play a significant role.

\subsubsection{$\bullet$ Single-Particle Levels and the SU(3) Matrix Elements}

\quad
The foundation of a microscopic type symmetry-adapted shell-model
calculation is the structure of the single-particle levels (SPL). The
single-particle levels should be related to a representation of the symmetry
group which is $SU(3)$ in our case. Therefore, a discussion focused on the
$SU(3)$ single-particle levels and matrix elements of the $SU(3)$ generators is
desirable and will be given in the following few paragraphs.

\paragraph{Single-particle Levels - Ordering Scheme:}

\begin{figure}[tbp]
\begin{center}
\leavevmode
\centerline {\includegraphics[width= \textwidth]{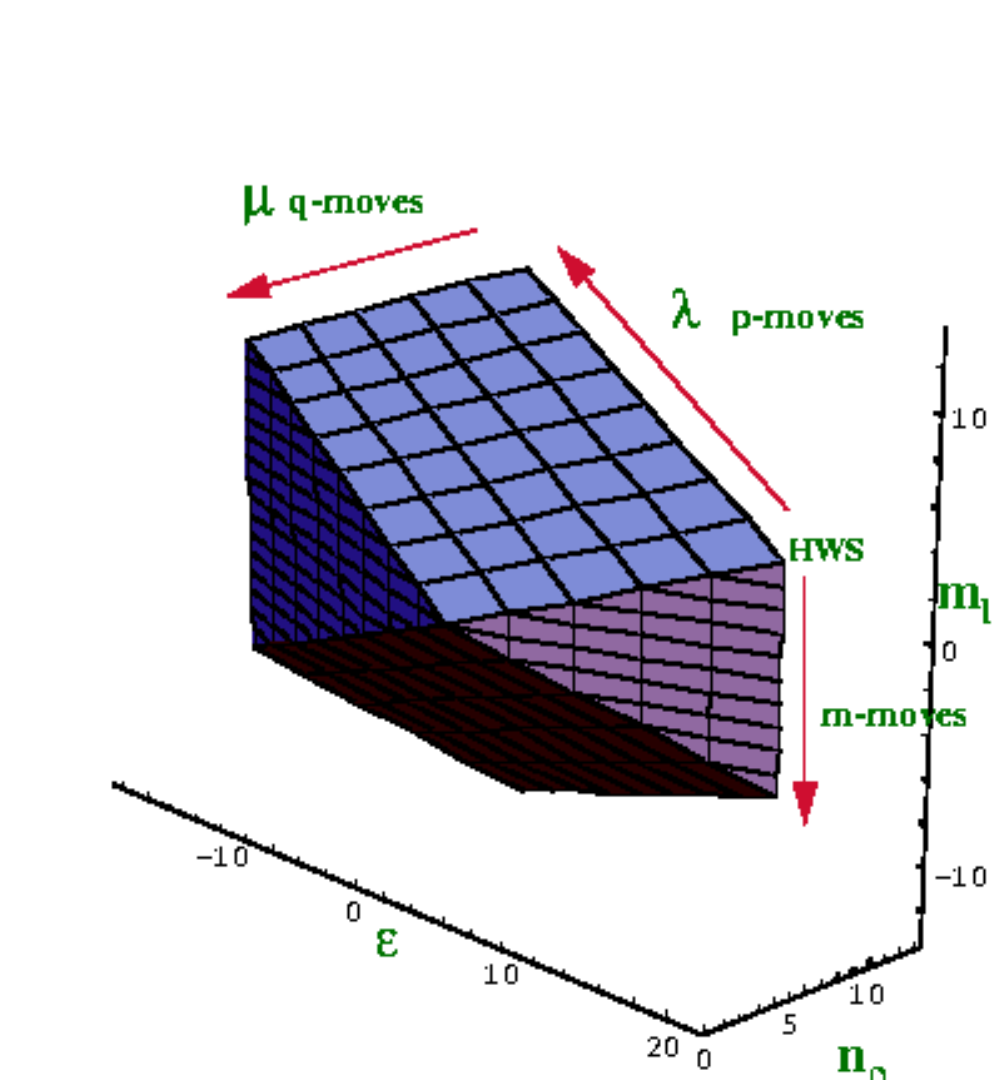}}

\end{center}
\caption{Three-dimensional view of the $(\lambda, \mu)$ $SU(3)$ irrep
together with the action of the $SU(3)$ step operators.}
\label{SU(3) Step Operators}
\end{figure}

\begin{figure}[tbp]
\begin{center}
\leavevmode
\centerline {\includegraphics[width= \textwidth]{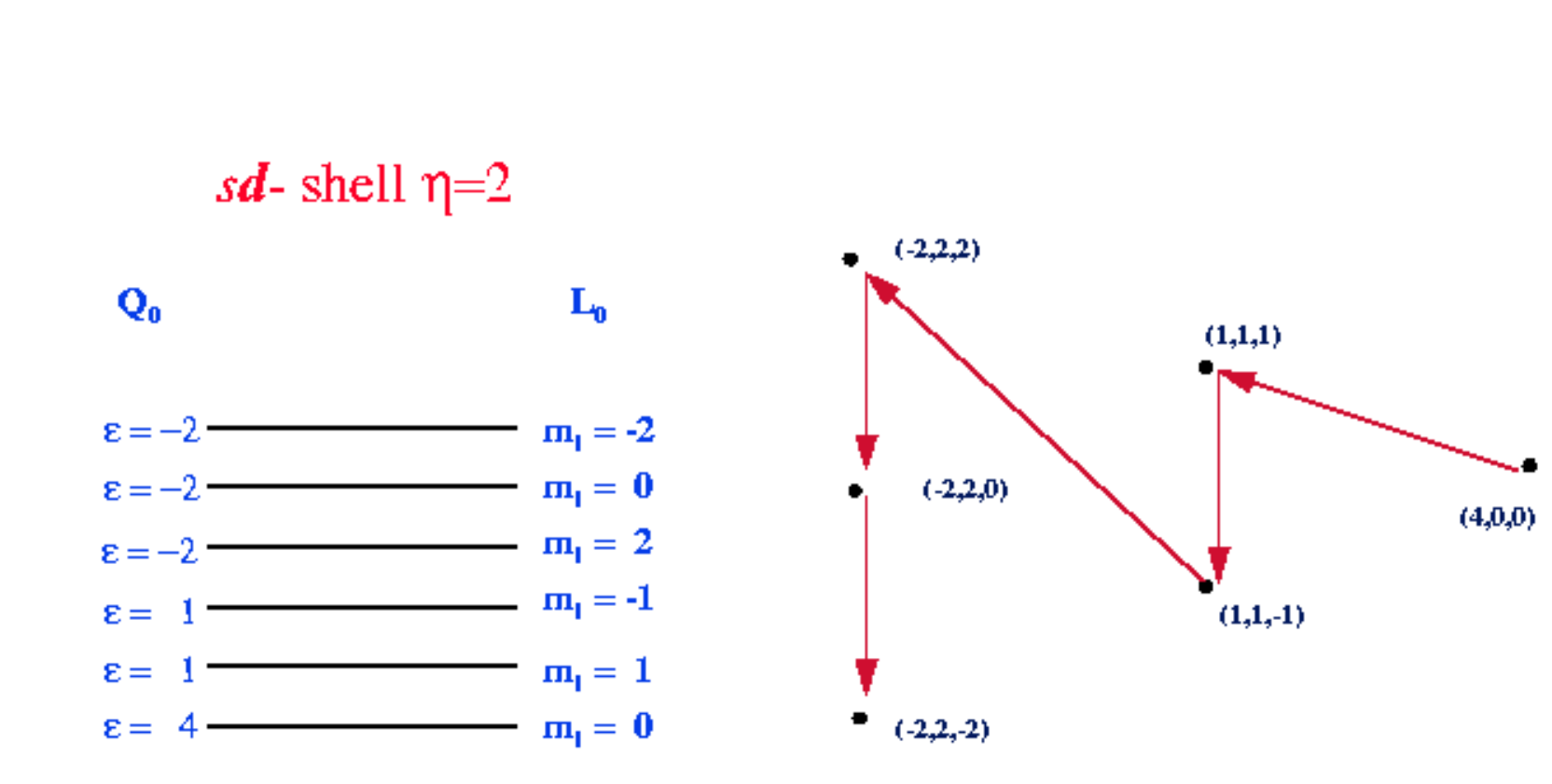}}

\end{center}
\caption{Ordering scheme of the single-particle levels.}
\label{Single Particle Levels}
\end{figure}

Single-particle levels of the $\eta =0,1,2,...$ $(s,p,sd,...)$ harmonic
oscillator shell belong to the symmetric $(\eta,0)$ irrep of $SU(3)$.
Because $\mu =0$, a typical three-dimensional representation of $SU(3)$
basis states, Fig. \ref{SU(3) Step Operators}, reduces to a special
two-dimensional triangular shape ($\varepsilon $ and $n_{\rho}$ become
linearly dependent), Fig. \ref{Single Particle Levels}. Also, because $SU(3)$
is a compact group, its irreps are finite dimensional, and many-particle
(fermion) configurations can be conveniently represented as binary strings
with a $1$ or $0$ symbolizing the presence or absence of a particle in the
corresponding single-particle level. (The latter, together with a ``sign
rule'' to accommodate fermion statistics, is a convenient computer
implementation of a Slater determinant representation of the basis states.)

Recall that the canonical reduction, $SU(3)\supset U(1)\otimes SU(2)$,
has two additive labels $\varepsilon $ ($Q_{0}$) and $m_{l}$ ($L_{0}$), and
the allowed values of these labels for fixed $SU(3)$ irrep $(\lambda,\mu)$
are given by \cite{Hecht}:

\begin{equation}
\varepsilon =2\lambda +\mu -3(p+q),\quad n_{\rho}=\mu +(p-q),\quad
m_{l}=n_{\rho}-2m \label{pqm-parametriztion again}
\end{equation}
where $0\leq p\leq \lambda $, $0\leq q\leq \mu $, and $0\leq m\leq n_{\rho}. $

A convenient ordering scheme (which tracks the arrows in Fig. \ref{Single
Particle Levels}) is set by requiring a simple representation of
many-particle configurations with maximum quadrupole deformation. This
objective can be achieved if the states are ordered by $\varepsilon $
(quadrupole moment) first and then by $m_{l}$ (third component of the angular
momentum).

\paragraph{Action of SU(3) Generators on Single-Particle States:}

\begin{figure}[tbp]
\begin{center}
\leavevmode
\centerline {\includegraphics[width= \textwidth]{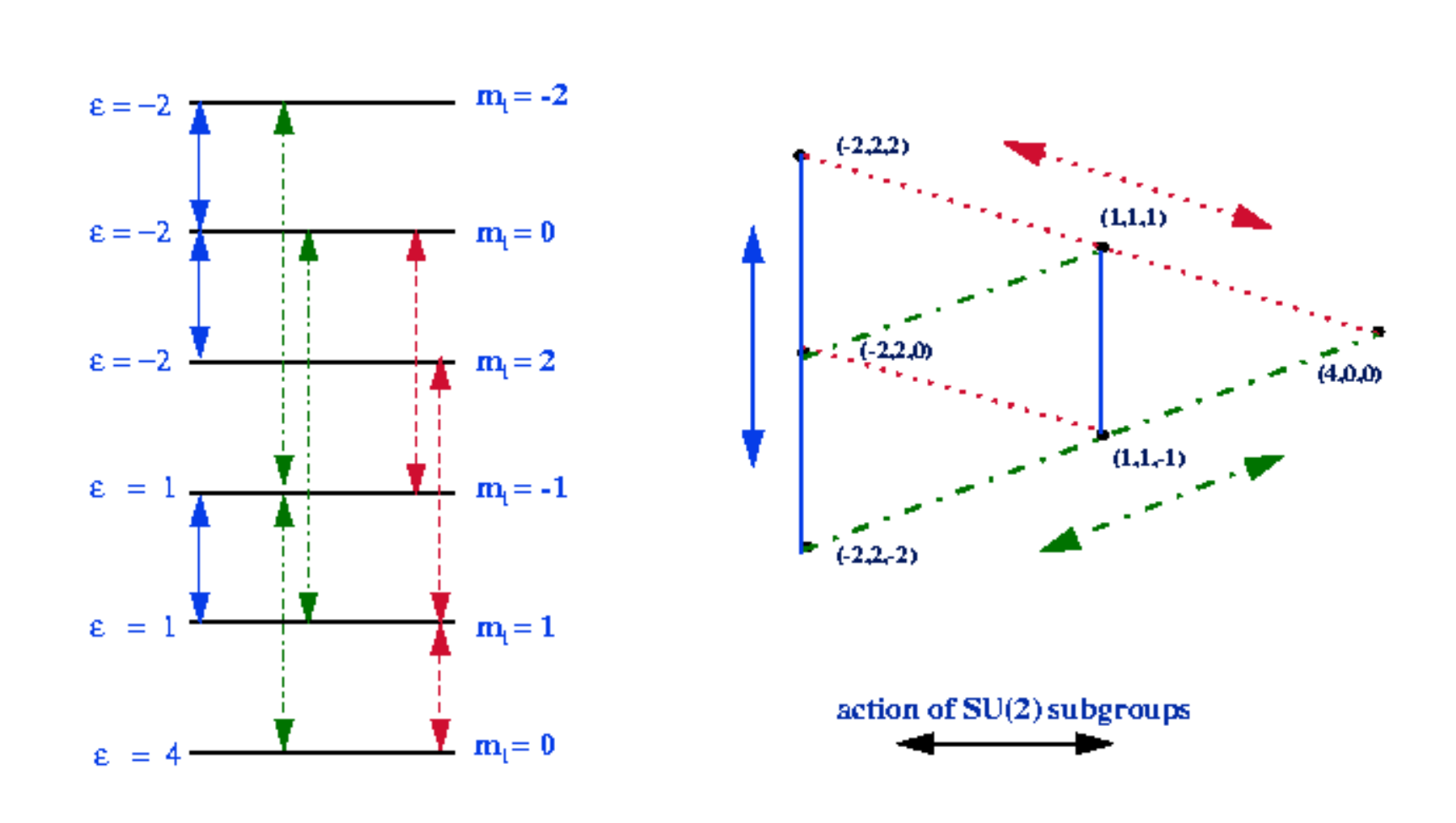}}
\end{center}
\caption{Action of the $SU(3)$ generators on the single particle levels. The
diagram on the left shows that applying an $SU(3)$ generator to a single-particle
state results in another single-particle state. The diagram on the right shows
the action of the six non-diagonal generators of $SU(3)$. The vertical solid
lines represent the action of the $SU(2)$ subgroup that enters the $SU(3)\supset
SU(2)\otimes U(1)$ chain.}
\label{SU(3) Matrix Elements}
\end{figure}

To be able to apply the $SU(3)$ generators on many-particle configurations,
it suffices to know the action of these generators on the single-particle
states. The eight generators of $SU(3)$ belong to the self-adjoint $(1,1)$
irrep of $SU(3)$. The operator structure should be chosen in the most
convenient form for the application under consideration. For the present
application, this choice is the same as used for the basis states, namely,
the $SU(3)\supset SU(2)\otimes U(1)$ reduction. The matrix elements of the $
SU(3)$ generators can be obtained either by using an application of the
appropriate Wigner-Eckart theorem or by using explicit expressions \cite
{Hecht} for determining the action of the operators on the basis states. For
computational purposes, it is better to adopt a direct solution. We use the
fact that the action of the operators is on a product of single-particle levels
each of which belongs to a symmetric $(\eta,0)$ irrep of $SU(3)$. This allows
the matrix elements of the $SU(3)$ generators to be calculated using properties
of the
$SU(2)$ only (Fig. \ref{SU(3) Matrix Elements}).

A key feature is the fact that the six non-diagonal generators of $SU(3)$
(recall that $L_{0}$ and $Q_{0}$ are diagonal) are rising or lowering
generators of $SU(2)$ subgroups of $SU(3)$. The three $SU(2)$ subgroups and
their respective actions are shown in Fig. \ref{SU(3) Matrix Elements}.
States that are collinear with one of the sides of the triangular shape
shown on the right in the figure form an irrep of the corresponding $SU(2)$
subgroup.

\subsubsection{$\bullet$ Action of SU(3) Generators on Many-Particle States}

\quad
Having been supplied with the single-particle levels and the action of the
$ SU(3)$ generators on them, we can construct many-particle states and extend
the action of the $SU(3)$ generators to these many-particle states as well.
Since one goal of the oblique-basis project is to include essential $SU(3)$
basis states in an $m$-scheme type calculation, we would briefly discuss the
maximally deformed HWS for protons (neutrons). These HWS are the leading
proton (neutron) irreps and can be coupled easily to the leading and other
proton-neutron irreps for a given nucleus. Once we have an $SU(3)$ HWS $
\otimes SU_{S}(2)$ LWS state, we can easily generate all the states with
good $M_{J}$ within this $SU(3)\otimes SU_{S}(2)$ irrep.

\paragraph{Highest Weight States of SU(3)$\otimes$SU$_{S}$(2) for Leading
Irreps:}

So far we have constructed single-particle states and evaluated matrix
elements of the generators of $SU(3)$ when they act on these states. The
next step is to construct many-particle HWS of $SU(3)\otimes SU_{S}(2)$. In
the chosen $SU(3)$ labelling scheme, there are seven extreme states which
correspond to the vertex points of the three-dimensional diagram
(Fig. \ref{SU(3) Step Operators}) of a general $(\lambda,\mu)$ irrep. We are
particularly interested in the vertex that has the maximum value for the
quadrupole moment of the system (Fig. \ref{SU(3) Step Operators}). Our HWS is
the state with $\varepsilon =2\lambda +\mu $, $n_{\rho}=\mu $, and $m_{l}=\mu
$. This HWS (maximum value of $m_{l}$ for maximum $\varepsilon $) can be easily
constructed by ensuring that the action of the $SU(3)$ rising generators
annihilates it. Indeed, for such a HWS, the values of $\lambda $ and $\mu $
can be determined from its $\varepsilon$ and $m_{l}$ labels.

Selecting the leading $(\lambda,\mu)$ irrep (HWS with maximum overall
value of $\varepsilon $) out of all possible irreps of an $N$ fermion system
with total system spin $S$ is very simple within the chosen scheme. This is
because the number of particles with spin up $n_{\uparrow}$ and spin down $
n_{\downarrow}$ is uniquely determined by the solution of two linear
equations:
\[
N=n_{\uparrow}+n_{\downarrow},\quad 2S=n_{\uparrow}-n_{\downarrow}.
\]
The second of these two equations expresses the fact that we also require the
state to be highest weight with respect to $SU_{S}(2)$. Further, maximizing the
value of $Q_{0}$ is achieved by filling the single-particle states of the $
(\eta,0)$ irrep (Fig. \ref{Single Particle Levels}) from bottom to top. The
chosen scheme ensures that this simple procedure gives maximum values for $
\varepsilon $ and $m_{l}$. The $SU(3)$ irrep labels $(\lambda,\mu)$ are
obtained by evaluating the quadrupole moment ($Q_{0}$) and angular momentum
projection ($L_{0}$) which are additive quantum numbers.

\begin{figure}[tbp]
\begin{center}
\leavevmode
\epsfxsize = 5in 
\centerline {\includegraphics[width= 5in]{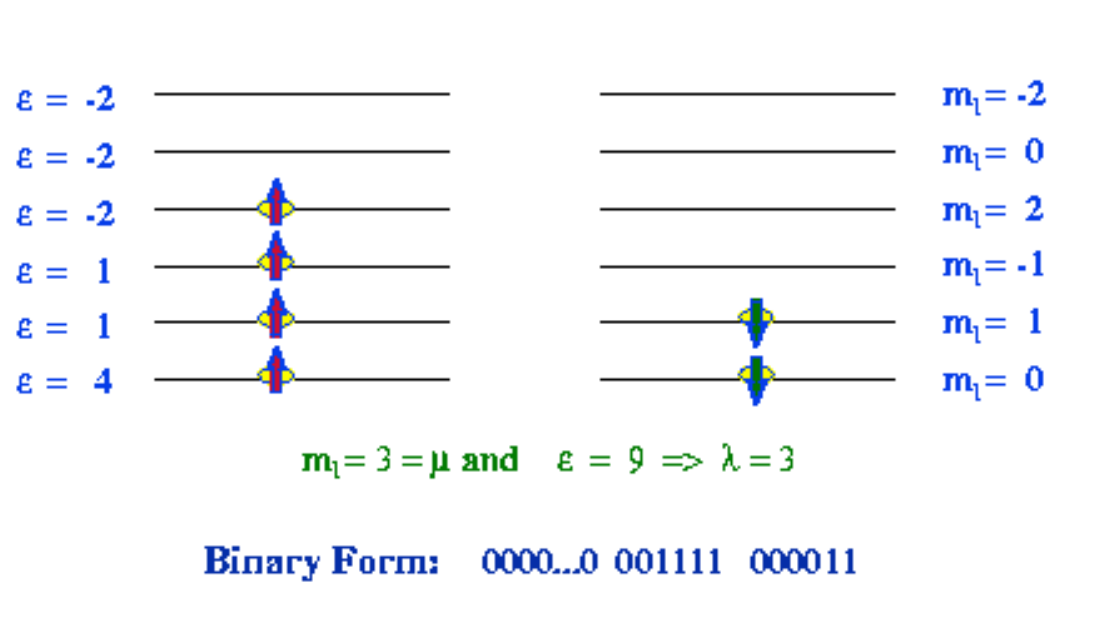}}

\end{center}
\caption{Highest weight state of the leading $(3,3)$ irrep for $N=6$ and $
S=1 $ in the $sd$-shell.}
\label{N=6&S=1}
\end{figure}

\begin{figure}[tbp]
\begin{center}
\leavevmode
\epsfxsize = 5in 

\centerline {\includegraphics[width= 5in]{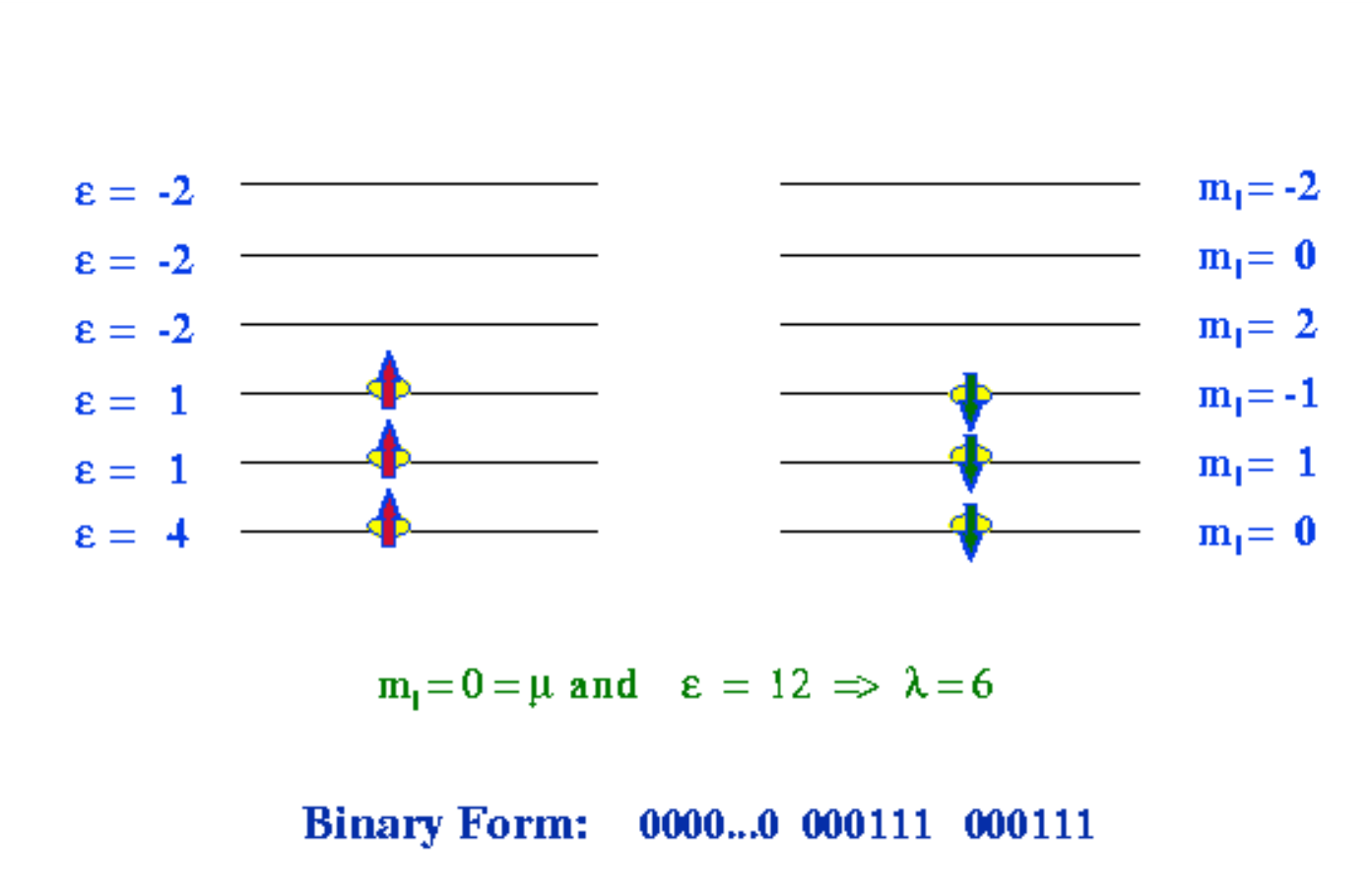}}

\end{center}
\caption{Highest weight state of the leading $(6,0)$ irrep for $N=6$ and $
S=0 $ in the $sd$-shell.}
\label{N=6&S=0}
\end{figure}

For example, in the $sd$-shell, there are six single-particle levels
corresponding to the $(2,0)$ irrep of $SU(3)$. The HWS of the leading irrep
for $N=6$ particles and total system spin $S=1$ is $(3,3)$ (Fig. \ref
{N=6&S=1}), whereas for $S=0$ the leading $SU(3)$ irrep is $(6,0)$
(Fig. \ref{N=6&S=0}). These many-particle configurations are HWS with respect to
$SU(3)
$ and $SU_{S}(2)$.


\paragraph{Generating SU(3) States with Step Operators:}

Once we have the HWS of $SU(3)$ $\otimes $ $SU_{S}(2)$, we can generate any
other state of the $SU(3)$ irrep $(\lambda,\mu)$ by applying step
operators similar to those given by Hecht. This process is needed when we
produce proton-neutron coupled $SU(3)$ irreps. By using the parameterization
(\ref{pqm-parametriztion again}) \cite{Hecht}, we can identify the
corresponding step operators, Fig. \ref{SU(3) Step Operators}.

It is important to note that applying $p$-move or $q$-move step operators
to states on the top surface yields other states (or zero) on that same
surface. Since the states on the top surface are HWS with respect to $SU(2)$
in the $SU(3)$ $\supset $ $SU(2)\otimes U(1)$ reduction, the $m$-move step
operator is an $SU(2)$ lowering operator which changes the third component of
the angular momentum ($m_{l}$). The $p$-move and $q$-move step operators can
be obtained by imposing the restriction that they generate only
transformations within the $SU(2)$ HWS space. From an algebraic perspective,
the $p$-move and the $m$-move operators are linear in $SU(3)$ generators
while the $q$-move operator is quadratic. Nevertheless, the state
generation process can be written in such a way that the $q$-move operator
effectively reduces to a linear action. These step operators can also be
obtained by a projection operator technique \cite{Tolstoy}.

\paragraph{Generating Proton-Neutron SU(3) Highest Weight States:}

In brief, the generation of the proton-neutron $SU(3)$ HWS is just a matter
of $SU(3)$ and $SU_{S}(2)$ couplings. However, the actual algorithm for the
generation process is somewhat backwards to the structure of the sentences
that we would use to describe it. What we mean by this is that when the
number of protons and the number of neutrons are given together with their
harmonic oscillator shells, then the algorithm does a loop over all possible
total proton-neutron spins ($S_{pn}$). For each total spin ($S_{pn}$), loops are
made over the possible proton spin ($S_{p}$) and neutron spin ($S_{n}$) that
can couple to the total spin ($S_{pn}$). This way the proton HWS can be
constructed, as described in the previous sections, by using the proton
number($N_{p}$) and spin ($S_{p}$) to get the $(SU(3)\otimes SU_{S}(2))_{p}$
HWS state. The same is done for the neutron HWS. This way all the major
labels, $\left| N(\lambda,\mu)\varepsilon,n_{\rho},m_{l}\right\rangle _{p}$ $
\otimes \left| SM_{S}\right\rangle _{p}$ for protons and $\left| N(\lambda,
\mu)\varepsilon,n_{\rho},m_{l}\right\rangle _{p}$ $\otimes \left|
SM_{S}\right\rangle _{p}$ for neutrons, have been determined and one can
proceed with the details of the coupling to $\left|
N(\lambda,\mu)\varepsilon,n_{\rho},m_{l}\right\rangle _{pn}$ $\otimes
\left| SM_{S}\right\rangle _{pn}$.
\[
(SU(3)\otimes SU_{S}(2))_{p}\otimes (SU(3)\otimes SU_{S}(2))_{n}\rightarrow
(SU(3)\otimes SU_{S}(2))_{pn}.
\]

One final detail on the generation process is that the proton HWS state is
actually an $SU(3)$ HWS but a spin LWS while for the neutrons it is an $SU(3)
$ and a spin HWS. Then they are coupled to a proton-neutron $SU(3)$ HWS with
a spin LWS structure. The reason behind this is that in the $SU(2)$ type
coupling one does a loop such that $m_{p}+m_{n}=m_{pn}.$ Hence, if $m_{p}$
is the minimal $m$-value of the proton spin such that $m_{p}+m_{n}=m_{pn}$
for a fixed $m_{pn}$, then the $m_{n}$ should be the maximal $m$-value of
the neutron spin. Therefore, the loop which satisfies $m_{p}+m_{n}=m_{pn}$
will have an increase of $m_{p}$ and simultaneous decrease of $m_{n}$ so
that $m_{pn}$ stays fixed. This is also the reason why the total
proton-neutron state is a LWS ($M_{S}=-S_{pn}$) so that the coupling to the
good $M_{J}=M_{l}+M_{S}$ is done in the same way since $M_{l}$ is at maximum
in the proton-neutron $SU(3)$ HWS.

\paragraph{Generating States of Good M$_{J}$:}

Since the action of the $SU(3)$ commutes with that of the spin group
($SU_{S}(2)$), it is not difficult to achieve the final goal of states with
good third component of the total angular momentum ($M_{J}$). Recall
that we have just generated proton-neutron EWS ($SU(3)^{HWS}$ $\otimes $
$SU_{S}(2)_{LWS}$). We also introduced the procedure to generate other states
of an $SU(3)$ irrep by applying step operators. Each of these states remains a
LWS of $ SU_{S}(2)$. Hence, after each move on the top surface  (see Fig.
\ref{SU(3) Step Operators}), using $p$-move or $q$-move step operators, we can
apply spin rising and angular momentum lowering to the corresponding state
toward $ M_{J}=M_{l}+M_{s}$. This way, we can generate all proton-neutron
states with labels:
$N$, $S$, $(\lambda,\mu)$, $\varepsilon $, $n_{\rho}$, $M_{l}$, and $M_{J}$.

%% file: VGGPhDThesisCh5.tex
\chapter{$^{24}$Mg Mixed-Symmetry Calculations}

\quad
The success and applicability of the oblique-basis approach to $^{24}$Mg,
which will be demonstrated in the following sections, can be related to the
fact that the spherical shell-model states are eigenstates of the one-body
Hamiltonian $(\sum \varepsilon _{i}a_{i}^{+}a_{i})$ while the two-body part
of the Hamiltonian ($\sum_{i,j}V_{kl,ij}a_{i}^{+}a_{j}^{+}a_{k}a_{l}$) is
strongly correlated with the quadrupole-quadrupole interaction ($Q\cdot Q$)
which is diagonal in the SU(3) basis \cite{QQ in sd-shell}. By combining
spherical shell-model states and SU(3) states, one accommodates, from the
onset, the dominant modes of the system.

In this chapter we discuss the oblique-basis technique as applied to $^{24}$Mg
\cite{VGG 24MgObliqueCalculations}. This is a strongly deformed nucleus with
well-known collective properties and is one of the best manifestations of the
Elliott's SU(3) symmetry \cite{Elliott's SU(3) model}. In terms of
dimensionality of the model space, adding a few leading SU(3) irreps to a
highly truncated spherical shell-model basis results in significant gains in
the convergence of the low-energy spectra towards the full space result. In
particular, the addition of leading SU(3) irreps yields the right placement of
the K=2 band and the correct order for most of the low-lying levels. Indeed, an
even more detailed analysis shows that the structure of the low-lying states is
significantly improved through the addition of a few SU(3) irreps. The
Hamiltonian used in our analysis is the Wildenthal interaction
\cite{Wildenthal}.

In the following sections we summarize some of the important features of the
spherical, SU(3), and mixed-symmetry type shell-model calculations. First we
discuss the dimensions of each model space. Then we consider the
ground-state energy as a function of the model space used. Next we focus on
the structure of the low-energy spectrum, and finally we discuss the
structure of the states as compared to the exact $sd$-shell results.

\section{Structure of the Model Spaces}

\quad
One important question in any computational study is the dimensions of
the matrices involved, as well as the structure of the model space used. In
this section, we address this question by briefly summarizing the space
structure and dimensions for the spherical, SU(3), and oblique shell-model
calculations.

\subsection{Model Space Dimensions}

\quad
Our model space for $^{24}$Mg consists of 4 valence protons and 4 valence
neutrons in the $0\hbar \omega $ $sd$-shell. The $m$-scheme dimensionality 
($M_{J}=0$) of this space is 28503. This space is denoted as FULL in the
figures that follow. To test the effects of truncations, calculations were
also carried out permitting $n$ particles to be excited out of the lowest 
$d_{5/2}$ orbit, i.e. $d_{5/2}^{8-n}(d_{3/2}s_{1/2})^{n}$, and are denoted as
SM(n). The SM(2) approximation is of particular interest since it allows one
to take into account the effect of pairing correlations (one pair maximum)
in the `secondary levels' ($s_{1/2}$ and $d_{3/2}$ for the $^{24}$Mg) with a
minimum expansion of the model space. The SU(3) part of the basis includes
two scenarios: one with only the leading representation of SU(3), which for 
$^{24}$Mg is the (8,4) irrep, with dimensionality 23 for the $M_{J}=0$ space
and denoted in what follows by (8,4); and another with the (8,4) irrep plus
the next most important representation of SU(3), namely the (9,2). The (9,2)
irrep occurs three times, once with $S=0$ ($M_{J}=0$ dimensionality 15) and
twice with $S=1$ ($M_{J}=0$ dimensionality $2\times 45=90$). All three (9,2)
irreps have total $M_{J}=0$ dimensionality of 15+90=105. The (8,4)\&(9,2)
case has total $M_{J}=0$ dimensionality of 23+105=128 and is denoted by
(8,4)\&(9,2). In Table \ref{Table1} we summarize the dimensionalities
involved.

\begin{table}[tbp]
\caption{Labels and $M_{J}=0$ dimensions for various $^{24}$Mg calculations.
The leading SU(3) irrep is denoted by (8,4) while (8,4)\&(9,2) implies that
(9,2) irreps have also been included. The SM(n) spaces correspond to
spherical shell-model partitions with $n$ valence particles excited out of
the $d_{5/2}$ shell into the $s_{1/2}$ and $d_{3/2}$ levels.}
\label{Table1}
\vskip 0.25cm 
\begin{center}
\begin{tabular}{rrrrrrrr}
\hline
Model space & $(8,4)$ & $(8,4)\&(9,2)$ & $SM(0)$ & $SM(1)$ & $SM(2)$ & $
SM(4) $ & $FULL$ \\ \hline
space dimension & $23$ & $128$ & $29$ & $449$ & $2829$ & $18290$ & $28503$
\\ 
\% of the full space & $0.08$ & $0.45$ & $0.10$ & $1.57$ & $9.92$ & $64.17$
& $100$
\end{tabular}
\end{center}
\vskip 0.25cm
\end{table}

\subsection{Visualizing the Oblique Basis for $^{24}$Mg}

\begin{figure}[tbp]
\begin{center}
\leavevmode
\epsfxsize = \textwidth
\centerline {\includegraphics[width= \textwidth]{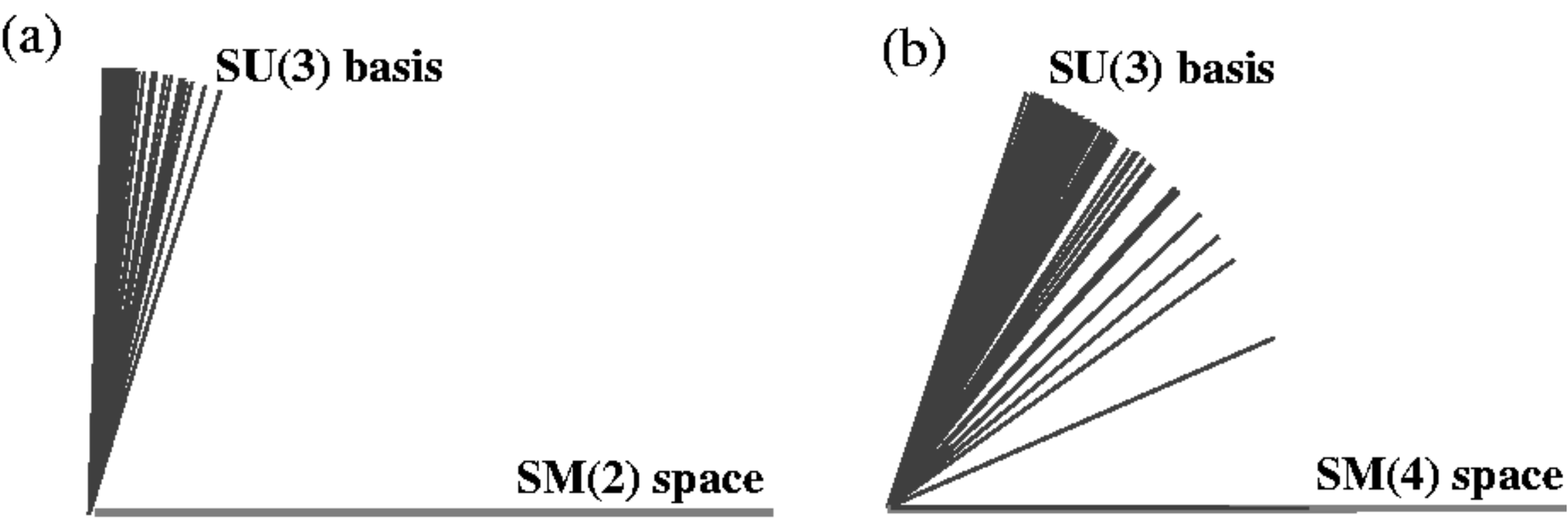}}
\end{center}
\caption{Orthogonality of the basis vectors in the oblique geometry. The
SU(3) space consists of (8,4)\&(9,2) basis vectors with the shell-model
spaces (SM(n) with n=2 and 4) indicated by a horizontal line. (a) SM(2) and
the natural SU(3) basis vectors and (b) SM(4) and the natural SU(3) basis
vectors. In the latter case (b), there are five SU(3) vectors that lie in
the SM(4) space.}
\label{SU3+ Relative To SM(2) and SM(4)}
\end{figure}

\quad
After obtaining an idea of the space dimensions involved, we now try to
visualize the oblique basis. The method described in the previous chapter
can be used to visualize the structure of the oblique basis. First, consider
the SM(2) space enhanced by the SU(3) irreps (8,4)\&(9,2). Since the SM(2)
and (8,4)\&(9,2) spaces are both relatively small (see Table \ref{Table1}),
we expect the basis vectors of these spaces to be nearly orthogonal. This
orthogonality is clearly seen from inset (a) in Fig. \ref{SU3+ Relative To
SM(2) and SM(4)}. Inset (b) in Fig. \ref{SU3+ Relative To SM(2) and SM(4)}
shows a loss of orthogonality between the SM(4) and the (8,4)\&(9,2) basis
vectors. This is due to the fact that SM(4) space is about 64\% of the full 
$sd$-space. Therefore, there is a relatively high probability that some
linear combinations of SU(3) basis vectors lie in the SM(4) space. Indeed,
it can be shown that there are five vectors from (8,4)\&(9,2) that lie
within the SM(4) space. Such redundant vectors are identified and excluded
from the calculation within the Cholesky algorithm when it is applied to the
overlap matrix ($\Theta \rightarrow UU^{T}$).

\section{Spectral Characteristics}

\quad
Reproducing the correct energy spectra of a nucleus is one of the goals
of any nuclear-structure study. Since we are trying to develop a new concept
for nuclear structure studies, the mixed-symmetry approach, we are currently
comparing our results only with full shell-model calculations. Therefore,
a computation-to-computation comparison is our reality check. In the next
sub-sections we compare the ground-state energy and energy spectrum for 
$^{24}$Mg as calculated with the Wildenthal interaction \cite{Wildenthal}
using spherical, SU(3), and mixed-symmetry shell-model bases.

\subsection{Ground-State Energy}

\begin{figure}[tbp]
\begin{center}
\leavevmode
\epsfxsize = \textwidth
\centerline {\includegraphics[width= \textwidth]{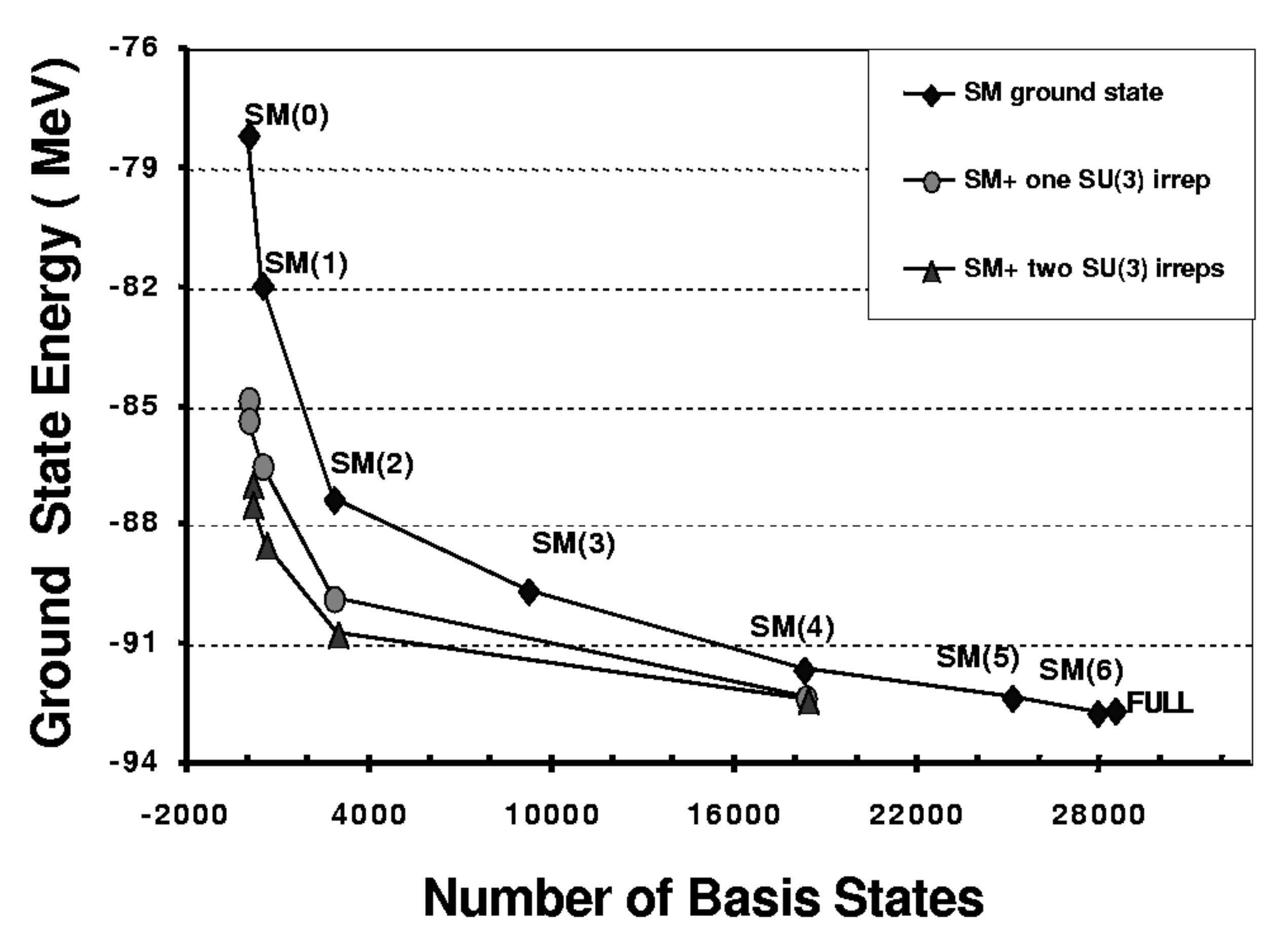}}
\end{center}
\caption{Calculated ground-state energy for $^{24}$Mg. Ground-state energy
as a function of the various model spaces. Note the dramatic increase in
binding (3.3 MeV) in going from SM(2) to SM(2)+two SU(3) irreps, (8,4)\&(9,2),
(a 0.5\% increase  in the dimensionality of the model space). Enlarging the
space from SM(2) to SM(4) (a 54\% increase in the dimensionality of the model
space) adds 4.2 MeV in binding energy.}
\label{Mg24DimConv}
\end{figure}

\quad
We now turn to the consideration of the main results of the oblique-basis
calculation, starting with ground-state convergence issues. The results
shown in Fig. \ref{Mg24DimConv} illustrate that the oblique-basis
calculation gives good dimensional convergence in the sense that the
calculated ground-state energy for the SM(2)+(8,4)\&(9,2) calculation is 3.3
MeV below the calculated energy for the SM(2) space alone. Adding the SU(3)
irreps only increases the size of the space from 9.9\% to 10.4\% of the full
space. Compare this 0.5\% increase in the size of the space with the huge
(54\%) increase in going from SM(2) to an SM(4) calculation. For the latter,
the ground-state energy is 4.2 MeV lower than the SM(2) result, somewhat
better than for the SM(2)+(8,4)\&(9,2) calculation but in 64.2\% rather than
10.4\% of the full model space.

The exponential fall-off of the ground-state energy in Fig. \ref{Mg24DimConv}
is striking. It has been observed many times, and has been recently
suggested as a possible extrapolation procedure for obtaining the
ground-state energy \cite{Zelevinsky and Volya}. An even more rigorous
extrapolation procedure has been suggested by Mizusaki and Imada \cite{Mizusaki
and Imada}. Within this procedure, one can also estimate the error of a given
calculation. Their procedure is also applicable to other observables, as well.

\subsection{Energy and Angular Momentum of the Low Lying States}

\begin{figure}[tbp]
\begin{center}
\leavevmode
\epsfxsize = \textwidth
\centerline {\includegraphics[width= \textwidth]{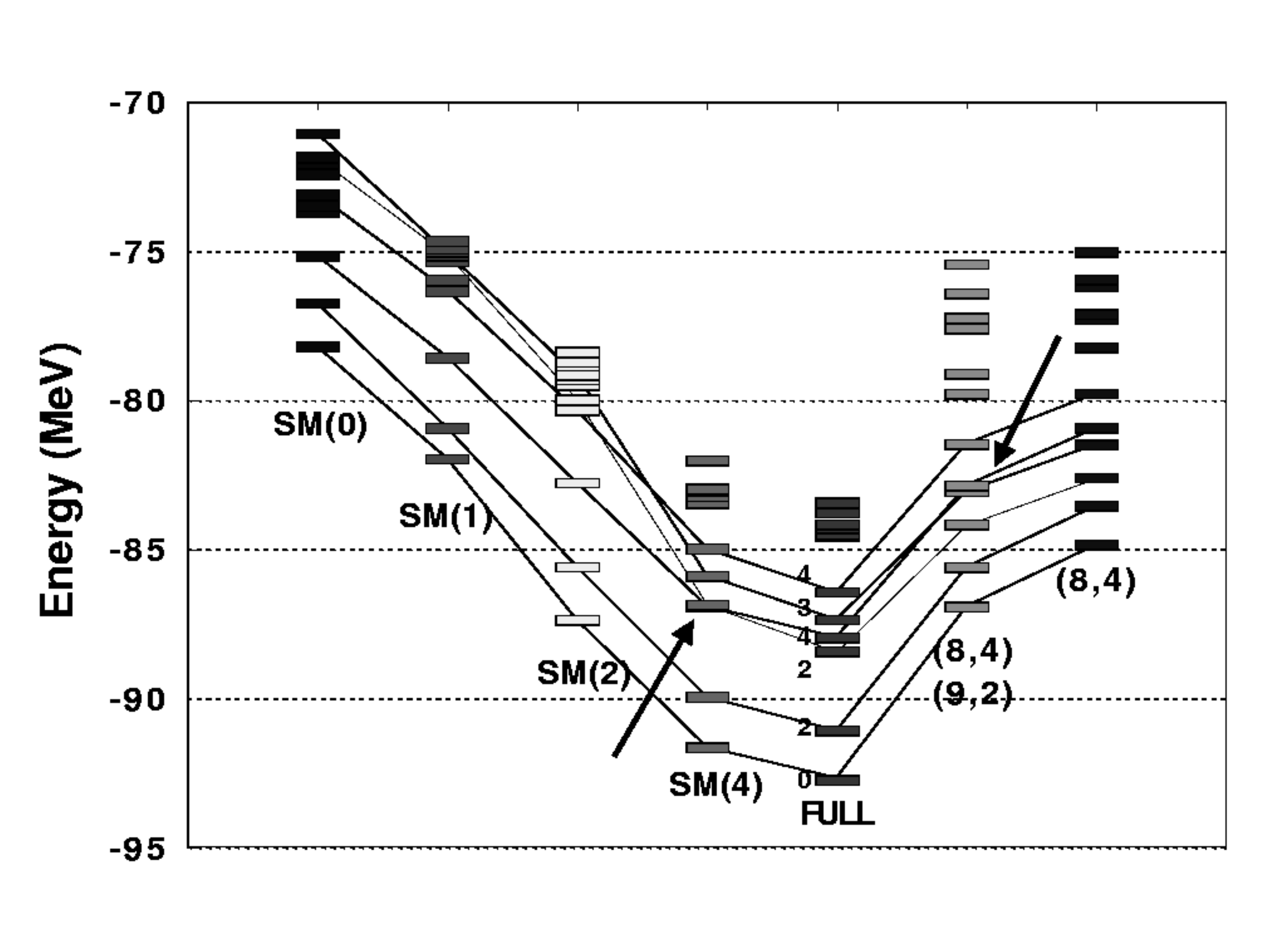}}
\end{center}
\caption{Structure of the energy levels for $^{24}$Mg for different
calculations. Pure $m$-scheme spherical basis calculations are on the
left-hand side of the graph; pure SU(3) basis calculations are on the
right-hand side; the spectrum from the FULL space calculation is in the
center.}
\label{LevelStructure}
\end{figure}

\begin{figure}[tbp]
\begin{center}
\leavevmode
\epsfxsize = \textwidth
\centerline {\includegraphics[width= \textwidth]{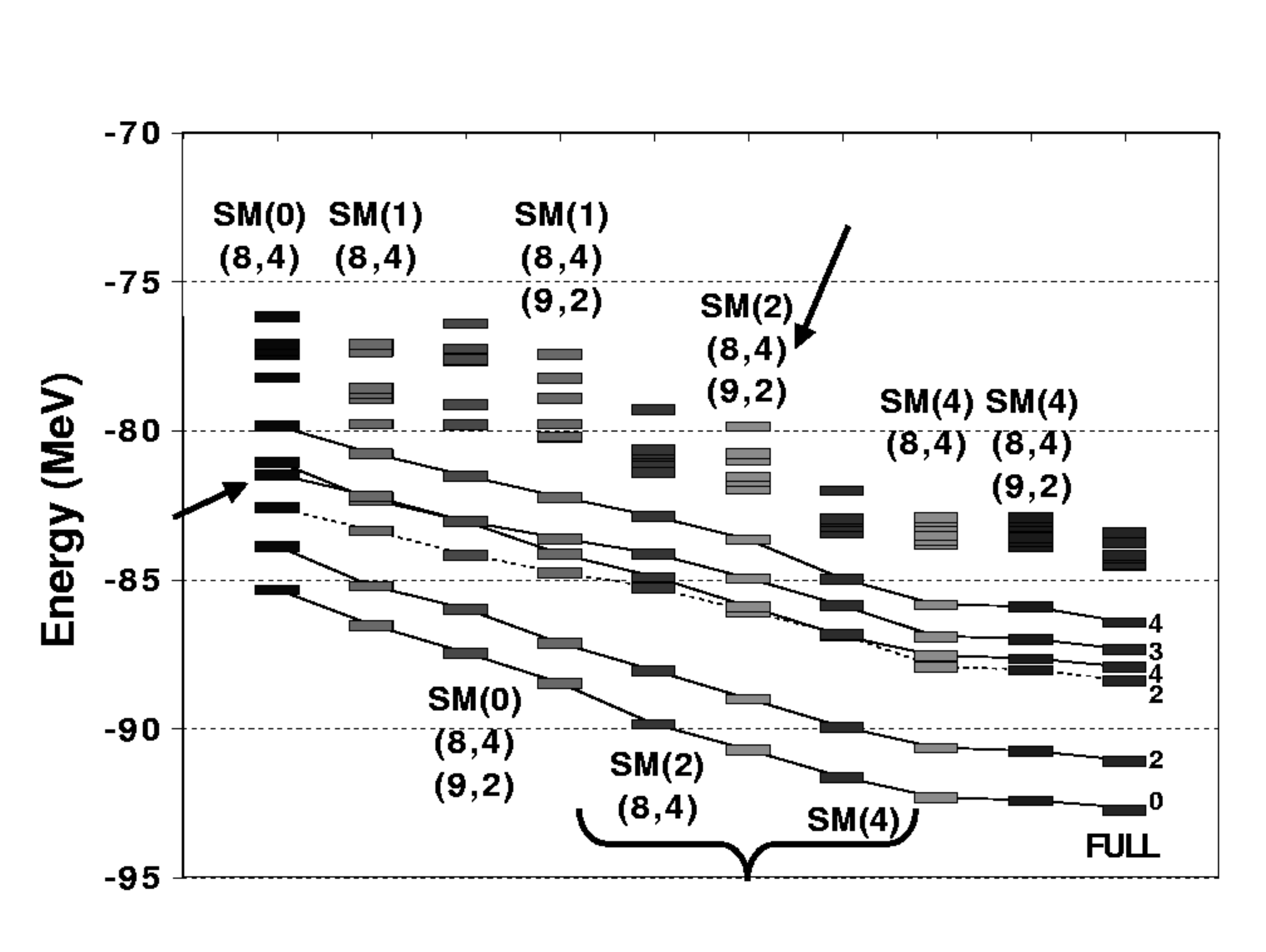}}
\end{center}
\caption{Energy levels for $^{24}$Mg as calculated for different oblique
bases. The SM(4) basis calculation is included for comparison.}
\label{RightLevelStructure}
\end{figure}

\quad
Fig. \ref{LevelStructure} and Fig. \ref{RightLevelStructure} show that
the oblique-basis calculation positions the K=2 band head correctly.
Furthermore, most of the other low-energy levels are also positioned
correctly. The results for pure spherical and pure SU(3) calculations are
shown in Fig. \ref{LevelStructure}. As can be seen from the results in Fig. 
\ref{LevelStructure}, an SM(4) calculation (64\% of the full model space) is
needed to get the ordering of the lowest angular momentum states correctly.
Also, notice that in this case the third and fourth energy levels are
practically degenerate. On the other hand, it only takes 0.5\% of the full
space to achieve comparable success with SU(3). In particular, Fig.
\ref{LevelStructure} shows that an SU(3) calculation using only the (8,4) and
(9,2) irreps gives the right ordering of the lowest levels. Note that the first
few low-energy levels for SM(2) are close in energy to the corresponding
low-energy levels for the (8,4)\&(9,2) result. Since these two spaces are
nearly orthogonal (see Fig. \ref{SU3+ Relative To SM(2) and SM(4)} ), these two
sets of levels mix strongly in an oblique calculation and yield excellent
results. The comparable ground-state energies of the SM(2) and (8,4)\&(9,2)
configurations can also be seen in Fig. \ref{Mg24DimConv}.

Compare the spectra shown in Fig. \ref{LevelStructure} with the results from
the oblique-basis calculations shown in Fig. \ref{RightLevelStructure}. From
this comparison one can see that the correct level structure can be achieved
by using 1.6\% (SM(1)+(8,4)) of the full $sd$-space. However, one should
also notice that for the SM(0)+(8,4) space, which is only 0.2\% of the full
space and the minimum oblique-basis calculation, the results are quite close
to the correct level structure. Despite the fact that the ground-state
energy of the oblique-basis calculation is higher than the ground-state
energy for the SM(4)-type calculation, the oblique calculations are
favorable in terms of dimensionality considerations and correctness of the
level structure.

\section{Overlaps with the Full $sd$-shell Calculation}

\begin{figure}[tbp]
\begin{center}
\leavevmode
\epsfxsize = \textwidth
\centerline {\includegraphics[width= \textwidth]{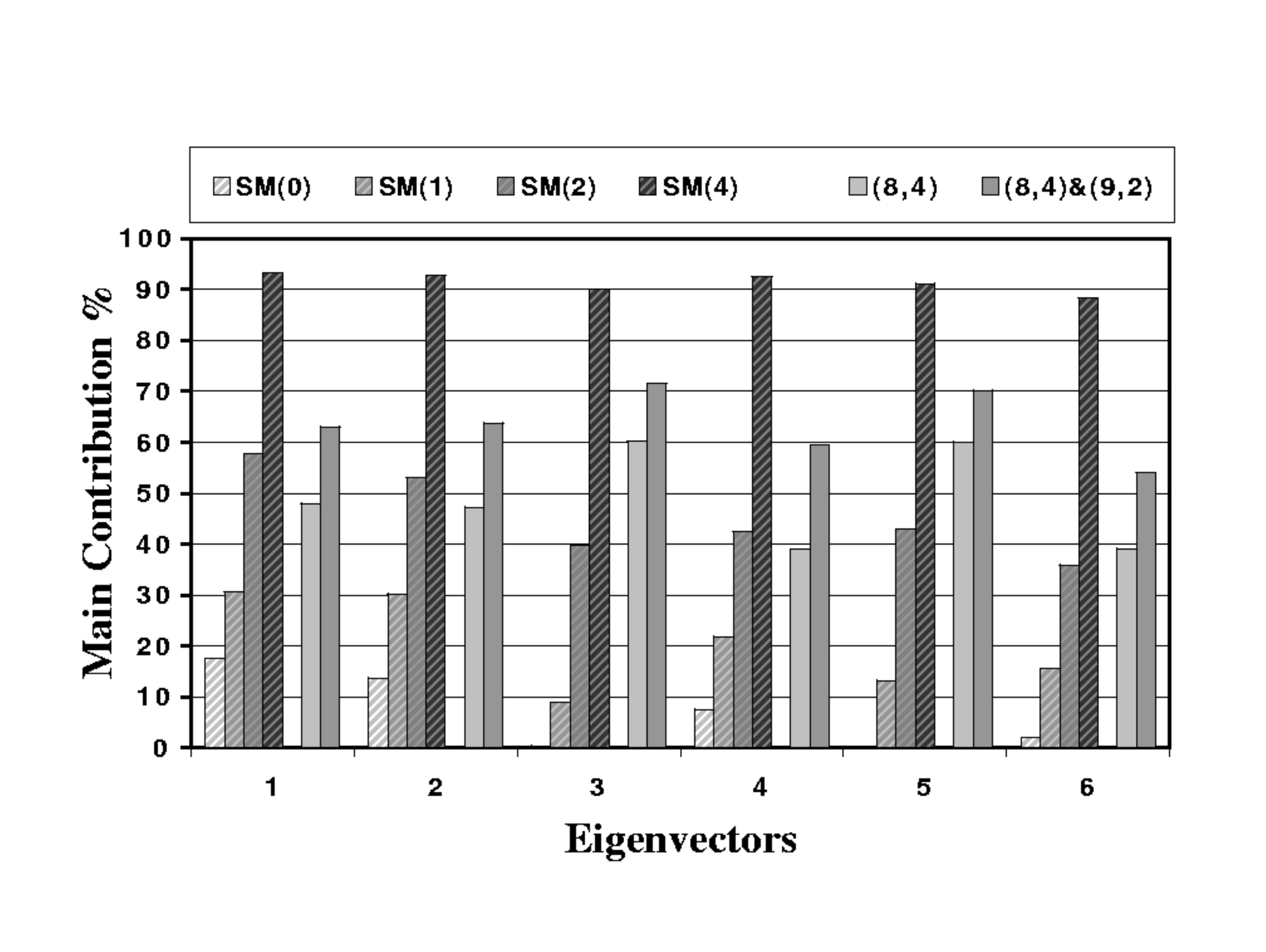}}
\end{center}
\caption{Overlaps of the pure spherical and SU(3) with the FULL states. The
first four bars represent the SM(0), SM(1), SM(2), and SM(4) calculations,
the next two bars represent SU(3) calculations, etc.}
\label{UsualCalculationOverlaps}
\end{figure}

\begin{figure}[tbp]
\begin{center}
\leavevmode
\centerline {\includegraphics[width= 5in ]{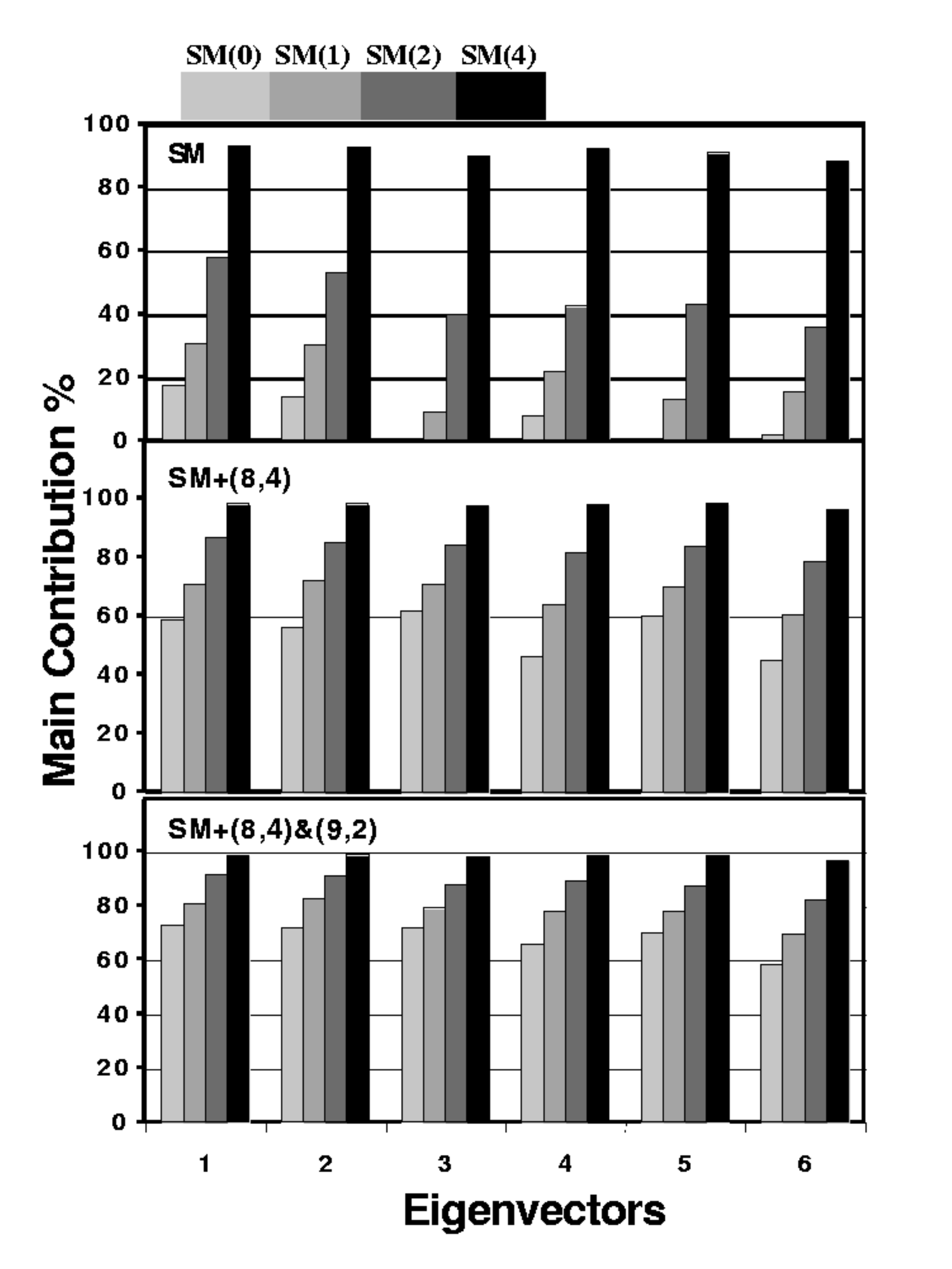}}
\end{center}
\caption{Overlaps of the oblique-basis states with the exact states (set I).
Inset SM contains the overlaps for the pure spherical shell-model basis
states only. Inset SM+(8,4) contains the overlaps of the SM basis enhanced
by the leading SU(3) irrep (8,4). Inset SM+(8,4)\&(9,2) has the (9,2) irreps
included as well.}
\label{MixedOverlaps}
\end{figure}

\begin{figure}[tbp]
\begin{center}
\leavevmode
\epsfxsize = \textwidth
\centerline {\includegraphics[width= \textwidth]{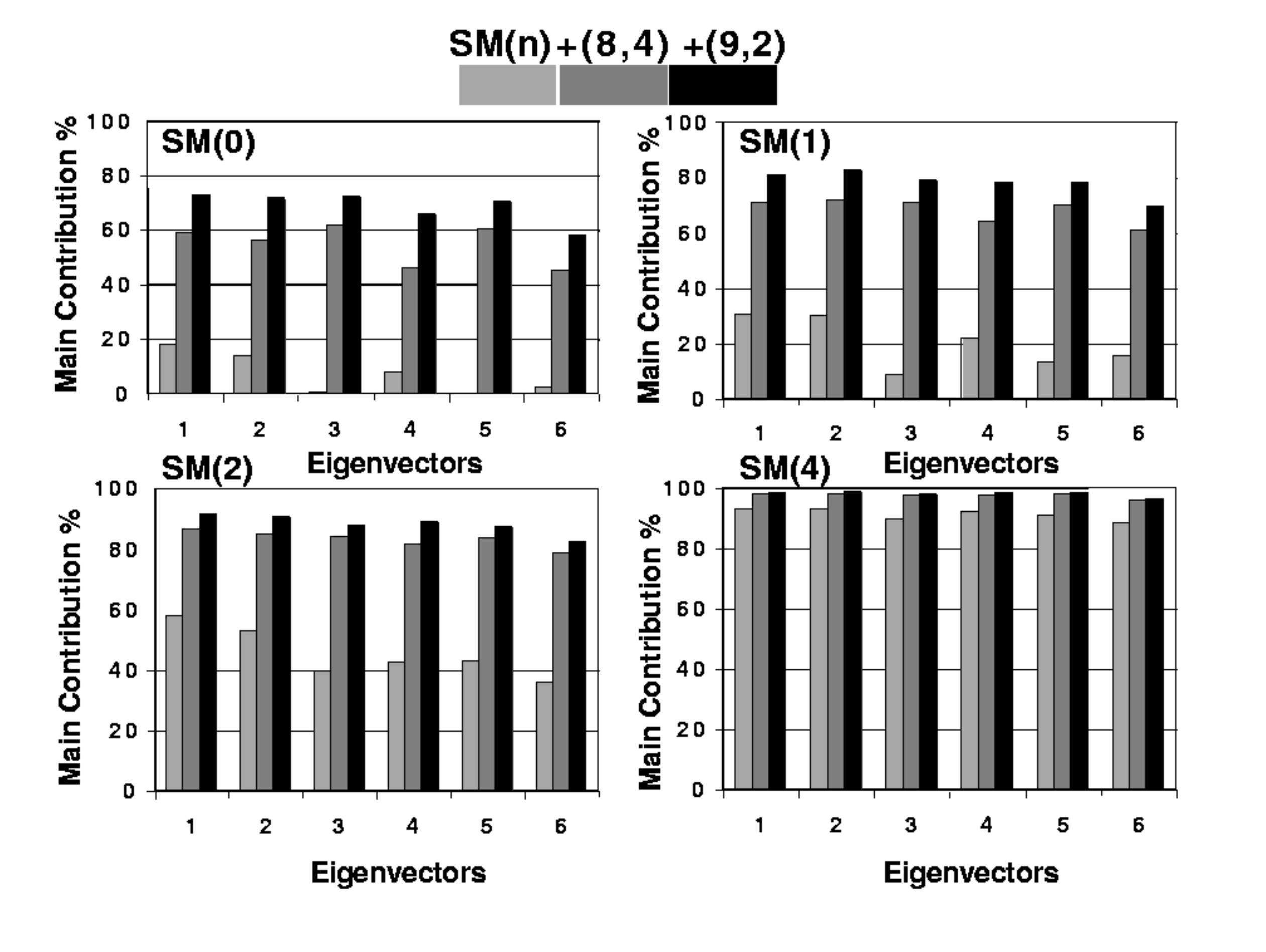}}
\end{center}
\caption{Overlaps of the oblique-basis states with the exact states (set
II). Each inset represents a particular SM(n)-type calculation, showing how
the overlaps change along the corresponding oblique-basis calculation.}
\label{MixedCalculationOverlaps}
\end{figure}

\begin{figure}[tbp]
\begin{center}
\leavevmode
\epsfxsize = \textwidth
\centerline {\includegraphics[width= \textwidth]{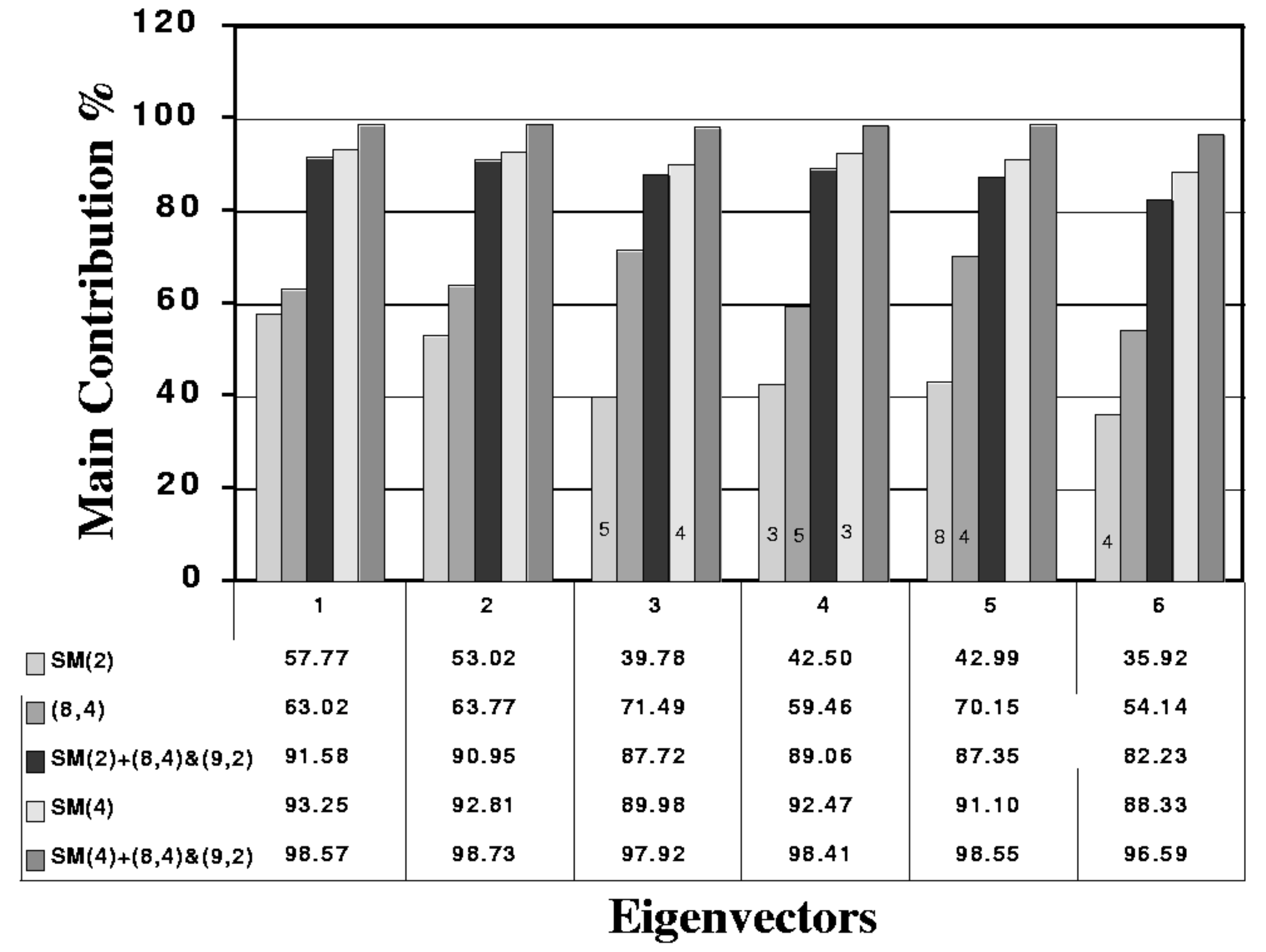}}
\end{center}
\caption{Representative overlaps of pure SM(n), pure SU(3), and
oblique-basis results with the exact full $sd$-shell eigenstates. A number
within a bar denotes the state with the overlap shown by the bar if it is
different from the number for the exact full-space calculation shown on the
abscissa. For example, for SM(2) the third eigenvector overlaps the most
with the fourth exact eigenstate, not the third, while the fifth SM(2)
eigenvector has the overlap shown with the third exact eigenstate.}
\label{SelectedOverlaps}
\end{figure}

\quad
Figs. \ref{UsualCalculationOverlaps}--\ref{SelectedOverlaps} focus on the
actual structure of the states by showing overlaps of eigenstates calculated
in the SM(n), SU(3), and oblique bases with the corresponding states of the
full space calculation. Specifically, in Fig. \ref{UsualCalculationOverlaps},
overlaps of states for pure SM(n) and pure SU(3)-type calculations are
given. Note that the SM(4) states have big overlap (90\%) for the first few
eigenstates. This should not be too surprising since SM(4) covers 64\% of
the full space.

The results in Fig. \ref{UsualCalculationOverlaps} show that in general
SU(3)-based calculations give much better results than the low-dimensional
SM(n)-type calculations. The SM(n)-based calculations have irregular
overlaps along the low-lying states and require SM(4), which is 64\% of the
full space, to get relatively well behaved overlaps. This can be seen most
clearly from the inset labeled SM in Fig. \ref{MixedOverlaps}. Note that the
SM(0) contributions to the third, fifth, and sixth states are very low while
SM(1) and SM(2) have varying contributions. The structure of the SU(3)-type
states leads to a stable picture for the oblique calculations as shown in
the inset SM(n)+(8,4) and SM(n)+(8,4)\&(9,2) in Fig. \ref{MixedOverlaps}.

In Fig. \ref{MixedCalculationOverlaps}, the improvement in the structure of
the calculated states is followed as the SU(3) states are added to the SM(n)
basis. From this graph, one can see that the improvement to the SM(0)- and
SM(1)-type calculation is due mainly to the goodness of SU(3) itself. The
improvement obtained in the oblique calculation is due to the SU(3)
enhancement of the SM(2) space. From this graph, one can also conclude that
there is only a small gain in going to the SM(4)-based oblique calculation.
However, this improvement can not be achieved by any other means with such a
small increase in the model space. This is clear from a careful examination
of Fig. \ref{Mg24DimConv} where one can see that the SM(5) result, which has
25142 basis vectors (88\% of the full $sd$-space), gives the same
ground-state energy as the SM(4)+(8,4)\&(9,2) result (64.6\% of the full
$sd$-space).

Finally, to compare the three schemes -- SU(3), SM(n), and the various
oblique-basis combinations -- representative overlaps are shown in Fig.
\ref{SelectedOverlaps}. From these results, it is very clear that SU(3)-type
basis states yield the right structure in a very low order. In particular, in
Fig. \ref{SelectedOverlaps}, it can be seen that a 90\% overlap with the exact
eigenvectors can be achieved by using only 10\% of the total space,
SM(2)+(8,4)\&(9,2). Furthermore, Fig. \ref{SelectedOverlaps} also shows that
SU(3) enhances the SM(4) results yielding eigenstates with overlaps that are
very close ($\approx$ 98\%) to the exact results.

%% file: VGGPhDThesisCh6.tex
\chapter{Study of Lower $pf$-shell Nuclei}

\quad
For $^{24}$Mg, the single-particle excitations, described by the spherical
shell model, and the collective excitations, described by the $SU(3)$ shell
model, are of comparable importance. In the previous chapter, we have shown
the relevance of the oblique-basis calculation for $^{24}$Mg. It is,
therefore, natural to seek other nuclear systems to apply the mixed-symmetry
method to. The even-even nuclei in the $sd$-shell are one place to start. For
the $sd$-shell nuclei, however, one can perform full $sd$-shell
calculations with modern computer codes.

The $pf$-shell nuclei are another option. For these nuclei, full $pf$-shell
calculations have just recently been achieved \cite{Ur et al, Caurier -full
pf shell}. However, adding only the leading and next to the leading irreps,
as it has been done for $^{24}$Mg, is not sufficient for the lower $pf$-shell
nuclei, Ti and Cr, to obtain results as good as those for $^{24}$Mg. This is
because the spherical shell model provides a significant part of the
low-energy wave functions of these nuclei within a few spherical shell-model
configurations, while in the $SU(3)$ shell-model basis one needs more than a
few $SU(3)$ irreps. This is due mainly to the strong breaking of
$SU(3)$ in the lower $pf$-shell induced by the spin-orbit interaction \cite{VGG
SU(3)andLSinPF-ShellNuclei}. When the spin-orbit splitting is removed,
the importance of the $SU(3)$ basis is restored. Although the usual $SU(3)$
structure of the states is lost, there is an adiabatic $SU(3)$ mixing which
gives rise to the coherent structure of the yrast states. We have already seen
this coherent mixing phenomenon in our toy model. In nuclei, however, this
coherent mixing can be interpreted as an illustration of the intrinsic state
idea where all the states within a given band can be projected out from a
particular intrinsic state \cite{Elliott's SU(3) model}. Even more, in nuclei
this coherent structure is assumed to be a result of an adiabatic $ SU(3)$
mixing which means that some observables stay close to the $SU(3)$ limit, that
is, as if there was a pure $SU(3)$ symmetry. Specifically, the $ B(E2)$ values
remain strongly enhanced with values close to the $SU(3)$ symmetry limit. It
is important to point out that there is a coherent mixing of the spherical
shell-model states as well.

In this chapter, we will discuss our study of the even-even lower $pf$-shell
nuclei $^{44-48}$Ti and $^{48}$Cr. First, we show what a few $SU(3)$ irreps
can do for us within a mixed-symmetry calculations for $^{44}$Ti nucleus.
Then, we focus on the spin-orbit interaction that strongly breaks the $SU(3)
$ symmetry. We conclude this chapter by discussing the coherent structure of
the states and the adiabatic $SU(3)$ mixing which produces enhanced $B(E2)$
values.

\section{$^{44}$Ti Oblique-Basis Results}

\quad
The simplest even-even nucleus in the $pf$-shell, from a computational
point of view, is $^{44}$Ti. In this section we discuss our oblique-basis
calculations for $^{44}$Ti. If one compares the spherical shell-model with the
$SU(3)$ shell-model results within the framework of a realistic interaction,
such as KB3 interaction \cite{KB3 interaction}, then $SU(3)$ seems to be
badly broken. Specifically, the ground-state energy and wave function are
poorly reproduced. This seems to be a common trend in the even-even $sd$-shell
nuclei as well
\cite{PHF and SU(3)}. Even the ground-state energies within the oblique-basis
calculations do not look prominent. However, at a closer examination one finds
that the oblique-basis idea still works. The results may be not as good as in
$^{24}$Mg, but there are some close analogies. For example, the SM(1) space in
$^{44}$Ti seems to be what SM(2) is for $^{24}$Mg, while the SM(2) space in
$^{44}$Ti seems to be what SM(4) is for $^{24}$Mg. By that we mean that the
$SU(3)$ enhanced SM(1) basis in $^{44}$Ti gives overlaps that are compatible
with the overlaps of the pure SM(2) calculation. In the next few sub-sections
we briefly illustrate these findings.

\subsection{Model Space Dimensions}

\quad
The model space for $^{44}$Ti consists of 2 valence protons and 2 valence
neutrons in the $pf$-shell. We use the same notation for the $m$-scheme
spherical bases as in \ref{Table1}. The SU(3) part of the basis
includes the leading irrep of SU(3), which for $^{44}$Ti is (12,0) with
$M_{J}=0$ dimensionality 7, and the next to the leading irrep, namely the
(10,1). The (10,1) irrep occurs three times, once with $S=0$ ($M_{J}=0$
dimensionality 11) and twice with $S=1$ ($M_{J}=0$ dimensionality $2\times
33=66$). All three (10,1) irreps have total $M_{J}=0$ dimensionality of
11+66=77. The (12,0)\&(10,1) case has total $M_{J}=0$ dimensionality of
7+77=84 and is denoted by (12,0)\&(10,1). In Table \ref{TableTi44} we
summarize the dimensionalities involved.

\begin{table}[tbp]
\caption{Labels and $M_{J}=0$ dimensions for various $^{44}$Ti oblique
calculations.}
\label{TableTi44}\vskip 0.25cm
\begin{center}
\begin{tabular}{rrrrrrrr}
\hline
Model space & $(12,0)$ & $(12,0)\&(10,1)$ & $SM(0)$ & $SM(1)$ & $SM(2)$ & $
SM(4)$ & $FULL$ \\ \hline
space dimension & $7$ & $84$ & $72$ & $580$ & $1908$ & $3360$ & $4000$ \\
\% of the full space & $0.18$ & $2.1$ & $1.8$ & $14.5$ & $47.7$ & $84$ & $100$
\end{tabular}
\end{center}
\vskip 0.25cm
\end{table}

As in the case of $^{24}$Mg, there are linearly dependent vectors within
some of the $^{44}$Ti calculations. For example, there is one such vector in
the SM(2)+(12,0) space, two in the SM(3)+(12,0), two in the
SM(1)+(12,0)\&(10,1), twelve in the SM(2)+(12,0)\&(10,1), and thirty-three
in the SM(3)+(12,0)\&(10,1). Each linearly dependent vector is handled as
discussed in the chapter devoted to $^{24}$Mg.

\subsection{Ground-State Energy}

\begin{figure}[tbp]
\begin{center}
\leavevmode
\epsfxsize = \textwidth
\centerline {\includegraphics[width= \textwidth]{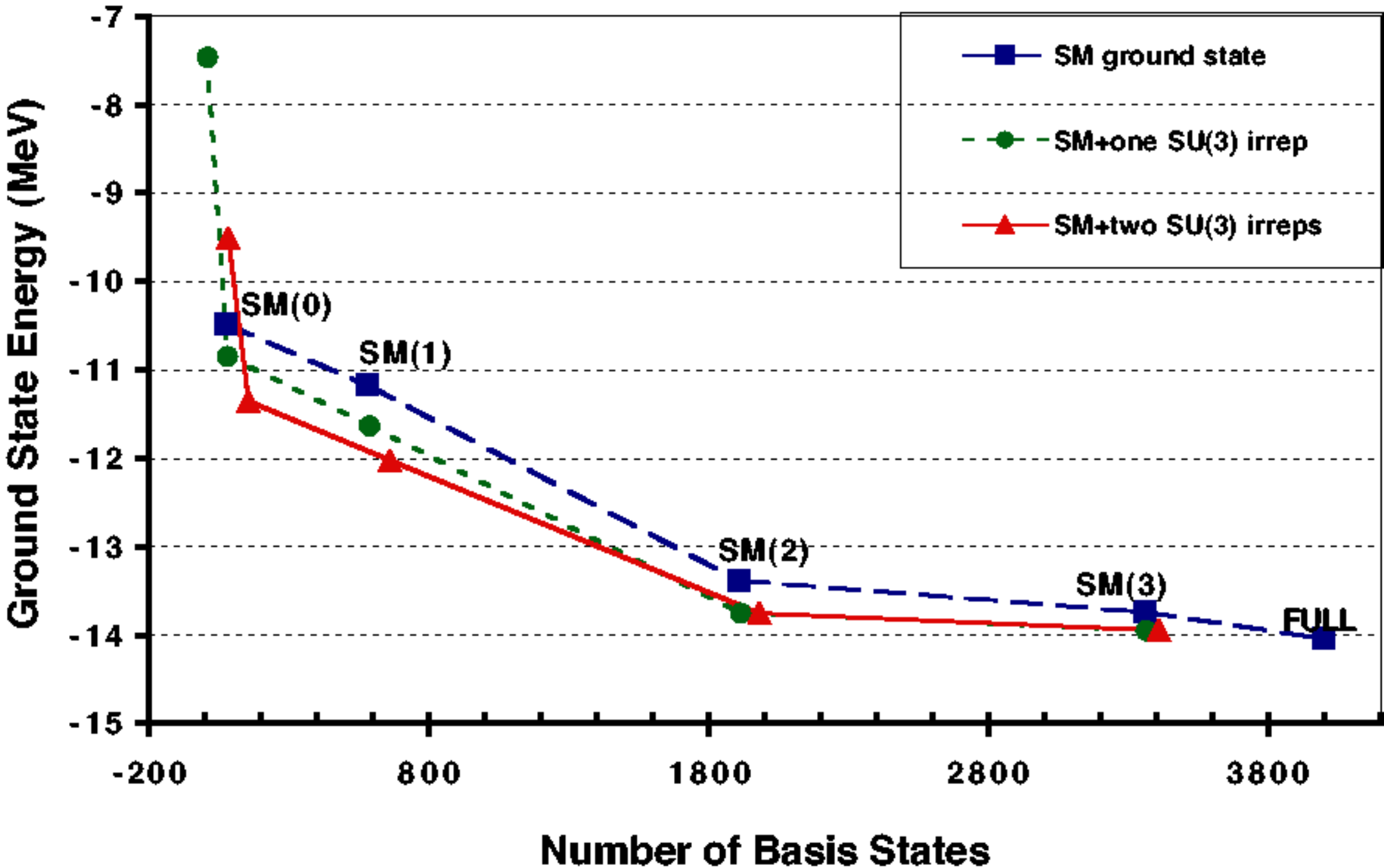}}
\end{center}
\caption{Ground-state energy for $^{44}$Ti as a function of the various
model spaces. The SU(3) irreps used are (12,0) and (10,1).}
\label{Ti44DimConv}
\end{figure}

\quad
Fig. \ref{Ti44DimConv} shows that the oblique-basis calculation of the
ground-state energy for $^{44}$Ti does not give results as good as for $
^{24} $Mg. The calculated ground-state energy for the SM(1)+(12,0)\&(10,1)
calculation is 0.85 MeV below the calculated energy for the SM(1) space
alone. Adding the two SU(3) irreps to the SM(1) space increases the size of
the space from 14.5\% to 16.6\% of the full space. This is a 2.1\% increase,
while going from the SM(1) to SM(2) involves an increase of 33.2\% in the model
space. For the latter, the ground-state energy is 2.2 MeV lower than the SM(1)
result.

\subsection{Energy Spectrum of the Low Lying States}

\begin{figure}[tbp]
\begin{center}
\leavevmode
\epsfxsize = 5in
\centerline {\includegraphics[width= 5in]{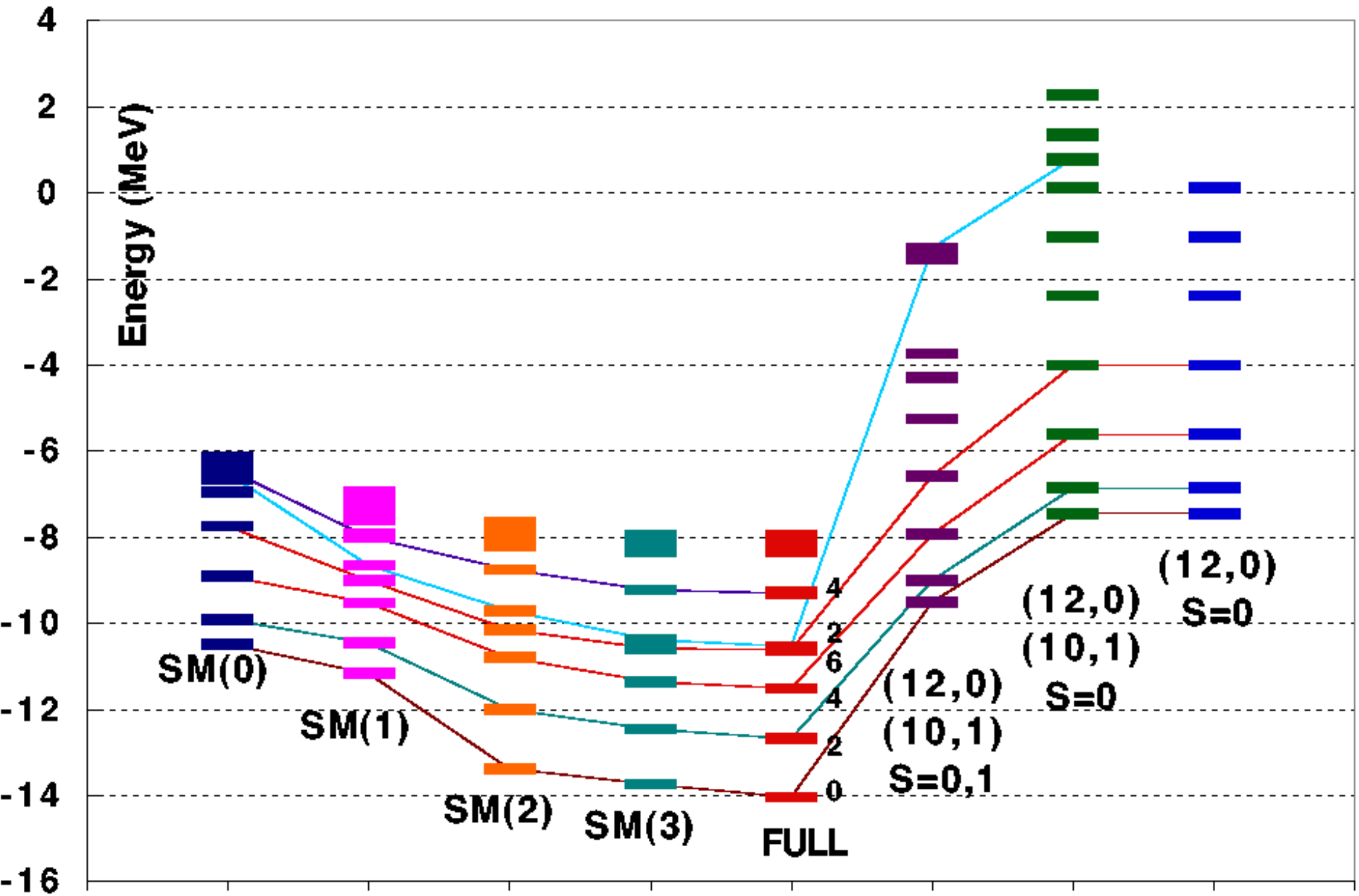}}
\epsfxsize = 5in
\centerline {\includegraphics[width= 5in]{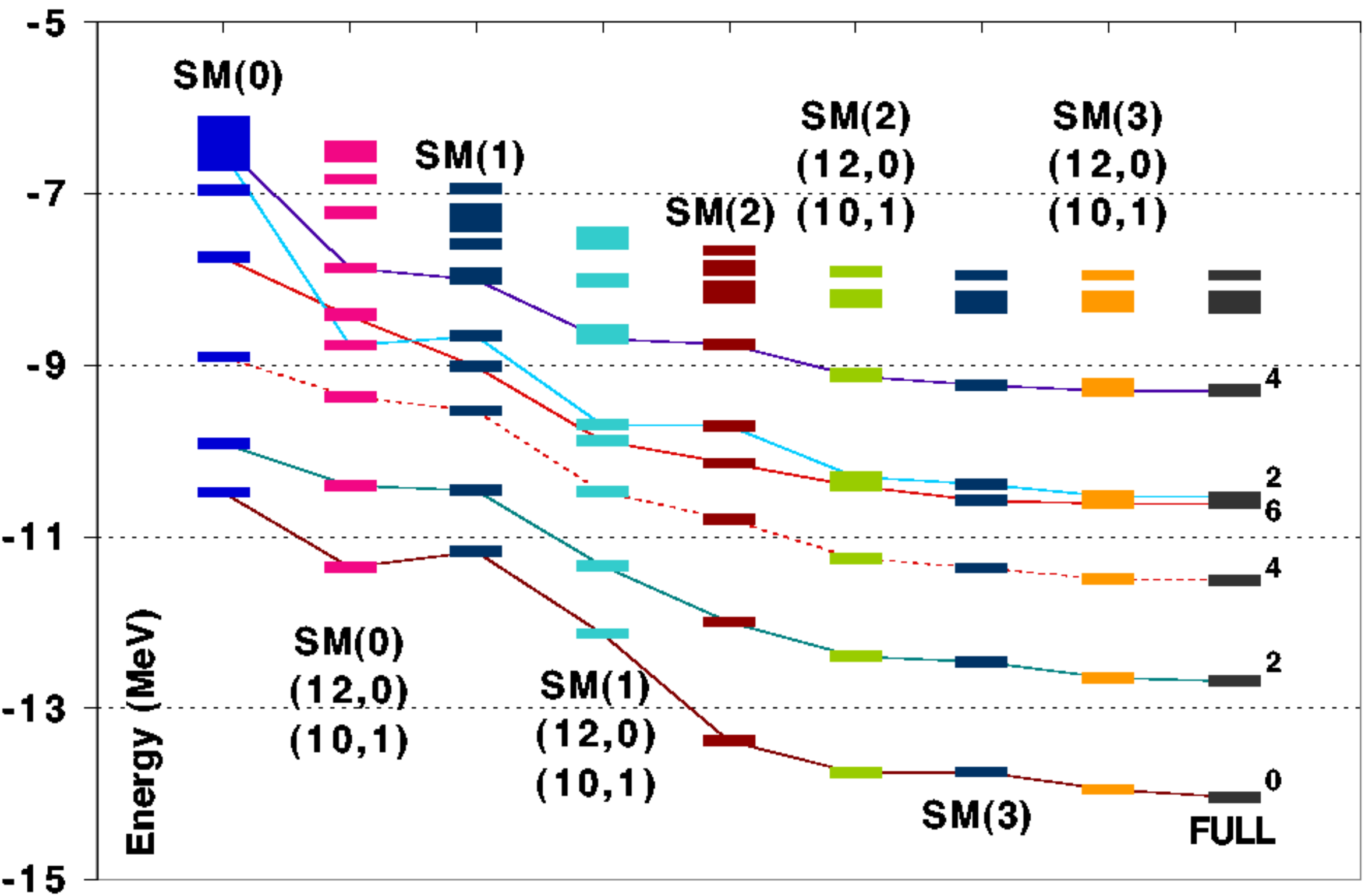}}
\end{center}
\caption[Structure of the energy levels for $^{44}$Ti for different
calculations.]{Structure of the energy levels for $^{44}$Ti for different
calculations. Pure $m$-scheme spherical-basis calculations are on the
left-hand side of the upper graph; pure SU(3)-basis calculations are on the
right-hand side; the spectrum from the FULL space calculation is in the
center. The spectra form oblique-basis calculations are in the lower graph.}
\label{Ti44LevelStructure}
\end{figure}

\quad
We have seen that the position of the K=2 band head for $^{24}$Mg is
correct for the SU(3)-type calculations but not for the low-dimensional SM(n)
calculations. In $^{44}$Ti this seems to be the opposite, the SM(n)-type
calculations reproduce the position of the K=2 band head while the SU(3) do
not, as shown in the upper graph in Fig. \ref{Ti44LevelStructure}.
Furthermore, most of the low-energy levels are much higher for the pure
SU(3) limit than for the pure SM(n) case. Thus, one may expect that these two
sets of levels (the SM(n) and the SU(3)) may not mix as strongly in an oblique
calculation as for the $^{24}$Mg case. Surprisingly, the oblique-basis
calculations seem to produce a good spectral structure as shown in the lower
graph in Fig. \ref{Ti44LevelStructure}. Notice that the SM(2)+(12,0)\&(10,1)
spectrum is very good and compatible with the SM(3) spectrum. This is
50\% compared to 84\% of the model space.

\subsection{Overlaps with the Exact Calculation}

\begin{figure}[tbp]
\begin{center}
\leavevmode
\epsfxsize = 5in
\centerline {\includegraphics[width= 5in]{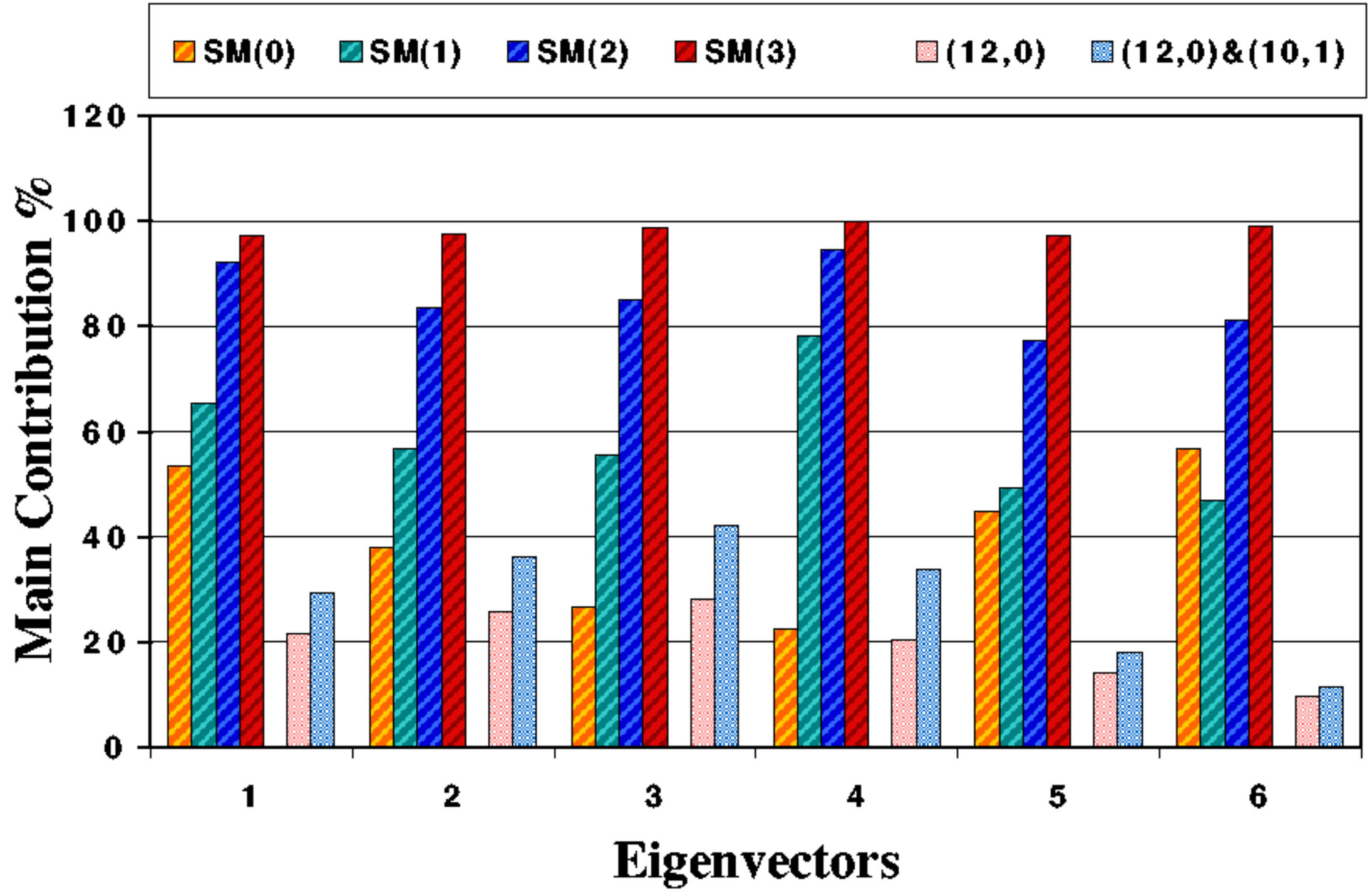}}
\epsfxsize = 5in
\centerline {\includegraphics[width= 5in]{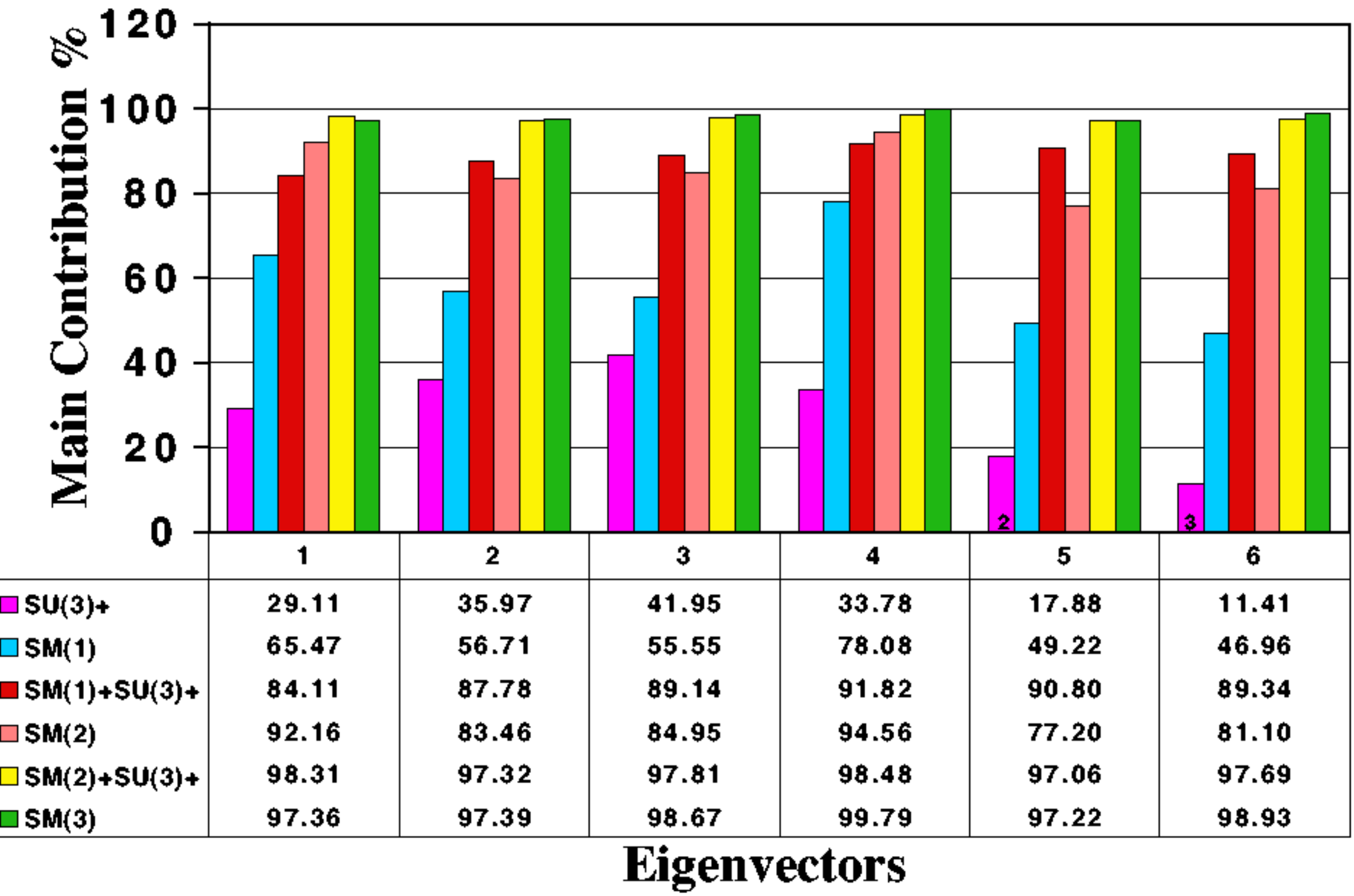}}
\end{center}
\caption{$^{44}$Ti wave function overlaps of pure spherical, SU(3), and
oblique states with the FULL states. The first four bars in the upper graph
represent the SM(0), SM(1), SM(2), and SM(3) calculations, the next two bars
represent the SU(3) calculations, etc. Representative overlaps of pure
SM(n), pure SU(3), and oblique-basis results with the exact full $pf$-shell
eigenstates are shown in the lower graph.}
\label{Ti44Overlaps}
\end{figure}

\quad
The top graph in Fig. \ref{Ti44Overlaps} shows overlaps of states for pure
SM(n) and pure SU(3)-type calculations while the lower part shows some
selected overlaps from the oblique calculations. Notice that the overlaps of
the pure SU(3)-type calculations are very small, often less than 40\%, while
the SM(n) results are far better with the SM(2)-type calculations having about
80\% overlap with the exact states. Note that the SM(3) states have big
overlap ($>$97\%) for the first few eigenstates. This is not surprising since
SM(3) covers 84\% of the full space. What is surprisingly good is that the
SM(2)+(12,0)\&(10,1)-type calculation is as good as the SM(3). Even more,
the SM(1)+(12,0)\&(10,1) overlaps seem to be often bigger than the SM(2)
overlaps.

\section{Set Up for the Study of the $SU(3)$ Breaking}

\quad
To understand better the results of the mixed-mode calculations described
in the previous section, we need to recall that the oblique-basis method is
expected to work well if we are dealing with two or more competing and
compatible modes. Therefore, if the Hamiltonian of the system is dominated
by its one-body term, then the effect of the two-body part will be
suppressed. However, if the single-particle energies are degenerate,
the importance of $SU(3)$ should reappear. In the next subsections, we
discuss the structure of the Hamiltonian, as well as some of the
computational methods used in our calculations.

\subsection{Interaction Hamiltonian}

\quad
To retain clarity of the discussion, we recall the structure and notations
of the one- plus two-body Hamiltonian:

\[
H=\sum_{i}\varepsilon _{i}a_{i}^{+}a_{i}+ \frac{1}{4}
\sum_{i,j,k,l}V_{kl,ij}a_{i}^{+}a_{j}^{+}a_{k}a_{l}.
\]

The summation indexes range over the single-particle levels included in the
model space. We only consider levels of the $pf$-shell which have the
following radial $(n)$, orbital $(l)$ and total angular momentum $(j)$
quantum numbers: $nl_{j}=\left\{ 0f_{7/2},0f_{5/2},1p_{3/2},1p_{1/2}\right\}
$. In what follows, the radial quantum number $(n)$ is dropped since the $
l_{j}$ labels provide a unique labelling scheme for single-shell
applications. It is common practice to replace the four single-particle
energies $\varepsilon _{i}$ by the $l^{2}$ and $l\cdot s$ interactions: $
\sum_{i}\varepsilon _{i}a_{i}^{+}a_{i}\rightarrow \epsilon (n_{i}-\alpha
_{i}l_{i}\cdot s_{i}-\beta _{i}l_{i}^{2})$, where $\epsilon $ is the average
binding energy per valence particle, $n_{i}$ counts the total number of
valence particles, and $\alpha $ and $\beta $ are dimensionless parameters
giving the interaction strength of the $l^{2}$ and $l\cdot s$ terms. For
realistic single-particle energies used in the KB3 interaction
(\ref{KB3 spe}), these parameters are $\epsilon =2.6$ $MeV,$ $\beta
=0.0096,$ $\alpha _{p}=1.3333,$ and $\alpha _{f}=1.7143.$ The small value
of $\beta $ signals small $l^{2}$ splitting (\ref{KB3p-f spe}) while the
values of $\alpha $ demonstrate the presence of a strong spin-orbit
splitting.

A significant part of the two-body interaction, $V_{kl,ij}$, maps onto the
quadrupole-quadrupole ($Q\cdot Q$) and the pairing ($P$) interactions. Since
$Q\cdot Q$ can be written in terms of $SU(3)$ generators, it induces no $
SU(3)$ breaking, as has been discussed in the first chapters. Hence $
Q\cdot Q$ serves to re-enforce the importance of the Elliott model \cite
{Elliott's SU(3) model}. However, the pairing interaction mixes different $
SU(3)$ irreps, but in our study it does not seem to cause any strong $SU(3)$
breaking. In this analysis the two-body part of the Hamiltonian ($V_{kl,ij}$
) is fixed by the Kuo-Brown-3 (KB3) interaction matrix elements, and the
single-particle energies, $\varepsilon _{i}$, are changed as described below.

The following single-particle energies are normally used with the KB3
interaction \cite{KB3 interaction}:

\begin{equation}
\mathrm{KB3\quad [MeV]}:\varepsilon _{p_{\frac{1}{2}}}=4,\quad \varepsilon
_{p_{\frac{3}{2}}}=2,\quad \varepsilon _{f_{_{\frac{5}{2}}}}=6,\quad
\varepsilon _{f_{\frac{7}{2}}}=0.  \label{KB3 spe}
\end{equation}

For the purposes of the current study, it is important to know the centroids
of the $p$- and $f$-shells. For example, the energy centroid of the $p$-shell
is given by:

\[
\varepsilon _{p}= \frac{ \varepsilon _{p_{\frac{1}{2}}} \dim(p_{\frac{1}{2}
})+ \varepsilon _{p_{\frac{3}{2}}} \dim (p_{\frac{3}{2}})} {\dim(p_{\frac{1}{
2}})+ \dim (p_{\frac{3}{2}})}.
\]

In what follows, we label by $KB3p\_f$ that Hamiltonian which uses the KB3
two-body interaction with single-particle $p$- and $f$-shell energies set to
their centroid values:

\begin{equation}
\mathrm{KB3}p\_f\quad [MeV]:\varepsilon _{p_{\frac{1}{2}}}=\varepsilon _{p_{
\frac{3}{2}}}=2.6670,\quad \varepsilon _{f_{_{\frac{5}{2}}}}=\varepsilon
_{f_{\frac{7}{2}}}=2.5710.  \label{KB3p-f spe}
\end{equation}

We use $KB3pf$ for the case when the single-particle energies are set to
their overall average:

\begin{equation}
\mathrm{KB3}pf\quad [MeV]:\quad \varepsilon _{p}=\varepsilon _{f}=2.6
\label{KB3pf spe}
\end{equation}

Due to the near degeneracy of the single-particle energies of the $KB3p\_f$
interaction (\ref{KB3p-f spe}), the results for the $KB3pf$ case are very
similar to those for $KB3p\_f$.

\subsection{Computational Procedures}

\quad
In our study, we have focused on $^{44}$Ti, $^{46}$Ti, $^{48}$Ti, and
$^{48}$Cr because these are $pf$-shell equivalents of $^{20}$Ne, $^{22}$Ne,
$^{24}$Ne, and $^{24}$Mg, respectively, which are known to be good $SU(3)$
$sd$-shell nuclei. Furthermore, data on these nuclei are readily available from
the National Nuclear Data Center (NNDC) \cite{NNDC} and full $pf$-shell
calculations are feasible \cite{Caurier -full pf shell}. The model
dimensionalities for full-space calculations increase very rapidly when
approaching the mid-shell region; those for the cases considered here are given
in Table \ref{pf space dimensions}.

\begin{table}[tbp]
\caption{Space dimensions for the $m$-scheme calculations in the full
$pf$-shell model space. We have used even parity and even isospin basis states
with no restrictions on the total angular momentum $J$ except for the $
M_{J}=0$ case where only states with even $J$ values have been selected.}
\label{pf space dimensions}
\begin{center}
\begin{tabular}{rrrrr}
\hline
Nucleus & $M_{J}=0$ & $M_{J}=6$ & $M_{J}=10$ & $M_{J}=14$ \\ \hline
$^{44}$Ti & 1080 & 514 & 30 & --- \\
$^{46}$Ti & 43630 & 32297 & 4693 & 134 \\
$^{48}$Ti & 317972 & 278610 & 57876 & 3846 \\
$^{48}$Cr & 492724 & 451857 & 104658 & 8997
\end{tabular}
\end{center}
\end{table}

The computational procedures and tools used in the analysis of the $SU(3)$
symmetry breaking are described in this section. In brief, the Hamiltonian
and other matrices are calculated using an $m$-scheme shell-model code \cite
{the M-scheme approach} while the eigenvectors and eigenvalues are obtained
by means of the Lanczos algorithm \cite{Whitehead-shell model}. All the
calculations are done in the full $pf$-shell model space.

First, the Hamiltonian $H$ for each interaction ($KB3$ (\ref{KB3 spe}), $
KB3p\_f$ (\ref{KB3p-f spe}), and $KB3pf$ (\ref{KB3pf spe})) is generated.
Then the eigenvalues and eigenvectors are calculated and the yrast states
identified. Next, the matrix for the second-order Casimir operator of $SU(3)$
, namely $C_{2}=(3L^{2}+Q\cdot Q)/4$, is generated using the shell-model
code, and a moments method \cite{Moments method} is used to diagonalize the $
C_{2}$ matrix by starting the Lanczos procedure with specific eigenvectors
of $H$ for which an $SU(3)$ decomposition is desired. Finally, $B(E2)$
values in $e^{2}fm^{4}$ units are calculated from one-body densities using
Siegert's theorem with a typical value for the effective charge \cite{Ur et
al, effective charges}, $q_{eff}$ = 0.5, so $e_{p}=(1+q_{eff})e=1.5e$ and $
e_{n}=(q_{eff})e=0.5e$.

Even though the procedure can generate the spectral decomposition of a state
in terms of the eigenvectors of $C_{2}$ of $SU(3)$, this alone is not
sufficient to determine uniquely all irrep labels $\lambda $ and $\mu $ of $
SU(3)$. For example, $C_{2}$ has the same eigenvalue for the $(\lambda ,\mu
) $ and $(\mu ,\lambda )$ irreps. Nevertheless, since for the first few
leading irreps (largest $C_{2}$ values) the $\lambda $ and $\mu $ values can
be uniquely determined \cite{Tabels of SU(N) to SU(3)}, this procedure
suffices for our study.

Usually, when considering full-space calculations, a balance between
computer time and accuracy has to be considered. While the Lanczos algorithm
\cite{Whitehead-shell model} is known to yield a good approximation for the
lowest or highest eigenvalues and eigenvectors, it normally does a
relatively poor job for intermediate states. This means, for example, that
higher states, in particular high total angular momentum states, may be
poorly represented or, in a worst case scenario, not show up at all when
these states are close to or beyond the truncation edge of the chosen
submatrix. An obvious way to maintain a good approximation is to run the
code for each $M_{J}$ value, that is, $M_{J}=0,2,4,6$\ldots . However, this
might be a very time consuming process, but nonetheless one which could be
reduced significantly if only a few $M_{J}$ values are used for each run.
For the calculations of this study, we have used $M_{J}=0,6,10,$ and $14$.
To maintain high confidence in the approximation of the intermediate states
which have $J=2,4,8,12,...$ we required that they be within the first half
of all the states produced. The code output was set for $29$ states. A
further verification of the accuracy of the procedure is whether the
energies of the same state calculated using different $M_{J}$ runs are close
to one another. For example, as a consistency check the energy of the lowest
$J=6$ state in the $M_{J}=0$ run was compared to the energy of the same
state obtained from the $M_{J}=6$ run.

\section{Measuring Symmetry Breaking Using $C_{2}$ of $SU(3)$}

\quad
In this section, we discuss the results of our study on the $SU(3)$
symmetry breaking in the $pf$-shell. In order to identify the $SU(3)$ structure
of an yrast state, we calculate the spectral distribution of the state along the
second Casimir operator ($C_{2}$) of $SU(3)$ as described in the previous
section. From the spectral distribution, we can clearly determine whether the
$SU(3)$ symmetry is broken or not. However, a graphic or table
representation of the data becomes very inelegant with growing space
dimensions. Thus, we have decided to use also average quantities, such as
centroid, width, and skewness of the distributions to illustrate the main
points one can deduce from a complicated spectral distribution.

\subsection{Spectral Distribution}

\begin{figure}[tbp]
\begin{center}
\leavevmode
\epsfxsize = \textwidth
\centerline {\includegraphics[width= \textwidth]{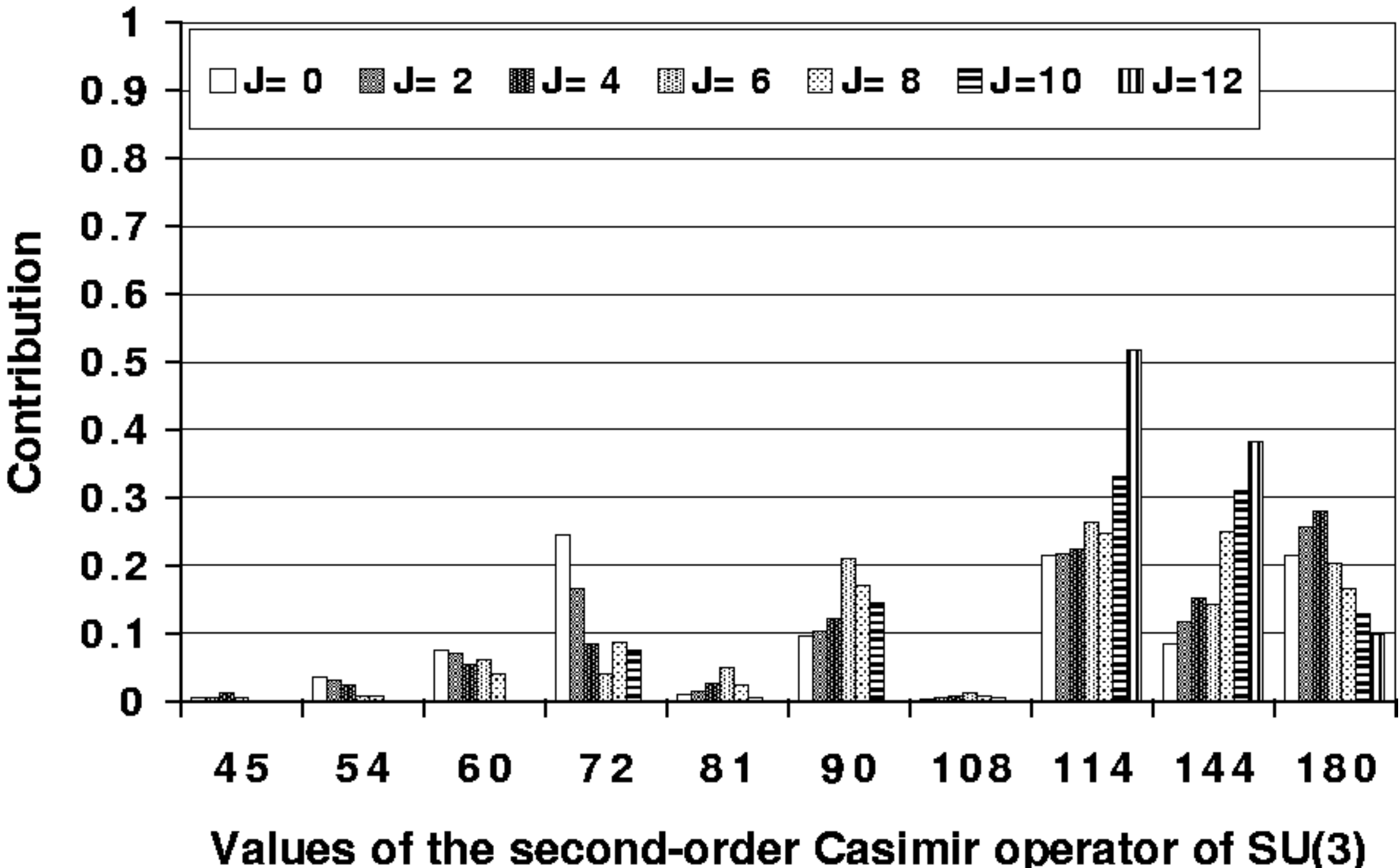}}
\end{center}
\caption{Strength distribution of $C_{2}$ of $SU(3)$ in yrast states of $
^{44}$Ti for realistic single-particle energies with Kuo-Brown-3 two-body
interaction ($KB3$).}
\label{C2 of SU(3) for KB3 in Ti44}
\end{figure}

\quad
The first set, Figs.\ref{C2 of SU(3) for KB3 in Ti44} and \ref{C2 of SU(3)
for KB3p_f in Ti44}, demonstrates the recovery of the $SU(3)$ symmetry as
the single-particle spin-orbit interaction is turned off, that is, in going
from the $KB3$ to the $KB3p\_f$ interaction. Corresponding results for the $
KB3pf$ interaction are similar to the $KB3p\_f$ results. In each graph, $
C_{2}$ values of $SU(3)$ are given on the horizontal axis with the
contribution of each $SU(3)$ state on the vertical axis. The bars within
each cluster are contributions to the yrast states starting with the ground
state ($J=0$) on the left. Hence the second bar in each cluster is for the $
J=2$ yrast state, etc.

\begin{figure}[tbp]
\begin{center}
\leavevmode
\epsfxsize = \textwidth
\centerline {\includegraphics[width= \textwidth]{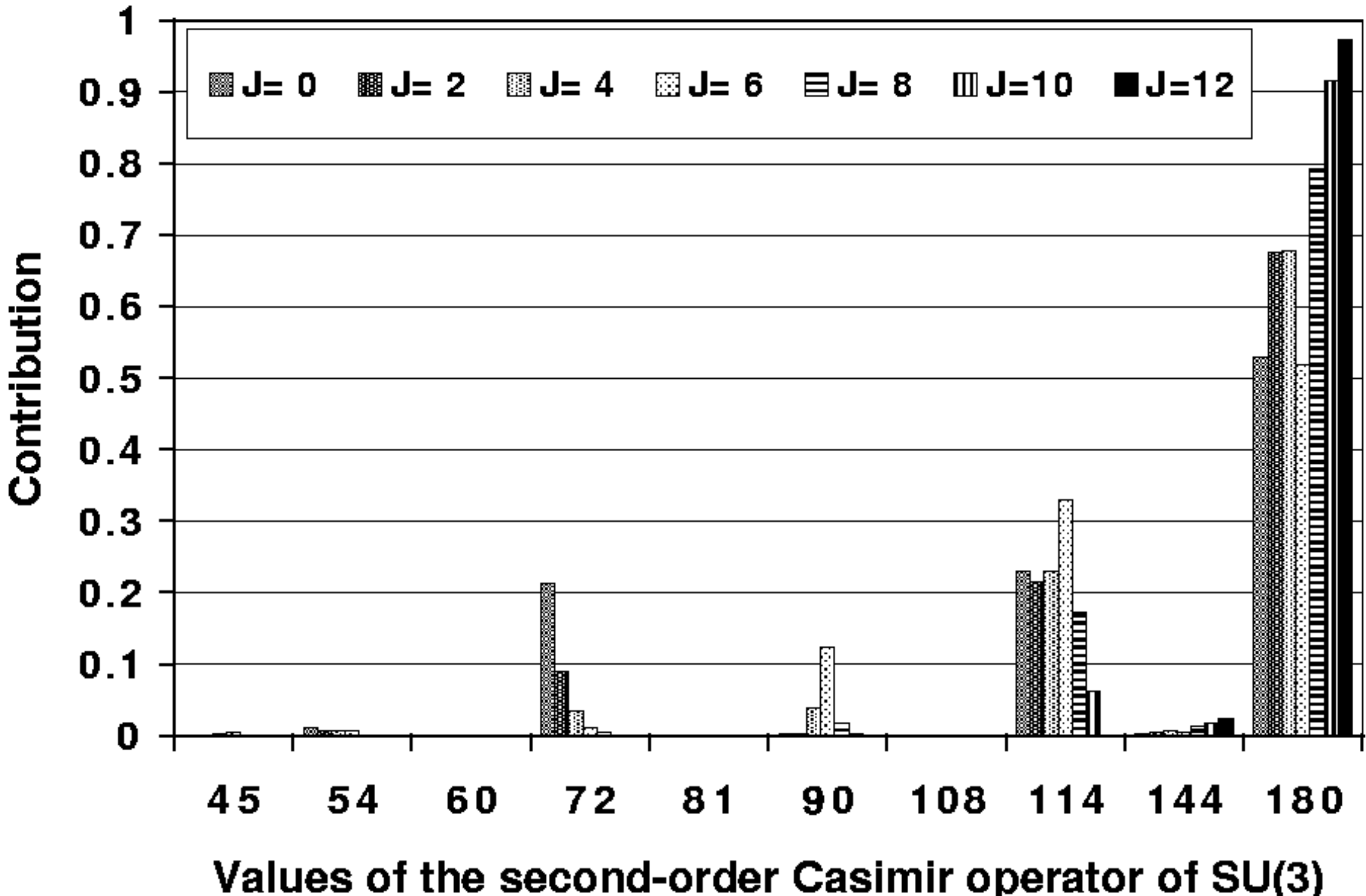}}
\end{center}
\caption{Strength distribution of $C_{2}$ of $SU(3)$ in yrast states of
$^{44}$Ti for degenerate single-particle energies with Kuo-Brown-3 two-body
interaction ($KB3p\_f$).}
\label{C2 of SU(3) for KB3p_f in Ti44}
\end{figure}

We have chosen $^{44}$Ti for an in-depth consideration of the fragmentation
of the $C_{2}$ strength in yrast states. The results for the nondegenerate
$KB3$ interaction are shown in Fig. {\ref{C2 of SU(3) for KB3 in Ti44}. In this
case the highest contribution (biggest bar) is more than $50\%$ which
corresponds to a $C_{2}$ value of 114 for the $J=12$ state. The $C_{2}=114$
value is for $(\lambda ,\mu)=(8,2)$ which is two $SU(3)$ irreps down from the
leading one, $(\lambda ,\mu)=(12,0)$ with $C_{2}=180$. The leading irrep only
contributes about 10$\%$ to the $J=12$ yrast state. The contribution of the
next to the leading irrep, $C_{2}=144$ for $(\lambda ,\mu)=(10,1)$, is slightly
less than 40$\%$. Thus, for all practical purposes, the first three irreps
determine the structure of the $J=12$ yrast state. This illustrates that the
high total angular momentum $J$ states are composed of only the first few
$SU(3)$ irreps. This is easily understood because high $J$ values require high
orbital angular momentum ($L$) states which are only present in
$SU(3)$ irreps with large $C_{2}$ values. The high $J$ states may therefore
be considered to be states with good $SU(3)$ symmetry. However, this is not
the case with the ground state of $^{44}$Ti which has very important
contributions from states with $C_{2}$ values 60, 72, 90, 114, 144, and 180
with respective percentages, 7.5, 25, 10, 21, 8, and 21$\%$. This shows that
the leading irrep is not the biggest contributor to the $J=0$ ground state;
there are two other contributors with about $20\%$, the third ($C_{2}=114$)
and seventh ($C_{2}=72$) $SU(3)$ irrep. }

When the spin-orbit interaction is turned off, which yields nearly
degenerate single-particle energies since the single-particle orbit-orbit
splitting is small, one has the $KB3p\_f$ interaction, and in this case the
structure of the yrast states changes dramatically, as shown in Fig. \ref{C2
of SU(3) for KB3p_f in Ti44}. In Fig. \ref{C2 of SU(3) for KB3p_f in Ti44}
one can see that the leading irrep plays a dominant role as its contribution
is now more than $50\%$ of every yrast state. As in the previous case, the
high total angular momentum $J$ states have the biggest contributions from
the leading irrep, for example, more than 97$\%$ for $J=12$, 91$\%$ for $
J=10 $, and 80$\%$ for $J=8$. The ground state is composed of few irreps
with $C_{2}$ values 72, 114, and 180, but in this case the leading irrep
with $C_{2}=180$ makes up more than 52$\%$ of the total with the other two
most important irreps contributing 21$\%$ [$C_{2}=72$, $(\lambda ,\mu
)=(4,4)]$ and 23$\%$ [$C_{2}=114$, $(\lambda ,\mu )=(8,2)].$

\subsection{Moments of the Spectral Distributions.}

\begin{figure}[tbp]
\begin{center}
\leavevmode
\epsfxsize = \textwidth
\centerline {\includegraphics[width= \textwidth]{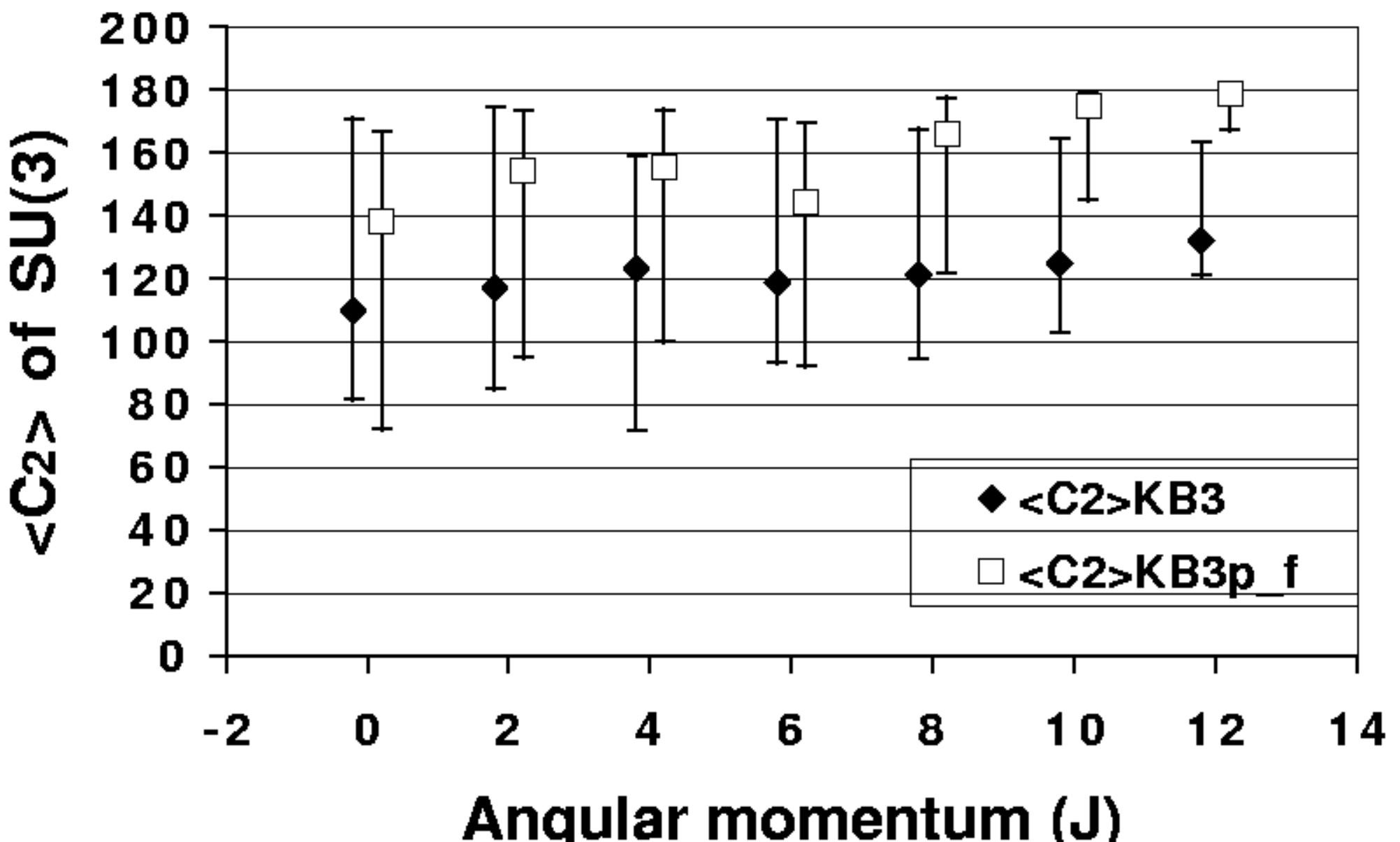}}
\end{center}
\caption{Average $C_{2}$ values for $KB3$ and $KB3p\_f$ interactions in $
^{44}$Ti.}
\label{<C2> for KB3 and KB3p_f in Ti44}
\end{figure}

\quad
An alternative way to show the recovery of the $SU(3)$ symmetry is given
in Fig. \ref{<C2> for KB3 and KB3p_f in Ti44} and Fig. \ref{<C2> for KB3 and
KB3p_f in Ti48}. These figures show the centroid, width, and skewness of the
$C_{2}$ distributions. The $J$ values are plotted on the horizontal axis
with the centroids given on the vertical axis. The width of the distribution
is indicated by the length of the error bars which is just the rms
deviation, $\Delta C_{2}=\sqrt{\left\langle \left(C_{2}-\left\langle
C_{2}\right\rangle \right) ^{2}\right\rangle }$, from the average value of
the second-order Casimir operator $\left\langle C_{2}\right\rangle $. The
third central moment, $\delta C_{2}=\sqrt[3]{\left\langle
\left(C_{2}-\left\langle C_{2}\right\rangle \right) ^{3}\right\rangle }$,
which measures the asymmetry, is indicated by the length of the error bar
above, $\Delta C_{2}+\frac{\delta C_{2}}{2}$, and below, $\Delta C_{2}-\frac{
\delta C_{2}}{2}$, the average value.

\begin{figure}[tbp]
\begin{center}
\leavevmode
\epsfxsize = \textwidth
\centerline {\includegraphics[width= \textwidth]{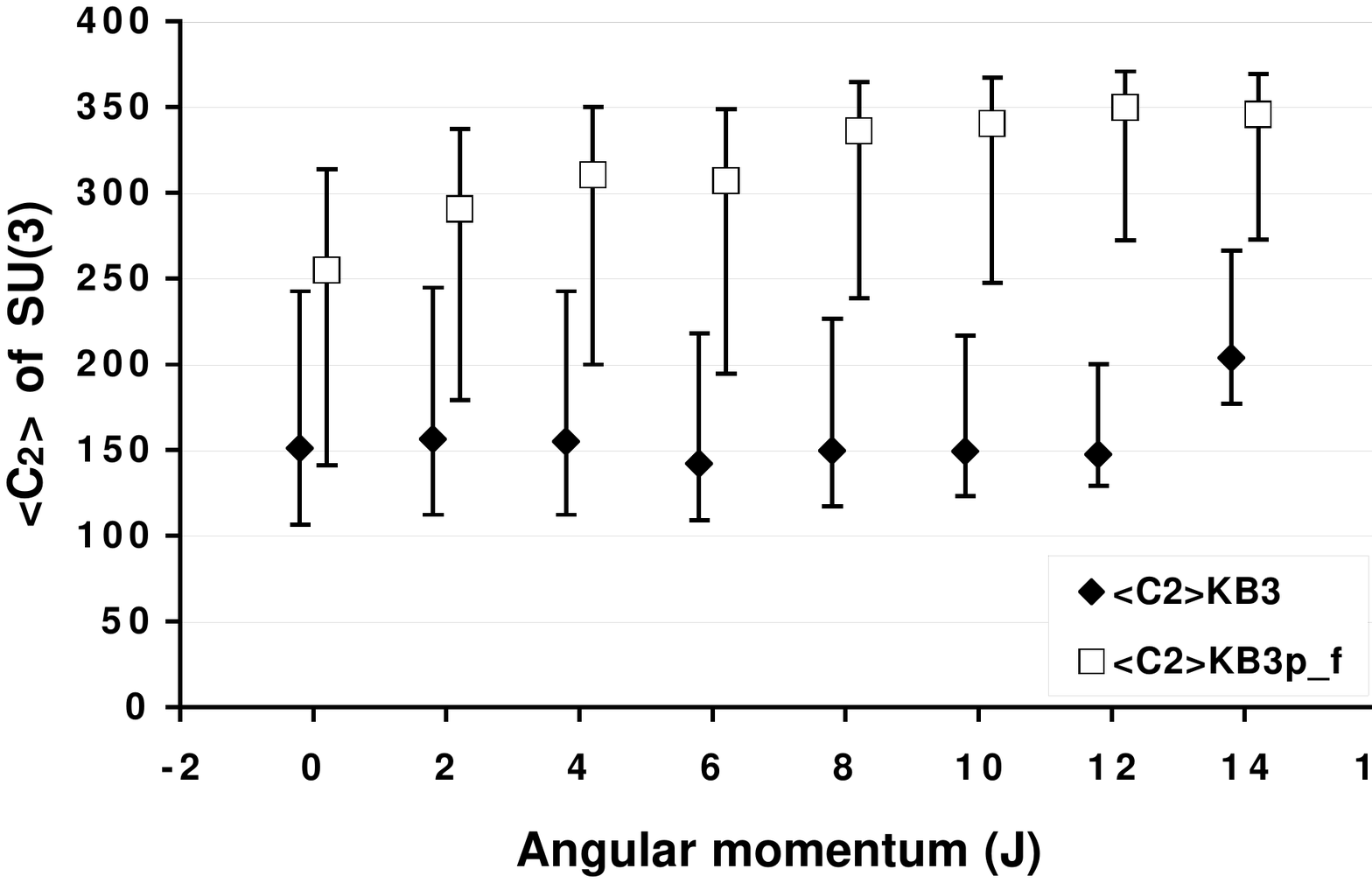}}
\end{center}
\caption{Average $C_{2}$ values for $KB3$ and $KB3p\_f$ interactions in $
^{48}$Ti.}
\label{<C2> for KB3 and KB3p_f in Ti48}
\end{figure}

Note that the recovery of the leading irrep when the spin-orbit interaction
is turned off is clearly signaled not only through an increase in the
absolute values of the first centroid $\left\langle C_{2}\right\rangle $ but
also through the skewness $\delta C_{2}$. For example, in $^{44}$Ti with the
$KB3$ interaction (spin-orbit interaction turned on) the ground state $J=0$
has $\left\langle C_{2}\right\rangle =110$ and skewness $\delta C_{2}=33$.
This changes for the $KB3p\_f$ interaction to $\left\langle
C_{2}\right\rangle =139$ and a skewness of $\delta C_{2}=-37$, as shown in
Fig. \ref{<C2> for KB3 and KB3p_f in Ti44}. The equivalent of the $^{44}$Ti
graph for the $^{48}$Ti case is shown in Fig. \ref{<C2> for KB3 and KB3p_f
in Ti48}. As for the $^{44}$Ti case, the results show the recovery of the $
SU(3)$ symmetry in $^{48}$Ti when the single-particle spin-orbit interaction
is turned off.

\subsection{Coherent Spectral Structure}

\quad
We now turn to a discussion of the coherence nature of the yrast states.
First, notice that the widths of the distributions as defined by $\Delta
C_{2}=\sqrt{\left\langle \left( C_{2}-\left\langle C_{2}\right\rangle
\right) ^{2}\right\rangle }$ are surprisingly unaffected (Fig. \ref{<C2> for
KB3 and KB3p_f in Ti44} and Fig. \ref{<C2> for KB3 and KB3p_f in Ti48}) when
turning the spin-orbit interaction on and off. This effect occurs in all
cases studied: $^{44}$Ti, $^{46}$Ti, $^{48}$Ti, and $^{48}$Cr. The more
detailed graphs, Fig. \ref{C2 of SU(3) for KB3 in Ti44} and Fig. \ref{C2 of
SU(3) for KB3p_f in Ti44}, offer an explanation in terms of the
fragmentation of the $C_{2}$ distribution. As can be seen from these graphs,
the irreps that are present in the structure of a given yrast state in the
presence of the spin-orbit interaction (Fig. \ref{C2 of SU(3) for KB3 in
Ti44}) remain present, even though with reduced strength, in the structure
of the state when the spin-orbit interaction is turned off (Fig. \ref{C2 of
SU(3) for KB3p_f in Ti44}). As a consequence, $\Delta C_{2}=\sqrt{
\left\langle \left( C_{2}-\left\langle C_{2}\right\rangle \right)
^{2}\right\rangle }$ which measures the overall spread of contributing
irreps, is more or less independent of the spin-orbit interaction. One can
see a sharp decrease in the width of the distribution only for high spin
states like $J=12$ in the graph for $^{44}$Ti in Fig. \ref{<C2> for KB3 and
KB3p_f in Ti44}.

\begin{figure}[tbp]
\begin{center}
\leavevmode
\centerline {\includegraphics[width= 5in]{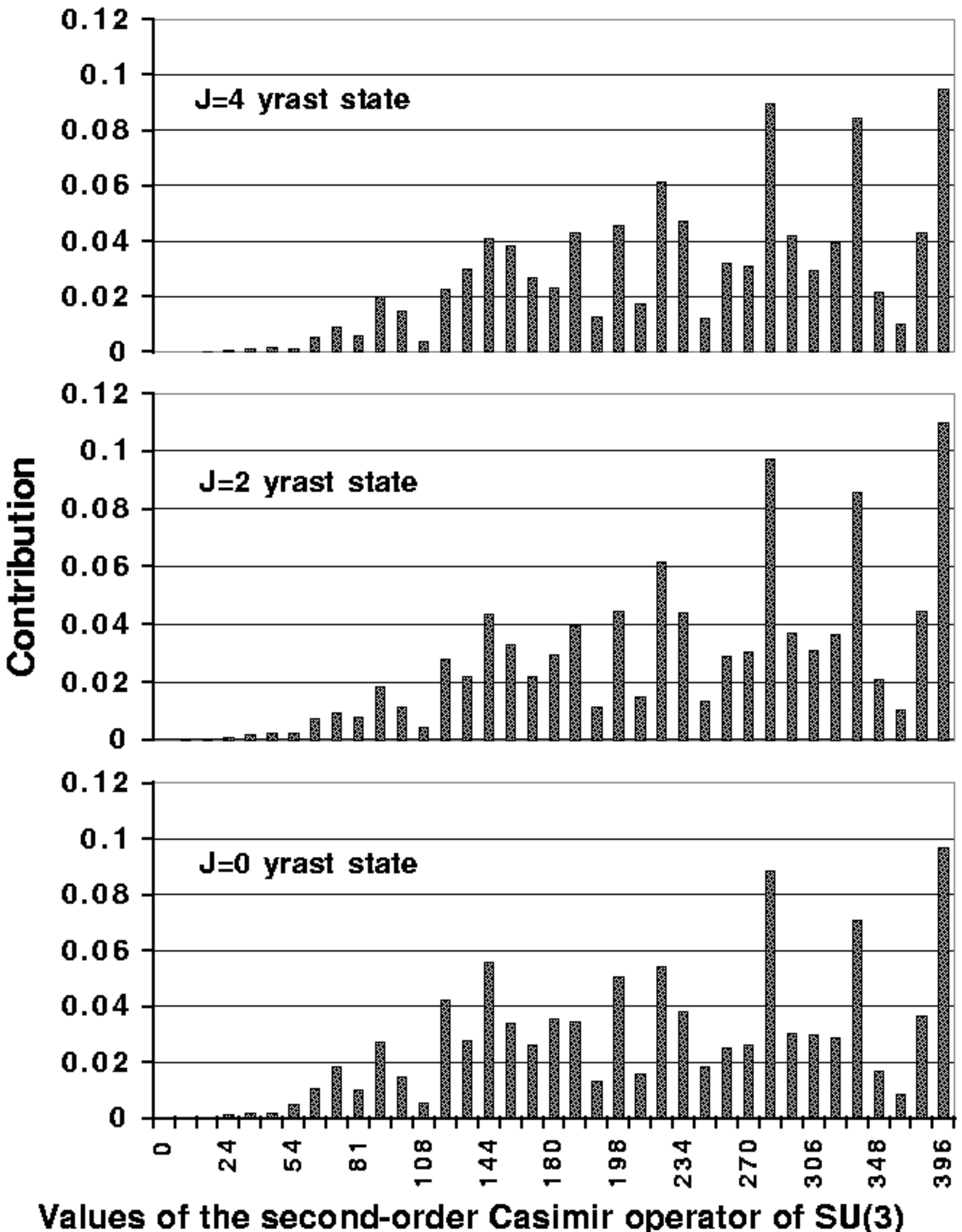}}
\end{center}
\caption{Coherent structure of the first three yrast states in $^{48}$Cr
calculated using realistic single-particle energies with Kuo-Brown-3
two-body interaction ($KB3$). On the horizontal axis is $C_{2}$ of $SU(3)$
with contribution of each $SU(3)$ state to the corresponding yrast state on
the vertical axis.}
\label{Coherence in Cr48 yrast band}
\end{figure}

Fig. \ref{Coherence in Cr48 yrast band} demonstrates the coherent nature of
the states within the yrast band. The three graphs shown give the spectrum
of the second-order Casimir operator $C_{2}$ of $SU(3)$ for the $J=0$, 2 and
4 yrast states in $^{48}$Cr. The axes are labelled the same way as in Figs.
\ref{C2 of SU(3) for KB3 in Ti44} and \ref{C2 of SU(3) for KB3p_f in Ti44},
but in this case all bars are for a single yrast state. In this figure there
are three peaks surrounded by smaller bars that yield a very similar
enveloping shape for the given yrast states. The fragmentation and spread of
$C_{2}$ values is nearly identical for these states with no dominant irrep,
indicative of severe $SU(3)$ symmetry breaking.

Graphs for the $KB3p\_f$ case, when the spin-orbit interaction is turned
off, are not shown since the results are similar to the results for $^{44}$Ti
shown in Fig. \ref{C2 of SU(3) for KB3p_f in Ti44}. For example, when the
spin-orbit interaction is on (KB3), the leading irrep for $^{48}$Cr has a $
C_{2}$ value of 396 and this accounts for only around 10$\%$ of the total
strength distribution (see Fig. \ref{Coherence in Cr48 yrast band}), but
when the spin-orbit interaction is off (KB3$p\_f$), the leading irrep is the
dominant irrep with more than 55$\%$ of the total strength.

\begin{figure}[tbp]
\begin{center}
\leavevmode
\centerline {\includegraphics[width= 5in]{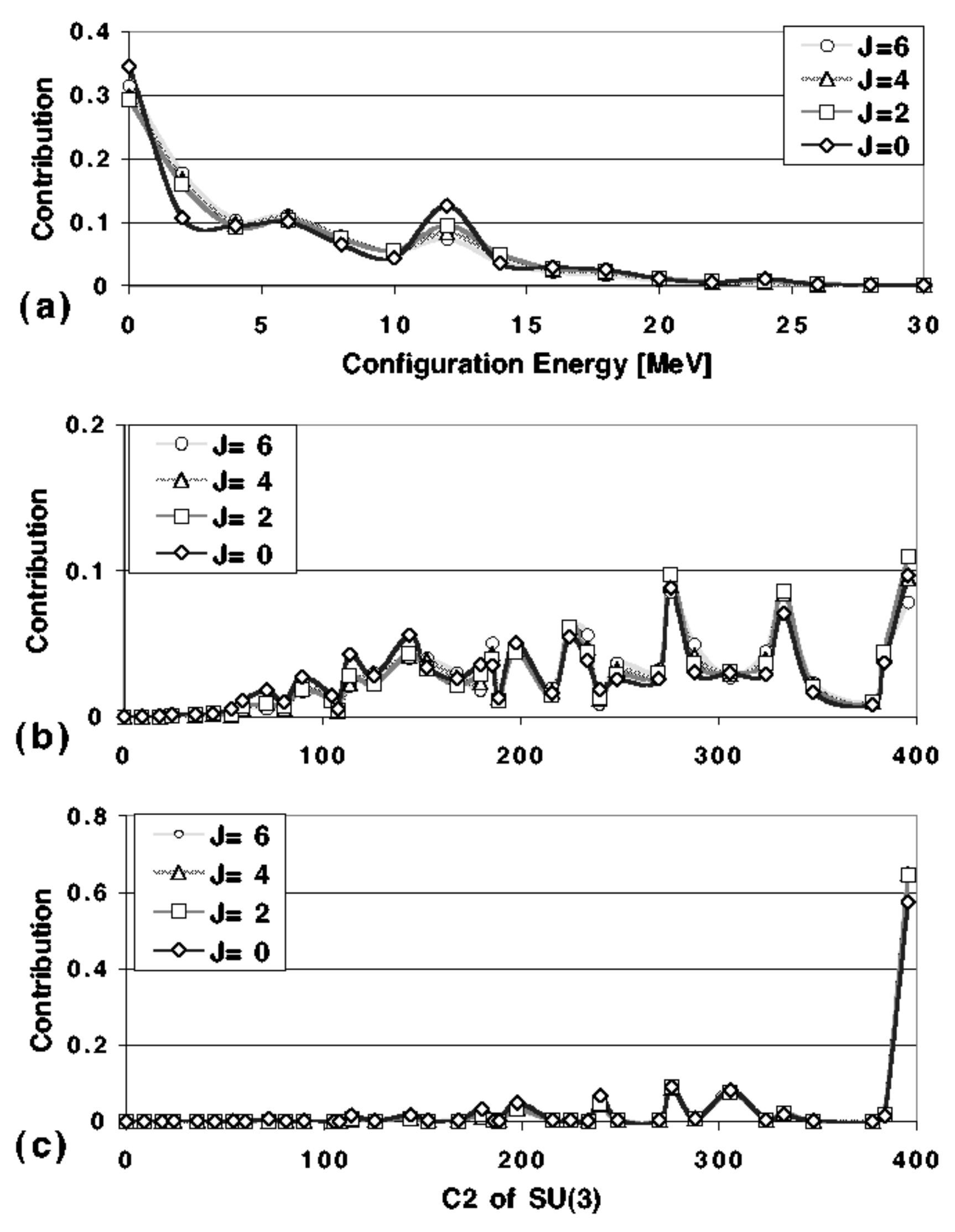}}
\end{center}
\caption{Coherent mixing and SU(3) breaking and recovery in $^{48}$Cr. Inset
(a) demonstrates the coherent structure of the yrast states with respect to
the spherical shell-model configuration basis (KB3); (b) coherent structure
of the yrast states with respect to the SU(3) basis (KB3); (c) recovery of
the SU(3) symmetry within the $KB3p_{f}$ interaction.}
\label{Coherent mixing in Cr48}
\end{figure}

We conclude the section with a discussion of the coherent structure of the
yrast states by an illustration of the coherent structure of the $^{48}$Cr
states within the spherical shell-model basis. The inset (a) of Fig. \ref
{Coherent mixing in Cr48} shows the spectral structure of the lowest yrast
states ($J=0,2,4,$ and $6$), as calculated with the KB3 interaction, with
respect to the spherical configuration basis. Notice the common spectral
distribution of these states. The distribution along the energy
configurations, related to excitation energies smaller than the harmonic
oscillator spacing ($<1\hbar \omega =10$ MeV), provides an illustration and
support of the energy-based configuration truncation scheme. The bump at 12
MeV is probably related to the fact that this is only a $pf$-shell
calculation ($0\hbar \omega $) which does not include the multi-shell
excitations that are at energies above $1\hbar \omega =10$ MeV.

\subsection{Enhanced Electromagnetic Transitions.}

\quad
Our results on the lower $pf$-shell nuclei, so far, have shown that $SU(3)$
symmetry breaking in this region is driven by the single-particle spin-orbit
splitting. However, even though states of the yrast band exhibit $SU(3)$
symmetry breaking, the yrast band $B(E2)$ values are insensitive to this
fragmentation of the $SU(3)$ symmetry; specifically, the quadrupole
collectivity as measured by $B(E2)$ transition strengths between low-lying
members of the yrast band remain high even though $SU(3)$ appears to be
broken.

\begin{figure}[tbp]
\begin{center}
\leavevmode
\epsfxsize = \textwidth
\centerline {\includegraphics[width= \textwidth]{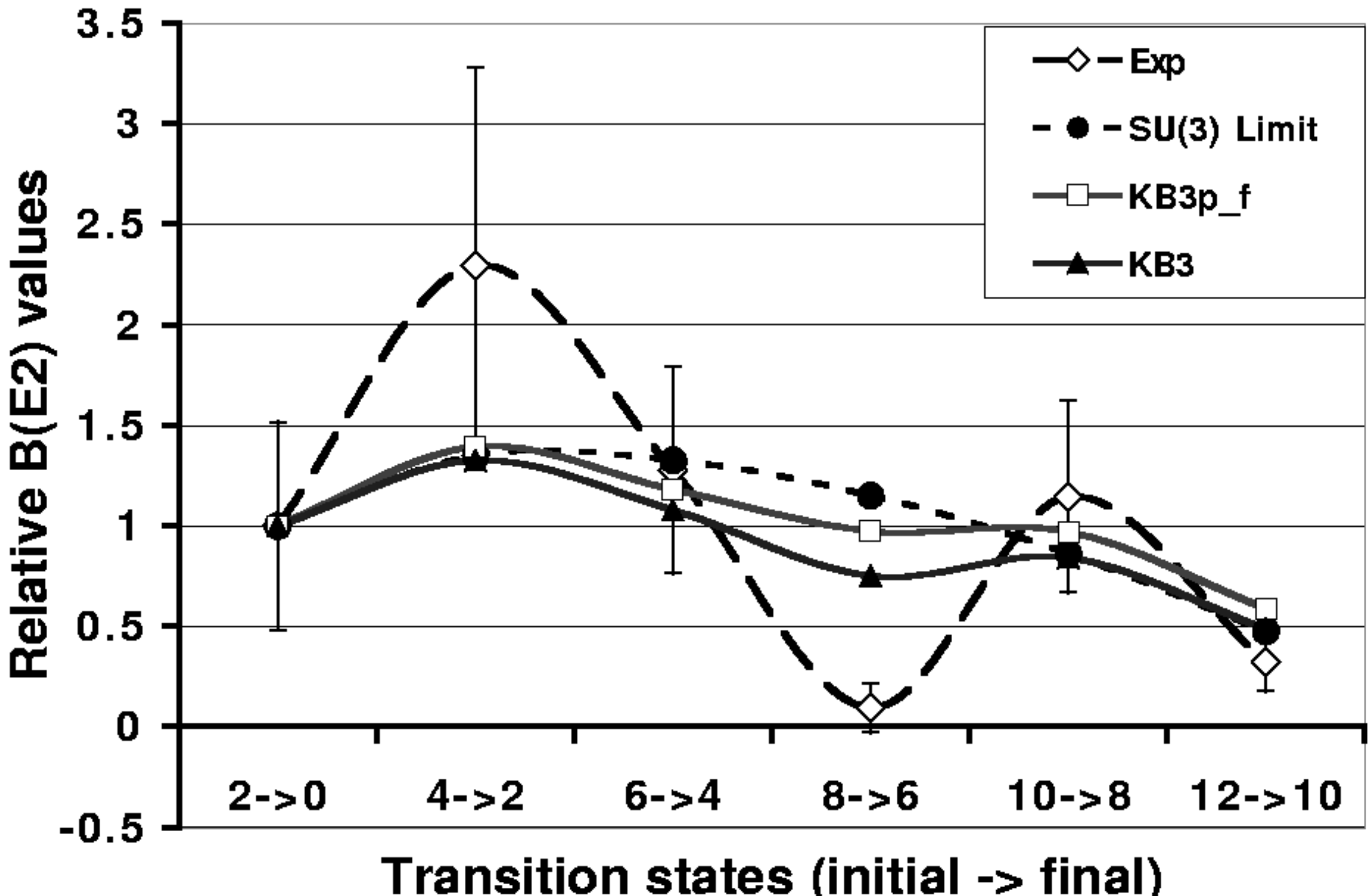}}

\end{center}
\caption{Relative $B(E2)$ values $\left(\frac{B(E2:J_{i}\rightarrow J_{f})}{
B(E2:2^{+}\rightarrow 0^{+})}\right) $ for $^{44}$Ti. The $
B(E2:2^{+}\rightarrow 0^{+})$ transition values are 122.69$e^{2}fm^{4}$ for
the experiment, 104.82$e^{2}fm^{4}$ for the KB3 interaction, and 138.58$
e^{2}fm^{4}$ for the $KB3p\_f$ case.}
\label{Relative B(E2) in Ti44}
\end{figure}

\begin{figure}[tbp]
\begin{center}
\leavevmode
\epsfxsize = \textwidth
\centerline {\includegraphics[width= \textwidth]{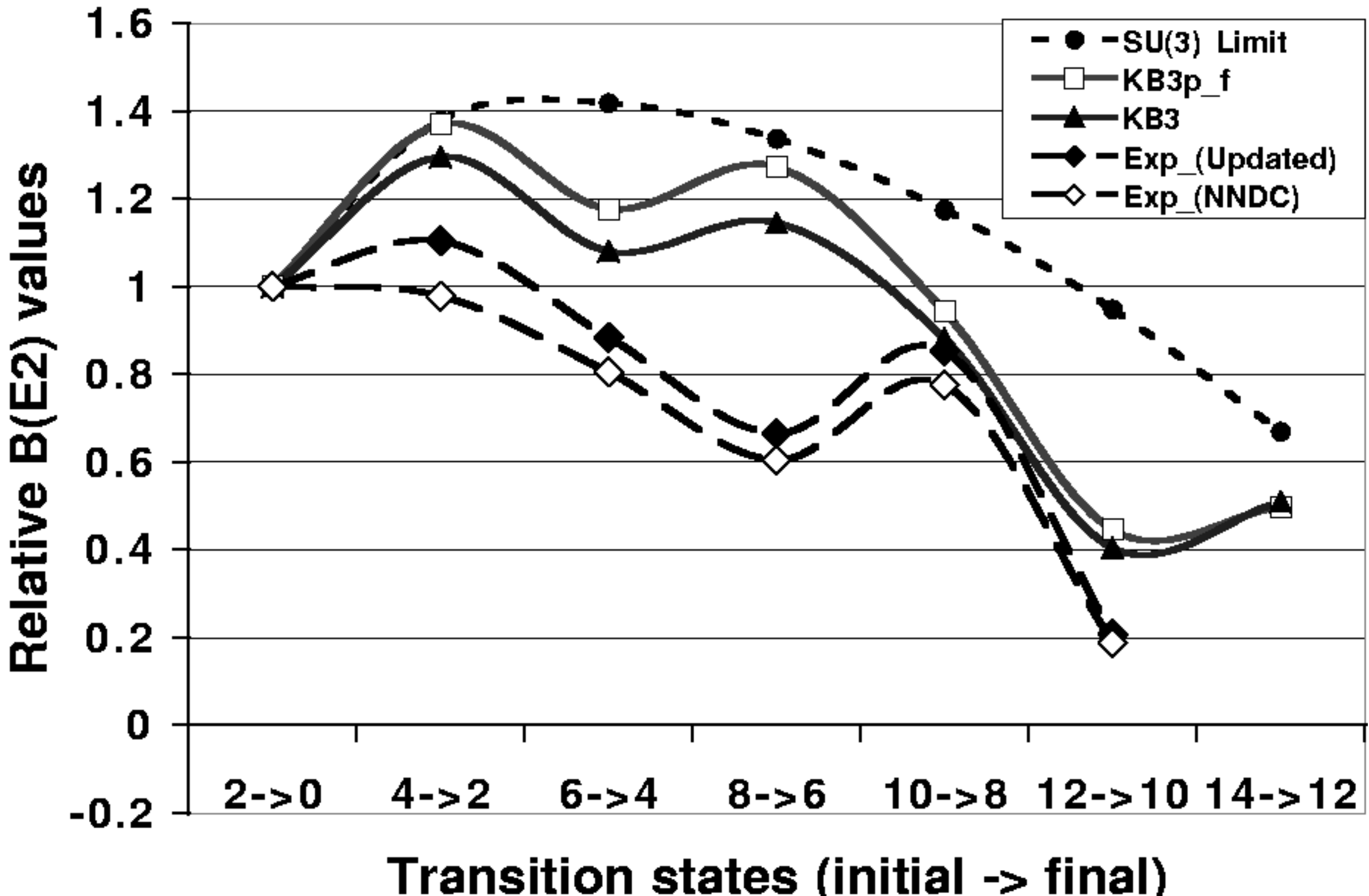}}
\end{center}
\caption{Relative $B(E2)$ values $\left(\frac{B(E2:J_{i}\rightarrow J_{f})}{
B(E2:2^{+}\rightarrow 0^{+})}\right) $ for $^{46}$Ti. The $
B(E2:2^{+}\rightarrow 0^{+})$ transition values are 199.82$e^{2}fm^{4}$ for
the experimental data, 181.79$e^{2}fm^{4}$ for the updated experimental
data, 208$e^{2}fm^{4}$ for KB3 interaction, and 299.83$e^{2}fm^{4}$ for $
KB3p\_f.$}
\label{Relative B(E2) in Ti46}
\end{figure}

\begin{figure}[tbp]
\begin{center}
\leavevmode
\epsfxsize = \textwidth
\centerline {\includegraphics[width= \textwidth]{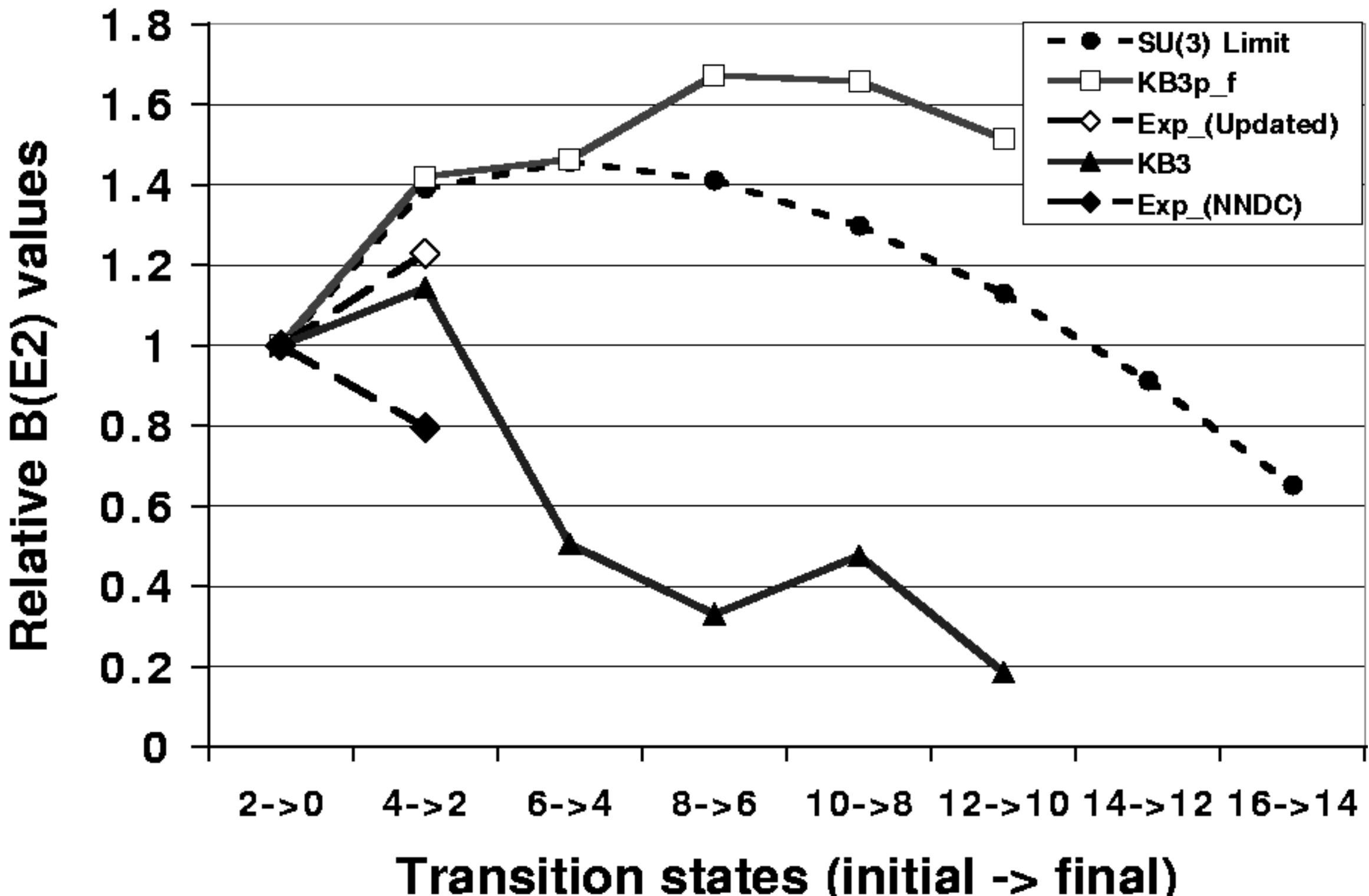}}
\end{center}
\caption{Relative $B(E2)$ values $\left(\frac{B(E2:J_{i}\rightarrow J_{f})}{
B(E2:2^{+}\rightarrow 0^{+})}\right) $ for $^{48}$Ti. The $
B(E2:2^{+}\rightarrow 0^{+})$ transition values are 144.23$e^{2}fm^{4}$ for
the experimental data, 155.5$e^{2}fm^{4}$ for the updated experimental data,
202.4$e^{2}fm^{4}$ for KB3 interaction, and 445.32$e^{2}fm^{4}$ for $
KB3p\_f. $}
\label{Relative B(E2) in Ti48}
\end{figure}

Relative $B(E2)$ values are shown in Figs. \ref{Relative B(E2) in Ti44},
\ref{Relative B(E2) in Ti46}, and \ref {Relative B(E2) in Ti48}, that is,
$B(E2)$ strengths normalized to the
$B(E2:2^{+}\rightarrow 0^{+})$ value. For isoscalar transitions, the relative
$B(E2)$ strengths are insensitive to the chosen effective charges which may be
used to bring the theoretical $ B(E2:2^{+}\rightarrow 0^{+})$ numbers into
agreement with the experimental values. Whenever absolute
$B(E2:2^{+}\rightarrow 0^{+})$ values are given, they are in $e^{2}fm^{4}$
units and the effective charges are
$1.5e$ for protons and $0.5e$ for neutrons ($q_{eff}$ = 0.5).

The first graph on relative $B(E2)$ values (Fig. \ref{Relative B(E2) in Ti44}
) recaps our results for $^{44}$Ti. Calculated relative $B(E2)$ values for
$^{44}$Ti corresponding to the spin-orbit interaction turned on (KB3) and
spin-orbit interaction off (KB3$p\_f$) are very close to the pure $SU(3)$
limit. The agreement with experiment is very satisfactory except for the $
4^{+}\rightarrow 2^{+}$ and $8^{+}\rightarrow 6^{+}$ transitions. However, the
experimental data \cite{NNDC} on the $8^{+}\rightarrow 6^{+}$ transition give
only an upper limit of 0.5 pico-seconds to the half-life. We have used the
worse case, namely a half-life of 0.5 ps, as a smaller value would increase the
relative $B(E2)$. For example, a half-life of 0.05 ps will agree well with the
relative $B(E2)$ value for the $KB3p\_f$ interaction. This example supports the
adiabatic mixing which seems to be present for all the yrast states of
$^{44}$Ti.

Fig. \ref{Relative B(E2) in Ti46} shows $B(E2)$ values for $^{46}$Ti. In
this case there are deviations from adiabatic mixing for the $
6^{+}\rightarrow 4^{+}$, $10^{+}\rightarrow 8^{+}$, and higher transitions.
Two experimental data sets are shown in Fig. \ref{Relative B(E2) in Ti46}:
data from the NNDC is denoted as Exp\_(NNDC), and updated data on $
2^{+}\rightarrow 0^{+}$ and $4^{+}\rightarrow 2^{+}$ transitions from \cite
{Recent Data on B(E2)} is denoted as Exp\_(Updated). For $^{46}$Ti the
agreement with the experiment is not as good as for $^{44}$Ti. However, the
experimental situation is also less certain. The adiabatic behavior is well
demonstrated for the first three yrast states $0^{+},$ $2^{+}$, and $4^{+}$
via relative $B(E2)$ values for the KB3 and $KB3p\_f$ interactions which are
very close to the $SU\left( 3\right) $ limit.

We conclude this section by showing the recovery of the $SU(3)$ symmetry;
this time via relative $B(E2)$ values as shown for $^{48}$Ti in Fig. \ref
{Relative B(E2) in Ti48}. In Fig. \ref{Relative B(E2) in Ti48} we see that
for the degenerate single particle case (KB3$p\_f$) the first few
transitions have relative $B(E2)$ values which follow the $SU(3)$ limit very
closely. On the other hand, the interaction involving spin-orbit splitting
(KB3) is far from the $SU(3)$ limit. The $B(E2:4^{+}\rightarrow 2^{+})$
transition is strongly enhanced due to the adiabatic mixing which is missing
in the higher than $J=4$ yrast states.

%% file: VGGPhDThesisCh7.tex
\chapter{Summary and Discussions}

\quad
The primary goal of the current work has been to study and apply a new
method--the mixed-symmetry approach--for large shell-model calculations. Our
aim was to combine two very successful computational methods: the $m$-scheme
spherical shell model and the $SU(3)$ shell model. In the process of this
study, we have realized a new computational paradigm: an oblique-basis
calculation that can be used to capture the mixed-mode structure of complex
systems, such as the atomic nuclei.

The two methods, the $m$-scheme and $SU(3),$ are closely connected to
the two dominant but often competing modes that characterize the structure
of atomic nuclei: the single-particle shell structure underpinned by the
validity of the mean-field concept, and the many-particle collective
behavior manifested through the nuclear deformation. This is reflected in
two dominant elements in the nuclear Hamiltonian: the single-particle term,
$H_{1}=\sum_{i}\varepsilon _{i}n_{i}$, and a collective two-body term $H_{2}.$
The collective term $H_{2}$ is dominated by the quadrupole-quadrupole
interaction, $H_{QQ}=Q\cdot Q$ which has good $SU(3)$ symmetry. It follows
that the simplified Hamiltonian $H=\sum_{i}\varepsilon _{i}n_{i}-\chi Q\cdot
Q$ has two exactly solvable limits and thus can be considered to be a
two-mode system.

To probe the nature of such a system, we have considered a simple toy
model: the one-dimensional harmonic oscillator in a box. As for real nuclei,
this system has a finite volume and a restoring force whose potential is of a
harmonic oscillator type. For this model, there is a well-defined energy
scale which measures the strength of the potential at the boundary of the
box, $E_{c}=\omega ^{2}L^{2}/2$. For this system, the use of two sets of
basis vectors, one for each of the two limits, has physical appeal,
especially at energies near $E_{c}$. One basis set consists of the harmonic
oscillator states; the other set consists of basis states of a particle in a
box. In the regime of strong mixing of the two modes at an energy scale
compatible with $E_{c}$, there is a coherent structure expressed through a
quasi-perturbational behavior of the system. Specifically, in this energy
region first-order perturbation theory is not appropriate since the 
zeroth order approximation to the wave function is very poor; nevertheless,
the first-order estimates of the energies are very close to the actual
results. Even more, the structure of the exact wave functions exhibits a
coherent mixing (Fig. \ref{States25to29}) similar to the one observed in
nuclei (Fig. \ref{Coherence in Cr48 yrast band}).

An application of the mixed-symmetry basis calculations to $^{24}$Mg, using
the realistic USD interaction of Wildenthal, has served to  demonstrate the
validity of the mixed-mode shell-model scheme. In this case, the oblique-basis
consists of the traditional spherical states, which yield a diagonal
representation of the single-particle interaction, together with collective
SU(3) configurations, which yield a diagonal quadrupole-quadrupole
interaction. The results obtained in a space that spans less than 10\% of the
full-space reproduce the correct binding energy, within 2\% of the full-space
result, as well as the low-energy spectrum, and the structure of the states
within 90\% overlap with the exact states. In contrast,  for an $m$-scheme
spherical shell-model calculation, one needs about 60\% of the full  space to
obtain results comparable with the oblique basis results. Calculations for 
$^{44}$Ti also support the mixed-mode shell-model scheme, even though
calculations using a few $SU(3)$ irreps  are not as good as the standard
spherical shell-model calculations. And, as the  results confirmed, the
combined basis yields less enhancements. For example, an oblique-basis
calculation in $50\%$ of the full $pf$-shell space is as good as a usual
$m$-scheme calculation in $80\%$ of space. These results show very clearly
that if the important modes can be isolated, then one can build an oblique
theory that incorporates leading configurations of each mode and could get
good convergence in a limited model space.

The study of the lower $pf$-shell nuclei $^{44-48}$Ti and $^{48}$Cr, using
the realistic Kuo-Brown-3 (KB3) interaction, has shown strong SU(3) symmetry
breaking due mainly to the single-particle spin-orbit splitting. When the
spin-orbit splitting is reduced, the importance of the $SU(3)$ as seen through
a growth in the dominance of the leading irrep is restored. Thus the KB3
Hamiltonian is at least a two-mode system. This is further supported  by the
behavior of the yrast band B(E2) values that seem to be insensitive to the 
fragmentation of the SU(3) symmetry. Specifically, the quadrupole collectivity
as  measured by the B(E2) strengths remains high even though the SU(3)
symmetry is rather  badly broken. This has been attributed to a quasi-SU(3)
symmetry where the  observables behave like a pure SU(3) symmetry while the
true eigenvectors exhibit a strong coherent structure with respect to each of
the two bases. This has been observed in all yrast states for the $^{44}$Ti
case; while for the other nuclei studied, this coherence breaks down after
the first few yrast states. In particular, even though the yrast states are
not dominated by a single $SU(3)$ irrep, the $B(E2:4^{+}\rightarrow 2^{+})$
values remain strongly enhanced with values close (usually within 10-20\%) to
the $SU(3)$ symmetry limit.

From a technical point of view, there are some other possible basis sets to
be studied.\footnote{In our study, SU(3) is shown to be good due to the
3D harmonic oscillator and the dominance of the Q.Q interaction in nuclei.
The cylindrical basis is just the easiest way to construct the SU(3) states
and seems to be most economical in terms of components. From computational
point of view, a good total angular momentum ($J$) and its third component
($M_J$) for the SU(3) states are essential. However, if one can find any
other basis set, besides the SU(3)-based one, with good $J$ and $M_J$, then
things may be as good, or even better.} For example, one can try to use
deformed Nilsson basis states, or a basis set generated from a Hartree-Fock
type procedure
\cite{PHF and SU(3)}. One can even try simple cylindrical basis states with
an appropriate procedure to maintain a complete set for good spin quantum
numbers. If good rotational symmetry is to be sacrificed, then one can try a
Lanczos algorithm which keeps only the big components during the iteration 
process.

Another further development of the theory and its application is a study of
other $sd$-shell nuclei as well as $pf$-shell nuclei. Such studies will
further test the theory and the codes that have been developed. In spite of
the results in the lower $pf$-shell, it is expected that in the mid-shell
region some sort of $SU(3)$ collective structure is important. Thus, the
oblique-basis calculation may be an important alternative for calculating
structure of nuclei, such as $^{56}Fe$ and $^{56}Ni.$ Another possibility is
to integrate the oblique basis concept into no-core calculations of the type
developed by \cite{Navratil'00}. Such an extension would involve the
symplectic group for multi-shell correlations rather than just SU(3) \cite
{Sp(6)models}.

An extension of the theory to a multi-mode oblique shell-model calculation
is also a possibility. An immediate extension of the current scheme might
use the eigenvectors of the pairing interaction \cite{Dukelsky et al-Pairing}
within the Sp(4) algebraic approach to the nuclear structure \cite
{Sviratcheva-sp(4)}, together with the collective SU(3) states and spherical
shell-model states. Even the three exact limits of the IBM \cite
{MoshinskyBookOnHO} can be considered to comprise a three-mode system.

Further, an even broader extension of the theory would involve a general
procedure for the identification of dominant modes from any one- and
two-body Hamiltonian along with a complementary partitioning of the model
space into physically relevant subspaces with small overlaps. One can then
start with eigenstates for an arbitrary subspace and constructively improve
the results by including corrections from the remaining subspaces. It should
be possible to do this by keeping only a small set of the calculated lowest
energy states at each iteration. Hamiltonian-driven basis sets can also be
considered. In particular, the method may use eigenstates of the very-near
closed shell nuclei obtained from a full shell-model calculation to form
Hamiltonian driven J-pair states for mid-shell nuclei \cite{Heyde's-shell
model}. This type of extension would mimic the Interacting Boson Model (IBM)
\cite{Iachello-1987} and the so-called broken-pair theory \cite
{Heyde's-shell model}. Nonetheless, the real benefit of this approach is
expected when the system is far away from any exactly solvable limit of the
Hamiltonian and the spaces encountered are too large to allow for exact
calculations.

In summary, we have studied a new computational method, the oblique-basis
method. The concept has been applied to a toy model, as well as to some
realistic nuclear systems. For realistic nuclei, we used spherical and
cylindrical single particle states to perform our mixed-symmetry
calculations. We have studied $^{24}$Mg in the $sd$-shell and $^{44}$Ti in the
$pf$-shell in a mixed-symmetry basis. For $^{24}$Mg, we have seen very
promising results with respect to the energy spectra and the structures of
the wave functions. When these results are translated into model space
dimensions, we see that an oblique-basis calculation in $10\%$ of the
full $sd$-shell space is as good as a usual $m$-scheme calculation in $60\%$
of the full $sd$-shell space. For $^{44}$Ti, the results are less pronounced
due to the dominance of the one-body over the two-body part of the
Hamiltonian. However, in model space dimensions, the results for $^{44}$Ti
state that an oblique-basis calculation in $50\%$ of the full $pf$-shell space
is as good as a usual $m$-scheme calculation in $80\%$ of the full $pf$-shell
space. Through a detailed study of $^{44}$Ti, $^{46}$Ti, $^{48}$Ti, and
$^{48}$Cr in the full $pf$-shell, we have confirmed the effect of the one-body
part of the Hamiltonian, that is, the strong $SU(3)$ symmetry breaking is due
to the spin-orbit interaction which splits the single-particle energies. For
degenerate single-particle energies, we have seen that one recovers  the
dominance of the leading $SU(3)$ irrep which is consistent with two-body
interaction dominated by the quadrupole-quadrupole interaction. Along our
study, we have seen some interesting coherent structures, such as coherent
mixing of basis states, quasi-perturbative behavior in the toy model, and an
enhanced $B(E2)$ strength toward the $SU(3)$ limit in nuclei. 

In concision, the main positive outcome of this work is a prove-of-principle
of the mixed-mode concept. We have shown that such calculations are doable
and may yield better results and lead to a clearer understanding of complex
systems. Problems yet to be solved are related mainly to the software package
and its development. First of all, a routine for the complete generation of
SU(3) shell model basis is needed. Basis sets other than the spherical
shell-model and SU(3) shell-model basis sets are also desirable; some
possible basis sets have been discussed. Another important software component
is a set of commonly used physical observables and their matrix elements. The
most important improvement, however, is to implement error estimate of the
final results and possible extrapolation procedure for estimating the exact
energy eigenvalues. Immediate further work should include a concentrated
study of other $sd$-shell nuclei, $pf$-shell nuclei, and multi-shell
calculations. Applications to atomic and molecular physics are also possible.

%% file: VGGPhDThesisAppendix.tex
\appendix
\chapter{On the Wave Function Spread and Localization}

\quad
In nuclear physics, we often use the three-dimensional harmonic-oscillator
(3D HO) potential as a zeroth order approximation to the nuclear mean-field
potential. This is usually done in the center-of-mass coordinate system,
assuming that all the nucleons experience the same attractive potential,
$ H_{0}=\sum \left( \frac{\vec{p}_{i}^{2}}{2m}+m\Omega ^{2}\vec{x}
_{i}^{2}\right) $\cite{MoshinskyBookOnHO}. If we assume the same
localization for the nucleons, there seems to be a
localization paradox since we are dealing with fermions that must obey the
Pauli exclusion principle. However, this apparent paradox is resolved by
using many-particle Slater determinant wavefunctions, constructed by
filling the single-particle levels of the three-dimensional
harmonic-oscillator potential.  The Slater determinant form satisfies the
Pauli principle requirements and yields different localization structure for
each nucleon.

A harmonic-oscillator potential is appropriate near stable equilibrium
where the interaction potential should have a local minimum. If rotational
invariance applies, then the potential near stable equilibrium should
actually be a three-dimensional harmonic-oscillator potential. However,
because of the Pauli principle, only the closed-shell nuclei have a spherical
shape, other nuclei have non-spherical ground state distributions that can be
characterized as oblate, prolate or tri-axial. As a consequence, for
non-closed shell nuclei, a deformed three-dimensional harmonic-oscillator
potential is a more appropriate ``mean field''. This idea is incorporated
in the deformed Nilsson model
\cite{Nilsson model}:
\[
H=\frac{\vec{p}^{2}}{2m}+m\Omega ^{2}\vec{x}^{2}+\varepsilon m\Omega
^{2}x_{3}^{2}+v_{ll}\vec{l}^{2}+v_{ls}\vec{l}\cdot \vec{s}.
\]
Here, $\varepsilon $ is a measure of the deformation when $\vec{l}^{2}$ and
$\vec{l}\cdot \vec{s}$ provide for the correct shell closures and magic
numbers in nuclei.

Large scale numerical calculations usually use basis functions of the
three-dimensional harmonic oscillator (3D HO). This way the wave function
parameter
$\omega
$ will not match the corresponding parameter $\Omega $ in the Hamiltonian. For
example,
$\Omega _{z}=\Omega \sqrt{1+\varepsilon}$ may be very different from $\omega
=\Omega
$ in super-deformed nuclei. It is therefore interesting to look at the
behavior of the fixed-basis calculations with respect to localization and
energy scale for the one-dimensional harmonic oscillator (1D HO).

Here we fix the parameters of the 1D HO Hamiltonian to be $m=\Omega =\hbar
=1$. Thus, its spectrum is simple: $E_{n}=n+\frac{1}{2}.$ The basis consists
of displaced and scaled harmonic-oscillator wave functions, $\Psi_{n}((q+\xi
)/\sigma);$ the $\xi =0$ states are squeezed/stretched states, and $\sigma =1$
states are coherent states \cite{Tapia-1993}. All the calculations are done
using the default settings for Mathematica 4.1 \cite {Mathematica 4.1}.

\begin{figure}[tbp]
\begin{center}
\leavevmode
\epsfxsize = \textwidth
\centerline {\includegraphics[width= \textwidth]{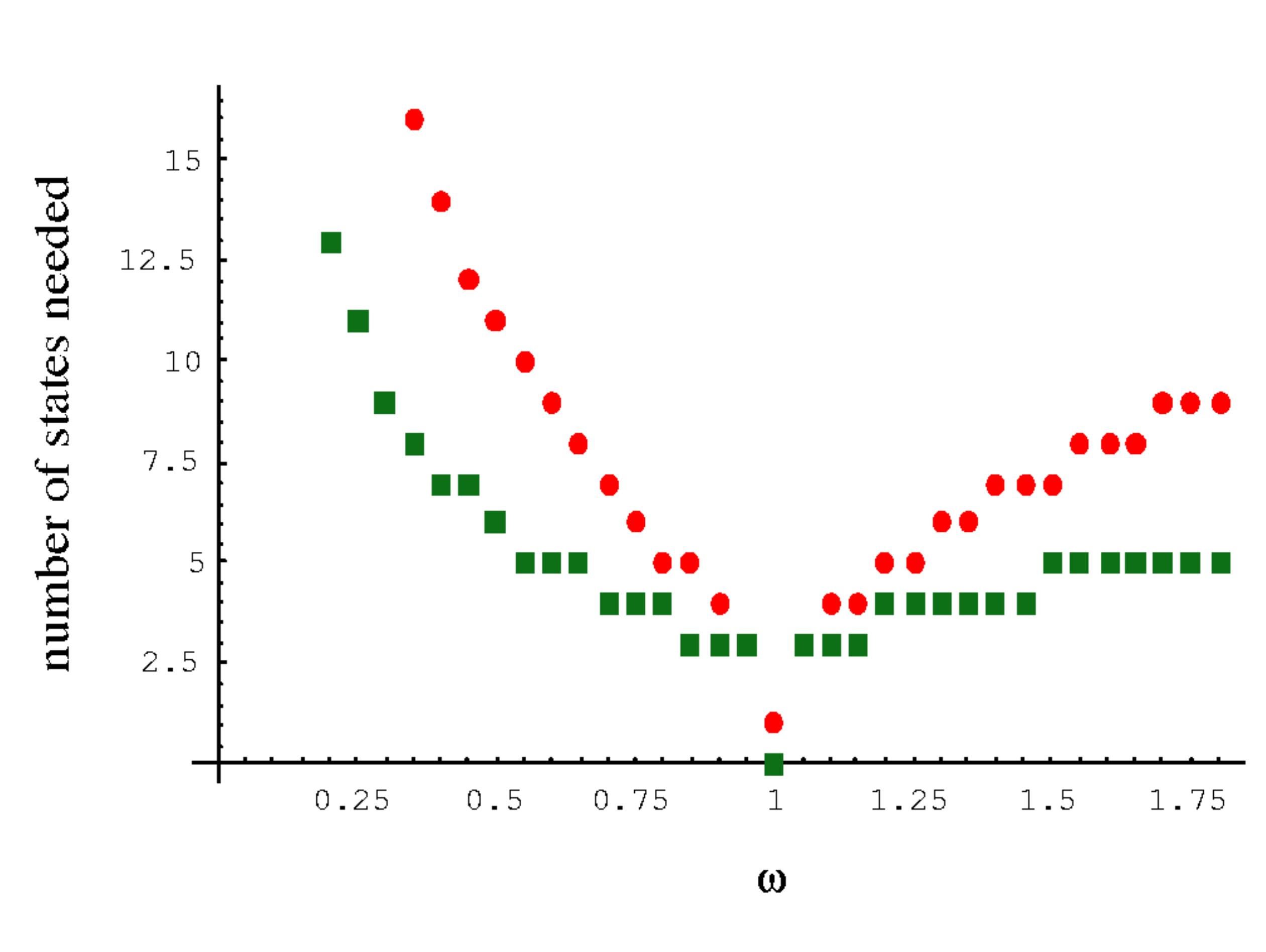}}
\end{center}
\caption{Ground-state convergence for the harmonic-oscillator problem with
$\Omega =1$ using squeezed basis states. The number of basis states needed
for $10^{-4}$ convergence accuracy of the ground-state eigenvalue is shown
in green squares. The red circles are for the ground-state
eigenfunction.}
\label{1DHOScaleConvergence}
\end{figure}

Fig. \ref{1DHOScaleConvergence} shows the number of the fixed-basis states
needed to achieve convergence to the $10^{-4}$ in the ground-state
eigenvalue and eigenvector as a function of the parameter $\omega$. The
convergence criteria for the eigenvalues requires two successive eigenvalues to
be less than the accuracy limit ($10^{-4}$) apart. The convergence criteria for
the eigenvectors is $\left|
\left| H\Psi -E\Psi
\right| \right| $ $<$ the accuracy limit. As expected, the
eigenvalues converge much earlier than the eigenvectors.

\begin{figure}[tbp]
\begin{center}
\leavevmode
\epsfxsize = \textwidth
\centerline {\includegraphics[width= \textwidth]{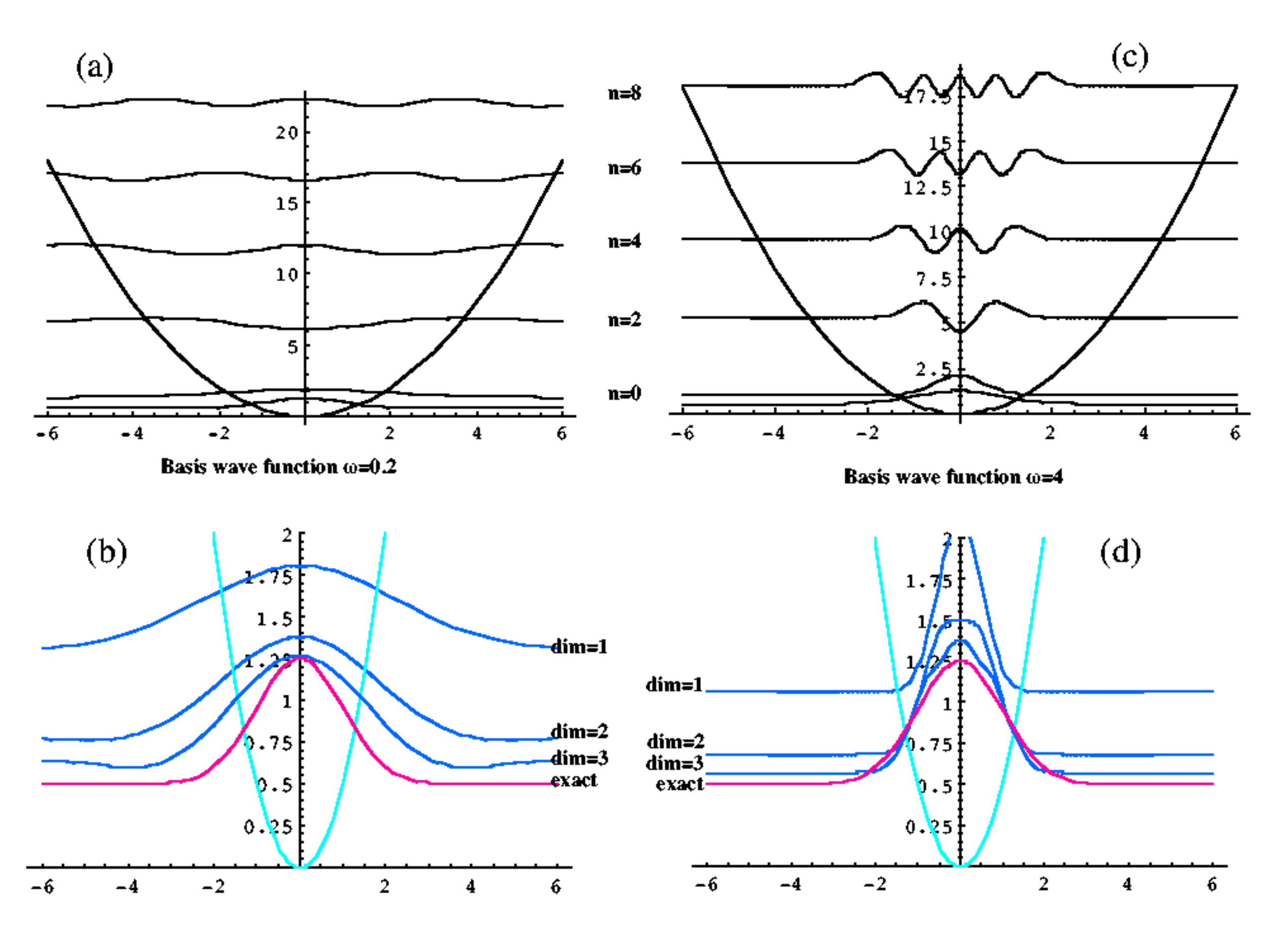}}
\end{center}
\caption{Role of the wave function spread. (a) Stretched basis states with
$\omega=0.2$ within the harmonic-oscillator potential $\Omega=1$. (b)
Consecutive approximations of the harmonic-oscillator ground state (red)
using the stretched basis states. (c) Squeezed basis states with $\omega=4$
within the harmonic-oscillator potential $\Omega=1$. (d) Consecutive
approximations of the harmonic-oscillator ground state (red) using the
squeezed basis states.}
\label{1DHOWFSpead}
\end{figure}

Fig. \ref{1DHOWFSpead} (a) shows the first few basis states ($\omega =0.2$),
the harmonic-oscillator potential ($\Omega =1$), and the true ground-state
wave function. Fig. \ref{1DHOWFSpead} (b) shows the calculated ground-state
wave functions at different dimensions of the basis states with $\omega
=0.2$. From these graphs, it is clear that when $\omega <\Omega =1,$ Fig.
\ref {1DHOWFSpead} (a) and (b), one uses more and more basis states to
produce the correct wave function behavior within the classically forbidden
region. When $\omega >\Omega =1,$ Fig. \ref{1DHOWFSpead} (c) and (d), the
focus is actually concentrated on getting the correct shape of the wave
functions within the potential well.

\begin{figure}[tbp]
\begin{center}
\leavevmode
\epsfxsize = \textwidth
\centerline {\includegraphics[width= \textwidth]{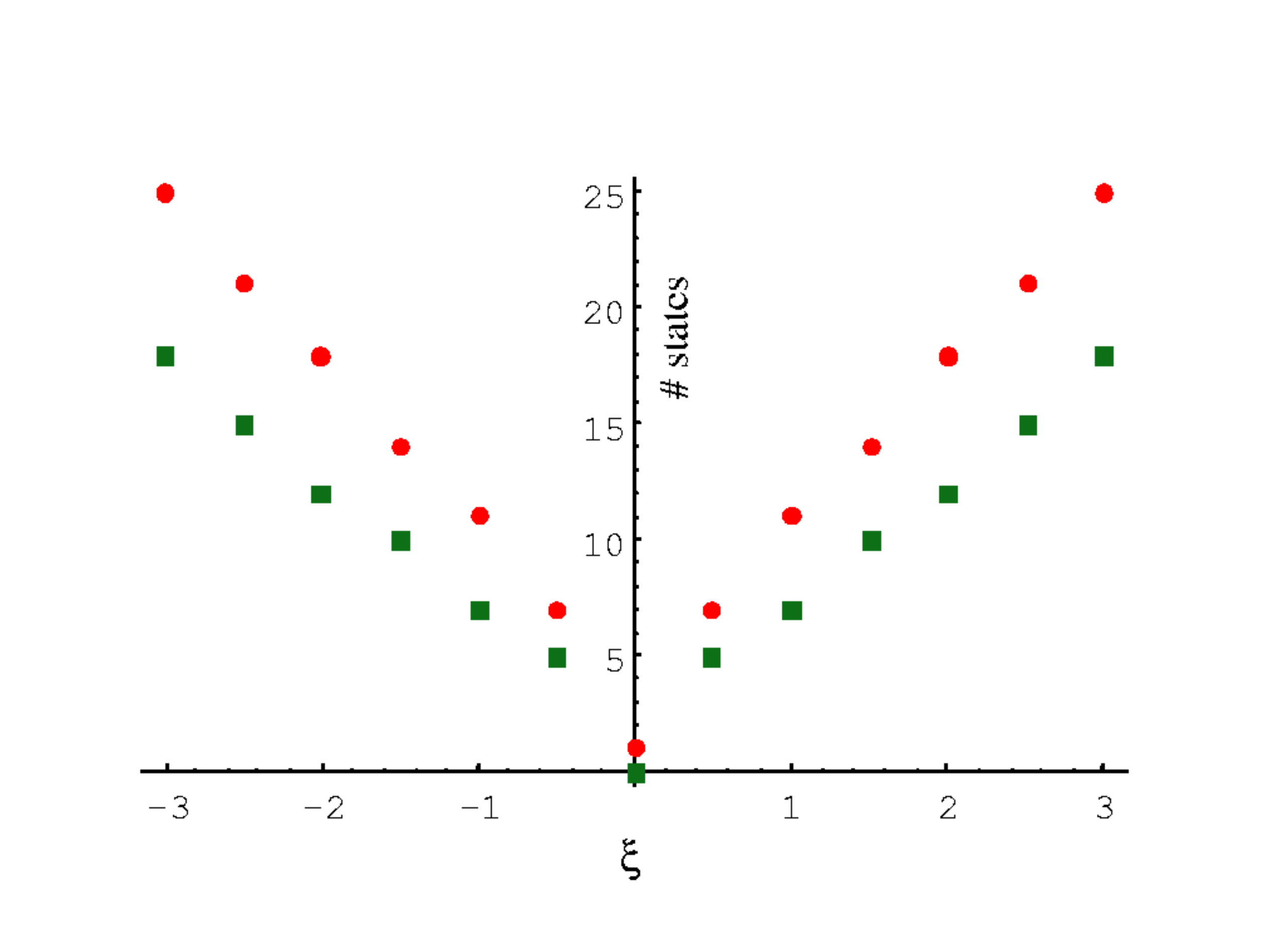}}
\end{center}
\caption{Ground-state convergence for the harmonic-oscillator problem with
$\Omega =1$ using coherent basis states. The number of basis states needed
for $10^{-4}$ convergence accuracy of the ground-state eigenvalue is shown
with the green squares. The red circles are for the ground-state
eigenfunction.}
\label{1DHOLocalizationConvergence}
\end{figure}

\begin{figure}[tbp]
\begin{center}
\leavevmode
\epsfxsize = \textwidth
\centerline {\includegraphics[width= \textwidth]{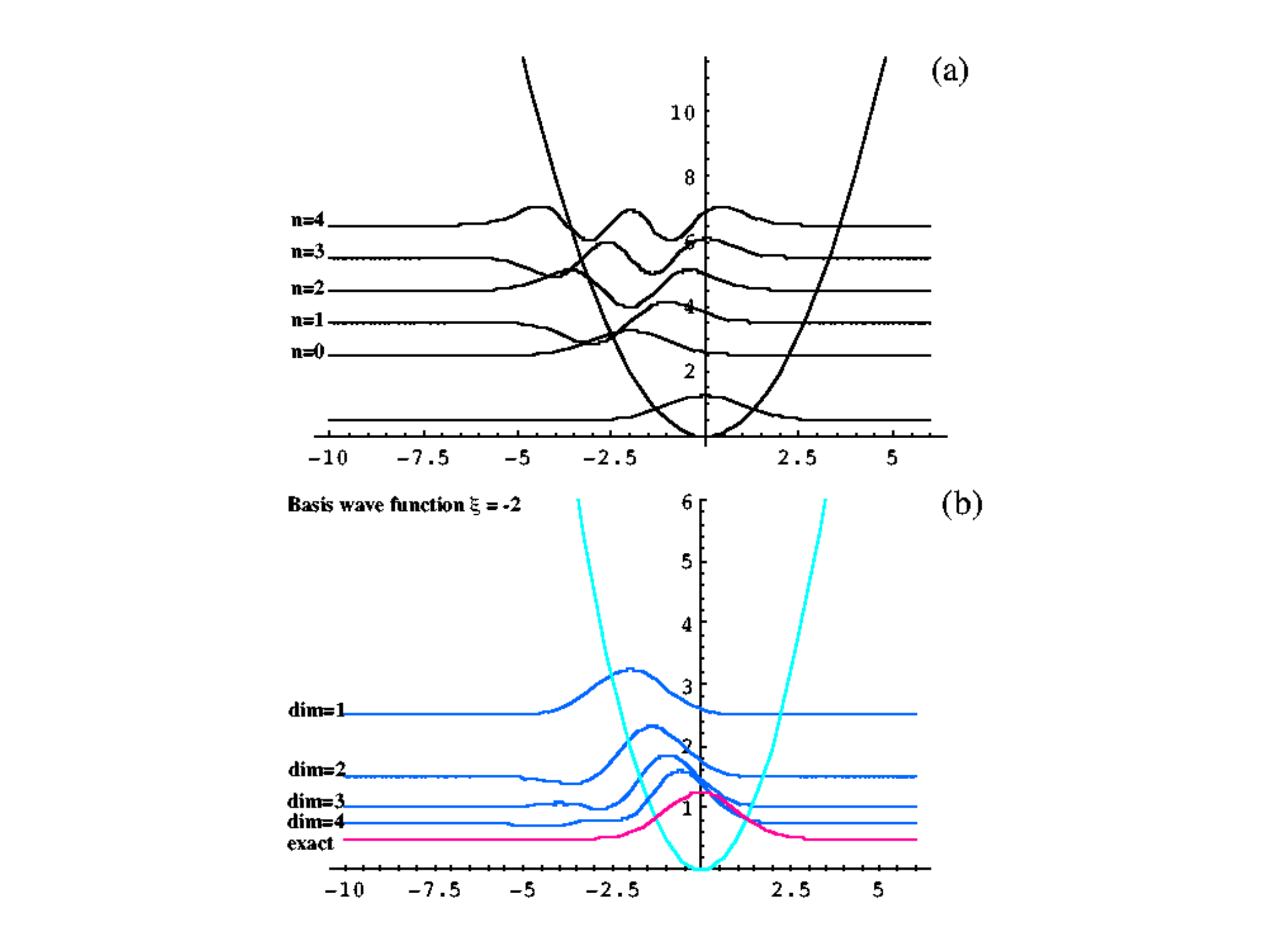}}
\end{center}
\caption{Role of the wave function localization. (a) Coherent basis states
with displacement $\xi =-2$ within the harmonic-oscillator potential $\Omega
=1$. (b) Consecutive approximations of the harmonic-oscillator ground state
(red) using the coherent basis states.}
\label{1DHOWFLocalization}
\end{figure}

Fig. \ref{1DHOLocalizationConvergence} is similar to Fig. \ref
{1DHOScaleConvergence} but shows the convergence within the displaced
(coherent states) harmonic-oscillator wave function basis ($\Psi _{n}(q+\xi
) $). Due to parity conservation, there is a good symmetry under the $\xi
\rightarrow -\xi $ transformation. An example of the basis structure and
convergence path similar to Fig. \ref{1DHOWFSpead} is shown in Fig.
\ref{1DHOWFLocalization} for $\xi =-2.$

Although one should be able to solve any problem in any arbitrarily chosen
orthonormal basis, the considerations presented here point to the need for
properly modified basis states to reduce the model-space dimension and thus
to avoid problems due to numerical noise. In the process of optimizing the
basis set for a particular Hamiltonian, the orthogonality of the basis would
inevitably be destroyed. In the toy model of a two-mode system, a few possible
types of basis-state refinements were considered. Originally, the
oblique-basis method was concerned with two or more basis sets as described
in the toy model and in nuclear physics applications. However, the idea
of the basis refinement can be extended in quite a general way as described
in the next section.

\chapter{Variationally-Improved Basis Method}

\quad
The usual fixed-basis method can be derived from the Rayleigh-Ritz variation
principle. If one considers minimization of $E\left(\vec{c}\right) $ with
respect to $\vec{c}$ for a Hamiltonian $(H)$ using the basis states $\phi
_{n}\left(x;\omega \right) $:
\[
E\left[ \vec{c}\right] =\left\langle \Psi \left| H\right| \Psi
\right\rangle-\lambda (\vec{c}\cdot \vec{c}-1),\quad \Psi \left(x\right)
=\sum_{n}c_{n}\phi _{n}\left(x;\omega \right)
\]
of a Hermitian operator $Y\left(\omega \right) $ such that
\[
Y\left(\omega \right) \phi _{n}\left(x;\omega \right) =Y_{n}\phi
_{n}\left(x;\omega \right),
\]
then $\delta E[\vec{c}]/\delta c_{n}^{*}=0$ is equivalent to solving the
matrix eigenvalue problem $\sum_{m}H_{nm}c_{m}=\lambda c_{n},$ where $
H_{nm}=\left\langle \phi _{n}\left| H\right| \phi _{m}\right\rangle,$ and
thus the set of $\lambda $s provides information for the eigenvalues of $H$.
If the set $\left\{\phi _{n}\right\} $ is taken to be non-orthogonal, then we
have a generalized eigenvalue problem $\sum_{m}\left(H_{nm}-\lambda \mu
_{nm}\right) c_{m}=0$ where $\mu _{nm}=\left\langle \phi _{n}|\phi
_{m}\right\rangle$.

Notice that there is a freedom that we have not yet specified: it is the
choice of $\omega $ and $Y\left( \omega \right)$. Here, $\omega $ is the
set of parameters characterizing the Hermitian operator $Y\left( \omega
\right) .$ Usually one fixes $\omega $ from experience or by simply
applying Rayleigh-Ritz variation with respect to $\omega $ in $E\left[
\omega \right] =\left\langle \phi _{0}\left( \omega \right) \left| H\right|
\phi _{0}\left( \omega \right) \right\rangle $. Thus, we have an orthonormal
basis $\phi _{n}$ with the same $\omega $ for any $n.$ One can try other
procedures for fixing $\omega $ and $Y\left( \omega \right) $ as well \cite
{Skyrme-1957 CinQM}.

All this seems fine as long as the spectrum of $H$ is expected to be similar
to the spectrum of $Y\left( \omega \right) ,$ but what if the potential for
$Y\left( \omega \right) $ does not match the ``landscape'' of the potential
for $H$. For example, would the harmonic-oscillator potential wave functions
be appropriate for solving an anharmonic potential problem or a double-well
potential? In principle, one should be able to use any basis, but it may be
at the expense of long and tedious calculations. Therefore, we may try to
let $\omega $ be a free parameter for different basis functions $\phi _{n}$.
Then we can find $\omega _{n}$ by variation of $E\left[ \omega \right]
=\left\langle \phi _{n}\left( \omega \right) \left| H\right| \phi _{n}\left(
\omega \right) \right\rangle $.

Often $\omega $ is related to the relevant energy scale. If we start with
the correct wave function, but with the `wrong' parameters, then clearly a
variational approach on the parameters will give us the right answer
immediately.

In the case of the harmonic-oscillator wave functions, one can argue that a
calculation with a varying $\omega $ is equivalent to a multi-shell
calculation with a fixed $\omega $ parameter. Since the harmonic-oscillator
basis is a complete basis, then each function $\phi _{n}\left(x,\omega
_{n}\right) $ can be expanded in the basis associated with $\Omega $, for
example:
\[
\phi _{n}\left(x,\omega _{n}\right) =\sum_{k}c_{n}^{k}\phi _{k}\left(x,
\Omega \right).
\]
Therefore, $\phi _{n}\left(x,\omega _{n}\right) $ can be viewed as the
result of a multi-shell calculation with the harmonic-oscillator parameter
$\Omega$.

Next, we discuss how in general one can refine any initial basis set so that
each basis vector in the new and improved basis is the optimal one with
respect to the Hamiltonian under consideration. Then,
instead of refining the basis vectors, one can effectively renormalize the
parameters in the Hamiltonian.

The main idea is to optimize each trial vector by applying the Rayleigh-Ritz
variational principle on $E\left[ \Psi ,\emph{A}\right] $:
\[
E\left[ \Psi ,\emph{A}\right] =\left\langle A\Psi \left| H\right| A\Psi
\right\rangle ,\quad \delta E\left[ \Psi ,\emph{A}\right] =0,
\]
where $\emph{A}$ represents the affine group $\emph{A}$ in $\mathbf{R}^{n}$.
An element $a$ of $\emph{A}$ has a rotational component $r$ and a
translational component $t,$ so that $(ax)_{j}=r_{i}^{j}x_{j}+t_{j}$.

If $G$ is the symmetry group of $H$, then:
\[
g^{-1}Hg=H,\quad g\in G.
\]
Therefore, the Rayleigh-Ritz variational principle should be applied with
respect to the homogeneous space $M=\emph{A/G}$ that excludes the symmetry
transformations $G$. For example, a translational symmetry of $H$ means that
scaling and rotation are the relevant transformations. Since the physical
systems usually have rotational and translational symmetry, then only
scaling is left as a relevant operation for constructing a
variationally-improved basis:
\[
\Psi \left( x_{1},...,x_{n}\right) \rightarrow \Psi \left(
sx_{1},...,sx_{n}\right) .
\]
The transformation $\left| \Psi \right\rangle \rightarrow \left| \emph{A}
\Psi \right\rangle $ can be defined to maintain the normalization of the
states:
\[
\Psi \left( x\right) \rightarrow A\Psi \left( x\right) =\sqrt{\det \left(
r\right) }\Psi \left( rx+t\right) .
\]
However, this transformation is not a unitary transformation in general, and
therefore, it will not map orthonormal states into a new set of orthogonal
states. Using scaling as a variational parameter has been done previously.
Specifically, in the context of the confined systems it was used by Martin
and Cruz to study hydrogen and helium enclosed in a spherical shell \cite
{Marin and Cruz-1991 AJP}, \cite{Marin and Cruz-1991 JPB}.

If $\int dy=\int \det \left( r\right) dx$ is used when $y=rx+t$, then the
variationally-improved basis can be treated as a renormalization problem for
the Hamiltonian $H$:
\begin{eqnarray*}
E\left[ \Psi ,\emph{A}\right] &=&\left\langle A\Psi \left| H\right| A\Psi
\right\rangle =\int \det \left( r\right) \Psi ^{*}\left( rx+t\right) H\left(
p,x\right) \Psi \left( rx+t\right) dx= \\
&=&\int \Psi ^{*}\left( y\right) H\left( rp,r^{-1}\left( y-t\right) \right)
\Psi \left( y\right) dy.
\end{eqnarray*}

Rescaling in the above way seems to be related more to the scaling methods
in condensed matter physics. For example, consider the one-dimensional
harmonic oscillator:
\[
H=\frac{1}{2}P^{2}+\frac{1}{2}Q^{2}.
\]
Then, the equation for the scale parameter $s$ from $E\left[ \Psi ,\emph{s}
\right] $ when $\Psi \left( x\right) \rightarrow \sqrt{s}\Psi \left(
sx\right) $ is:
\begin{eqnarray*}
E\left[ \Psi ,\emph{s}\right] &=&\frac{1}{2}s^{2}\left\langle
P^{2}\right\rangle +\frac{1}{2}\frac{1}{s^{2}}\left\langle
Q^{2}\right\rangle , \\
\frac{\partial E\left[ \Psi ,\emph{s}\right] }{\partial s^{2}}
&=&0\Rightarrow s^{2}=\sqrt{\frac{\left\langle Q^{2}\right\rangle }{
\left\langle P^{2}\right\rangle }}=\frac{\Delta q}{\Delta p}.
\end{eqnarray*}
Here, $\left\langle Q^{2}\right\rangle =\left\langle \Psi \left|
Q^{2}\right| \Psi \right\rangle $ and $\Delta q=\sqrt{\left\langle
Q^{2}\right\rangle }$. It is assumed that $\Psi $ is such that $\left\langle
Q\right\rangle =\left\langle P\right\rangle =0$, which means that the
localization of the wave function has been selected. Evaluating $E\left[
\Psi ,\emph{s}\right] $ at the extremum $s^{2}=\Delta q/\Delta p$ gives:
\[
E\left[ \Psi \right] =\Delta q\Delta p.
\]
Finally, using $\left[ p,q\right] =-i\Rightarrow \Delta q\Delta p\geq \frac{1
}{2}$, we find that the minimum of the energy is exactly the zero point
energy for the harmonic oscillator $E_{0}=\frac{1}{2}.$ Notice that quantum
mechanics was only used to provide us with a constraint on the fluctuations
of the observables $q$ and $p$; other than that, we can consider the system
as purely statistical. Thus, different $\Delta q\Delta p$ will give
different value of $E\left[ \Psi \right] $. Turning this argument around, we
would expect $\Delta q\Delta p\geq \left( n+\frac{1}{2}\right) $ when $
\Delta q$ and $\Delta p$ are evaluated in the space of wave functions with
$n$-nodes.

Another interesting way to obtain the same result is to use $H$ expressed in
terms of the operators $a^{+}$ and $a$. Then, by using coherent states as
trail wave functions $a\left| z\right\rangle =z\left| z\right\rangle $ we
have $E\left[ z\right] =\left| z\right| ^{2}+\frac{1}{2}$ and thus
$E_{0}= \frac{1}{2}$.

\chapter{On the Loss of Hermiticity}

\quad
When the choice of the basis is not carried out with appropriate attention,
an operator, supposedly Hermitian, may acquire a non-hermitian matrix
realization within this basis. For example, a wrong basis may produce a
non-Hermitian matrix for the Hamiltonian under consideration. Although this is
unlikely to be encountered within the finite shell-model calculations using an
occupation number representation, it is an obstacle when one wishes to use a
hard core potential and a harmonic-oscillator basis
\cite{MoshinskyBookOnHO}.

Here we discuss the problem of a free particle in a one-dimensional box in
the harmonic-oscillator basis. In order to proceed, we notice that the
Hilbert space for the harmonic oscillator is not quite the same as for the
free particle in a one-dimensional box. This is clear from the domains of
the wave functions. The harmonic-oscillator wave functions are defined on
the whole real axis $\mathbf{R}^{1},$ when the wave functions for a free
particle in a box are defined on a finite interval $[-L,L].$ This
discrepancy is easily fixed by projecting the harmonic-oscillator wave
functions onto the interval $[-L,L]$, which changes the inner product for
the wave functions:
\[
\left( f,g\right) =\int_{-\infty }^{\infty }f^{*}\left( x\right) g\left(
x\right) dx\rightarrow \int_{-L}^{L}f^{*}\left( x\right) g\left( x\right)
dx.
\]
However, in this basis the matrix corresponding to $H$ will be nonhermitian
in general.

To understand the loss of hermiticity, we look at the off-diagonal matrix
elements of the momentum operator ($P=-i\hbar \frac{\partial }{\partial q}$):
\begin{eqnarray*}
(\Psi _{m},P\Psi _{n}) &=&\int\limits_{-L}^{L}\Psi _{m}^{*}(q)(P\Psi
_{n}(q))dq=\int\limits_{-L}^{L}\Psi _{m}^{*}(q)(-i\hbar \frac{\partial \Psi
_{n}(q)}{\partial q})dq= \\ &=&i\hbar \int\limits_{-L}^{L}\frac{\partial (\Psi
_{m}^{*}(q)\Psi _{n}(q))}{\partial q}dq+i\hbar
\int\limits_{-L}^{L}\frac{\partial \Psi _{m}^{*}(q)}{
\partial q}\Psi _{n}(q)dq= \\
&=&i\hbar \left. (\Psi _{m}^{*}(q)\Psi _{n}(q))\right|
_{-L}^{L}+\int\limits_{-L}^{L}\left( -i\hbar \frac{\partial }{\partial q}
\Psi _{m}(q)\right) ^{*}\Psi _{n}(q)dq= \\
&=&i\hbar \left. (\Psi _{m}^{*}(q)\Psi _{n}(q))\right| _{-L}^{L}+(P\Psi
_{m},\Psi _{n}).
\end{eqnarray*}

It is clear from the above expression that the \textit{hermiticity will be
maintained only when all of the basis functions are zero\footnote{Wave functions
with the same value at $\pm$ L is a necessary condition; wave functions should
be zero at $\pm$ L only for an infinite potential.} at the boundary of the
interval} [-L,L]. This condition is essential for solving exactly the
quantization of a free particle in a one-dimensional box.

\chapter{Coherent Behavior, Quasi-Symmetries and Quasi-Labels}

\quad  Recently the notion of a quasi-symmetry and adiabatic mixing has been
introduced in nuclear physics \cite{Adiabatic mixing}. The toy model of a
harmonic oscillator in a one-dimensional box can be used to introduce and
illustrate one possible definition of a quasi-symmetry, an asymptotic label
(quasi-label), and a coherent behavior associated with a quasi-symmetry.

First, we define a similarity relation of two states $\left| \Phi
\right\rangle $ and $\left| \Psi \right\rangle $ with respect to some
Hermitian operator $\mathcal{H}$, and denote it as:
\[
\left| \Phi \right\rangle \stackrel{\mathcal{H}}{\sim }\left| \Psi
\right\rangle .
\]
In this approach, the operational definition of such a similarity relation
uses the eigenvectors $\left| \mathcal{H};\Lambda \right\rangle $ and the
eigenvalues $\Lambda $ of $\mathcal{H}$:
\[
\mathcal{H}\left| \mathcal{H};\Lambda \right\rangle =\Lambda \left| \mathcal{
\ H};\Lambda \right\rangle .
\]
We would say that $\left| \Phi \right\rangle $ and $\left| \Psi
\right\rangle $ are $\mathcal{H}$ similar ($\left| \Phi \right\rangle
\stackrel{\mathcal{H}}{\sim }\left| \Psi \right\rangle $), if there is a
function $f$, eventually monotonic, that maps the distribution
$\rho _{\Phi }\left( \Lambda \right) =\left| \left\langle \mathcal{H};\Lambda
|\Phi
\right\rangle \right| ^{2}$ to $\rho _{\Psi }\left( \Lambda \right) =\left|
\left\langle \mathcal{H};\Lambda |\Psi \right\rangle \right| ^{2}$ so that:
\[
\left| \rho _{\Phi }\left( \Lambda \right) \right| ^{2}\approx \left| \rho
_{\Psi }\left( f\left( \Lambda \right) \right) \right| ^{2}.
\]

In simple words, this means that the shape of the probability distribution
$\rho _{\Phi}$ is similar to the shape of the probability distribution
$\rho_{\Psi}$.

If $G$ is a symmetry group for the operator $\mathcal{H}$, then the
eigenspace for a given $\Lambda $ may be degenerate, and any function
$\left| \Phi \right\rangle $ obtained from $\left| \Psi \right\rangle $ by a
unitary transformation $U\in G$ ($\left| \Phi \right\rangle =U\left| \Psi
\right\rangle $) will be $\mathcal{H}$ similar to $\left| \Psi \right\rangle
$. Thus, $G$ defines an intrinsic symmetry for the wave functions that are
similar to $\left| \Psi \right\rangle $. Therefore, $\left| \Psi
\right\rangle $ can be viewed as an ``intrinsic state'' with respect to the
symmetry of $\mathcal{H}$.

In the case when $\mathcal{H}$ is one of the exact limits of a Hamiltonian
$H$, i.e. $H=\mathcal{H}+\lambda ^{-1}V$, then one can define an adiabatic
mixing of the states $\left| \mathcal{H};\Lambda \right\rangle $ due to the
interaction $V$. An asymptotic label (quasi-label) $\Lambda $ can be
assigned to each eigenvector $\left| \Psi ;\lambda \right\rangle $ of
$H=\mathcal{H}+\lambda ^{-1}V$ in the limit $\lambda \rightarrow $ $\infty$:
\[
\mathcal{H}\left| \Psi ;\Lambda ,\lambda \rightarrow \infty \right\rangle
=\Lambda \left| \Psi ;\Lambda ,\lambda \rightarrow \infty \right\rangle
\]
thus,
\[
\left| \Psi ;\Lambda ,\lambda \right\rangle {\rightarrow }\left| \mathcal{H}
;\Lambda \right\rangle ,\quad when\quad \lambda \rightarrow \infty .
\]
Assigning $\Lambda $ by using the natural order of the levels must be done
carefully by tracing the sign of each level crossing.

Once the asymptotic label $\Lambda $ has been assigned for a state $\left|
\Psi \right\rangle $, then a coherent behavior with respect to an observable
can be defined as well. There is a quasi-symmetry $\mathcal{H}$ for the
observable $O:\left| \Psi \right\rangle \rightarrow \mathbf{R}$, if its
value $O\left[ \left| \Psi ;\Lambda ,\lambda \right\rangle \right] $ does
not depend much on the parameter $\lambda $:
\[
O\left[ \left| \Psi ;\Lambda ,\lambda \right\rangle \right] \approx O\left[
\left| \mathcal{H};\Lambda \right\rangle \right] .
\]
Some common functions for $O\left[ \left| \Psi ;\Lambda ,\lambda
\right\rangle \right] $ are related to the expectation values of $O:$
\[
\left\langle \Psi ;\Lambda ,\lambda \rightarrow \infty \right| O\left| \Psi
;\Lambda ,\lambda \rightarrow \infty \right\rangle \approx \left\langle
\Lambda \right| O\left| \Lambda \right\rangle .
\]
In general, a coherent behavior with respect to other quantities can be
defined as well. For example, a relative transition rate from a state $
\left| \Psi ;J\right\rangle $ to a state $\left| \Psi ;J+2\right\rangle$
due to a transition operator, say $E2,$ can be defined, say by using
$B\left( E2,\Psi ,J\right) $, where $\Psi $ is the ``intrinsic state'' upon
which the band is built. All the states within the band should actually be
within the class of $\mathcal{H}$ equivalent states. In particular, the
asymptotic label $\Lambda $ can be used as a band label.

Notice that, as in the toy model studied, it may happen that at finite
$\lambda$ the wave function $\left| \Psi \right\rangle $ has been assigned
label $\Lambda $ while its components along the space $\left| \mathcal{H}
;\Lambda \right\rangle $ are practically missing. Following the results from
the toy model, we can define some possible types of spectral structures that
may exhibit such coherent behavior with respect to a Hamiltonian $H=\mathcal{
H}+V,$ and thus to specify a quasi-symmetry.

Specifically, setting $\lambda =1,$ $H=\mathcal{H}+V$ has a quasi-symmetry
if:
\begin{eqnarray*}
H\left| \Psi ;\Lambda _{n}\right\rangle  &=&E\left( \Lambda _{n}\right)
\left| \Psi ;\Lambda _{n}\right\rangle , \\
\mathcal{H}\left| \Lambda _{n}\right\rangle  &=&\Lambda _{n}\left| \Lambda
_{n}\right\rangle ,\quad \Lambda _{n+1}>\Lambda _{n}, \\
\Lambda _{n} &>&\left\langle \Lambda _{n}\left| V\right| \Lambda
_{n}\right\rangle >\Lambda _{n+1}-\Lambda _{n}, \\
E\left( \Lambda _{n}\right)  &\approx &\left\langle \Lambda _{n}\left|
H\right| \Lambda _{n}\right\rangle .
\end{eqnarray*}
Here, $\Lambda _{n}$ is the corresponding asymptotic label of the state
$\left| \Psi ;\Lambda _{n}\right\rangle .$ The term $\left\langle \Lambda
_{n}\left| V\right| \Lambda _{n}\right\rangle >\Lambda _{n+1}-\Lambda _{n}$
means that $V$ mixes strongly different $\left| \Lambda _{n}\right\rangle$
states. Therefore, \textit{perturbation theory cannot be applied in the
usual small perturbation regime}. However, $\Lambda _{n}>\left\langle
\Lambda _{n}\left| V\right| \Lambda _{n}\right\rangle $ together with $
E\left( \Lambda _{n}\right) \approx \left\langle \Lambda _{n}\left| H\right|
\Lambda _{n}\right\rangle =\Lambda _{n}+\left\langle \Lambda _{n}\left|
V\right| \Lambda _{n}\right\rangle $ means that the spectral structure of $H$
in this region is similar to the spectral structure of $\mathcal{H}$ within
a few percent:
\[
\frac{E\left( \Lambda _{n}\right) -\Lambda _{n}}{E\left( \Lambda _{n}\right)
}\approx \frac{\left\langle \Lambda _{n}\left| V\right| \Lambda
_{n}\right\rangle }{\Lambda _{n}+\left\langle \Lambda _{n}\left| V\right|
\Lambda _{n}\right\rangle }\approx \frac{\left\langle \Lambda _{n}\left|
V\right| \Lambda _{n}\right\rangle }{\Lambda _{n}}<1.
\]

This seems to be the situation discussed in the toy model case. Since
$\Lambda _{n}\sim n^{2},$ then $2\Lambda _{n}>\Lambda _{n+1}$ gives
$(n-1)^{2}>8$ which is always satisfied for $n>4.$

\chapter{Guide to the Oblique-Basis Package}

\quad 
Although most of the routines\footnote{The \textit{Oblique Basis package
2002 is }available from the author upon request.} in the \textit{Oblique-Basis
Package 2002} can be compiled by using the \textit{makefile} routines provided,
there is a need to follow a few simple steps in order to be able to carry out
oblique-basis calculations. Here we describe some of the technical problems and
their solutions that one may face in using the \textit{Oblique-Basis package
2002}. The process of running an oblique-basis calculation consists of four main
steps which are described below:

\begin{itemize}
\item[(1)]  selecting and generating basis states ($nuke$, $PNGGMJ$),

\item[(2)]  preparing interaction file(s) ($IsoInt2pn,$ $MakeInteractions$),

\item[(3)]  evaluating matrix elements for the chosen interactions ($su3pn$),

\item[(4)]  solving the generalized eigenvalue problem which includes:
obtaining the eigenvectors of the Hamiltonian, calculating expectation
values and transition probabilities for some desired operators ($GLanczos$).
\end{itemize}

In the parentheses are given the names of some relevant routines. Some
current limitations of the \textit{Oblique-Basis package 2002} are related
to the single-particle basis currently employed in the computations. In its
present form, the main routine $su3pn$, which is used to evaluate the matrix
elements of the operators (step 3 above), is set to operate on spherical and
cylindrical single-particle states. This clearly restricts the basis
generation process (step 1 above) to the same type of states (spherical and
cylindrical). Even though the spherical and cylindrical single-particle
states are of special interest, the code could, at least in principle, be
changed to operate on other desirable single-particle states, such as those one
can obtain via Hartree-Fock procedure.

\section{Generating Basis States}

\quad
Since the oblique-basis idea is to combine two or more basis sets, the
oblique-code package has been designed to use basis states generated and
used by other shell-model packages. The two main codes in mind are: a
version of the Glasgow code, which performs calculations in the spherical
shell-model basis, and the SU(3) RME code, which performs calculations in
the SU(3) shell-model basis using cartesian (cylindrical) single-particle
states.

\paragraph{Spherical Shell-Model States.}

To generate spherical shell-model basis states one needs to run the Glasgow
code ($nuke$ located in the folder $Glasgow-code$) with the desired
configuration limits and with $IBASIS=-1$ in the appropriate interaction
file ($*.int$). For more details see the example file $BasisOnly.int$ and
files $Glasgow.int-instructions$ or \textit{Instructions Glasgow test9.doc}.
There is a script file, $RunGlasgow.sc$, which one may find useful when
running $nuke$. This script file uses head files (*.head) and an interaction
file (*.int) to construct different input files (*.inp) for $nuke.$

The wave functions from $nuke$ are stored in the file $fort.60$ (see
$Basis-states.info$ for its structure). This file is used by
$GlsgwBasis2Redstick$, located in the folder $Oblique-Glasgow,$ to produce
file $fort.35$ which contains the spherical shell-model basis states and the
single-particle states data. If $nuke$ (Glasgow) is used to calculate the
eigenvectors of a Hamiltonian, then these eigenvectors are extracted from
file $fort.60$ into file $fort.36$.

\textbf{Important note}: File $fort.11$ is very important pre-generated
file. This file ($fort.11$) is used by $nuke$ and $GlsgwBasis2Redstick$.
\textbf{Do not lose the file }$fort.11$!

\textbf{Tip}: In oblique runs of the $su3pn$ code, consider using sorted
spherical shell-model states. Such states are produced by the $EpsSorting$
code located in the folder $Oblique/SphToCyl$.

\paragraph{SU(3) Shell-Model States.}

Even though the C. Bahri's SU(3) RME code may provide the SU(3) highest
weight states in the near future, the \textit{Oblique -Basis package 2002}
has its own highest-weight state generator for some of the most important
SU(3) irreps. The SU(3) related fortran codes are in the folder $SU3Generator
$.

The current highest weight state generator $SU3\_HWS\_GEN$ is located in a
folder with the same name. Before using $SU3\_HWS\_GEN,$ one may need to run
the auxiliary code $SU3Lister$ which would help in the HWS selection
process. The $SU3Lister$ code also generates the necessary cylindrical
single-particle states (file $cylin.sps$) that are needed in the evaluation
of matrix elements of operators through the $su3pn$ code.

\textbf{Important note}: Before doing any runs, compile file $fort.4$ using
$SU3GENBK$ located in the folder $ProjectLibrary$.

\textbf{Tip}: By comparing the output of the $SU3Lister$ with the output of
the $genwsirl$, one can determine the irreps that are not generated by the
generator $SU3\_HWS\_GEN$.

Once the highest weight states are created, they are stored in a file with
extension $*.hws.$ This file is used as input file to the SU(3) code $PNGGMJ$
located in the sub-folder $PNGGMJ.$ The code $PNGGMJ$ generates two files
containing basis states $*.su3$ and $*.bas$. It also generates a file
$PNGGMJ\_Brief\_Info.log$ which contains information that may be used to set
the parameters \textbf{max\_nbas\_su3} and \textbf{maxmpc} for the $su3pn$
code. The $*.su3$ file has some of the SU(3) related information and is
usually used in testing tools which are located in the sub-folder
$ProjectTools$. The $*.bas$ file is mainly for use by the $su3pn$ code.

\section{Generating Interaction Files}

\quad
Even though there are a few the realistic interactions commonly used (KB3,
Wildenthal...), their format files may differ significantly. Interactions given
in the isospin format as used by the Glasgow code can be transformed into the
proton-neutron format by using the code $IsoInt2pn$ located in the folder
$Oblique-Glasgow$.

Often used schematic interactions are also available through a package
called $MakeInteractions$ made by Dr. C. Johnson. $MakeInteractions$ allows
one to generated combinations of frequently studied nuclear interactions.
The menu of the currently available interactions is:

\begin{itemize}
\item[(0)]  Random noise (TBRE, two-body random ensemble),
\item[(1)]  Pairing,
\item[(2)]  Multipole-multipole (you choose L),
\item[(3)]  S\symbol{94}2 (total spin),
\item[(4)]  L\symbol{94}2 (total orbital angular momentum),
\item[(5)]  J\symbol{94}2 (total angular momentum),
\item[(6)]  L*S (spin-orbit) 1+2 body.
\end{itemize}

\section{Running the Main Routines}

\paragraph{Evaluating Matrix Elements of Operators.}

Usually, the running of programs takes considerable time. Thus, it is better
to use a script file for such runs. Some example script files are
$RunSu3pn+GLDriver.sc$ and $RunSu3pnGLanzos.sc.$ Basically, the input of the
$ su3pn$ code requires the following entries to be specified:

\begin{itemize}
\item[$>$]  name of the file containing single-particle levels with
extension *.sps,
\item[$>$]  scaling of the two-body matrix elements $\left( A/B\right)
\symbol{94}X$,
\item[$>$]  interaction file (*.int),
\item[$>$]  name of the file containing cylindrical single-particle levels
(*.sps),
\item[$>$]  name of the file containing the su3 basis states (*.bas)
\item[$>$]  name of the file containing the ssm basis states (*.bas)
\item[$>$]  desired name for the output files (*.ham and *.ovr)
\end{itemize}

\paragraph{Eigenvectors, Expectation Values and Transition Matrix Elements.}

After generation of the Hamiltonian and operator matrices, and the overlap
matrices (if needed), which are stored in files with extensions *.ham and
*.ovr, one has to run the generalized Lanczos code ($GLanczos$) to obtain
the eigenvalues, eigenvectors, expectation values, and transition matrix
elements. There is a script file $RunGLDriver.sc$ which may be used.